\documentclass[preprint,showkeywords,preprintnumbers,amsmath,amssymb,aps]{revtex4}
\usepackage{graphicx}
\usepackage{amssymb}
\usepackage{epstopdf}
\usepackage{bm}% bold math

\begin{document}
    
\title{Feynman-Weinberg Quantum Gravity and the Extended Standard Model as a Theory of Everything}
    
\author{F.~J.~Tipler}
\affiliation{Department of Mathematics and Department of Physics, Tulane University, New Orleans, LA 70118}

\date{\today}

\begin{abstract}
I argue that the (extended) Standard Model (SM) of particle physics and the renormalizable Feynman-Weinberg theory of quantum gravity comprise a theory of everything.   I show that imposing the appropriate cosmological boundary conditions make the theory finite.  The infinities that are normally renormalized away and the series divergence infinities are both eliminated by the same mechanism.  Furthermore, this theory can resolve the horizon, flatness, and isotropy problems of cosmology.  Joint mathematical consistency naturally yields a scale-free, Gaussian, adiabatic perturbation spectrum, and more matter than antimatter.  I show that mathematical consistency of the theory requires the universe to begin at an initial singularity with a pure $SU(2)_L$ gauge field.  I show that quantum mechanics requires this field to have a Planckian spectrum whatever its temperature.  If this field has managed to survive thermalization to the present day, then it would be the CMBR.  If so, then we would have a natural explanation for the dark matter and the dark energy.  I show that isotropic ultrahigh energy (UHE) cosmic rays are explained if the CMBR is a pure  $SU(2)_L$ gauge field.  The  $SU(2)_L$ nature of the CMBR may have been seen in the Sunyaev-Zel'dovich effect.  I propose several simple experiments to test the hypothesis.
\par
\end{abstract}

%\showkeys{dark matter, dark energy, cosmological constant}
%Use showkeys class option if keyword display desired
\maketitle

\setcounter{equation}{0}

\section{Introduction}

	Can the structure of physical reality be inferred by a pure mathematician? As Einstein posed it, ``Did God have any choice when he created the universe?''  Or is mathematics a mere handmaiden to the Queen of the Sciences, physics?  Many Greeks, for instance Plato, believed that the world we see around us was a mere shadow, a defective reflection of the true reality, geometry.  But the medieval university was based on the primacy of the physics of Aristotle over mere mathematics.  Galileo, a poor mathematician, had to live on a salary of 520 ducats at the University of Padua, while Cesare Cremonini, the university's natural philosopher (physicist) had a salary of 2,000 ducats (Tipler, 1994, pp. 372--373).  Recently, mathematics has regained some of its primacy as theoretical physicists and mathematicians have struggled to determine if there is a Brane theory picked out by mathematical consistency.
	
\medskip
I shall investigate the idea that physical reality is pure number in the second section of this paper.  I shall point out that quantum mechanics --- more precisely the Bekenstein Bound, a relativistic version of the Heisenberg uncertainty principle --- implies that the complexity of the universe at the present time is finite, and hence the entire universe can be emulated down to the quantum state on a computer.  Thus it would seem that indeed the universe is a mere expression of mathematical reality, more specifically an expression of number theory, and of integers to boot.

\medskip
I shall challenge this conclusion in the third section of this paper.  I shall point out that even though quantum mechanics yields integers in certain cases (e.g. discrete eigenstates), the underlying equations are nevertheless differential equations based on the continuum.  Thus if we consider the differential equations of physics as mirroring fundamental reality, we must take the continuum as basic, not the integers.  I review the field of mathematical logic, and point out the implications for pure mathematics of taking the continuum as fundamental.

\medskip
But if we take the continuum as fundamental, we are faced with the infinities of quantum field theory, and the curvature singularities of general relativity.  I shall argue in the fourth section of this paper that taking proper account of the latter allows us to avoid the former.  In particular, I shall argue that the mathematical difficulties of the most successful quantum field theory, the Standard Model (SM) of particle physics --- all experiments carried out to date confirm the Standard Model --- naturally disappear if one requires that the SM be consistent with quantum gravity.  

\medskip
One might object that there is no consistent quantum gravity theory.  On the contrary, there is a qualitatively unique quantum gravity theory based on the continuum, on the metric of general relativity.  In fact, this theory has been in effect independently discovered by Feynman, DeWitt, and Weinberg among others, but because this theory has a ``philosophical problem'', a problem which arises from taking the integers as fundamental rather than the continuum, these great physicists did not realize that they had solved the problem of quantizing gravity.  They also did not realize that the correct quantum gravity theory is consistent only if a certain set of boundary conditions are imposed, which I shall describe.  Quantum gravity stabilizes the SM, but this stabilization forces the constants of the SM to depend on cosmic time.  Salam and Strathdee (1978) and Isham {\em et al} (1971) long ago suggested that gravity might eliminate the infinities of quantum field theory.  I shall argue that they were correct.

\medskip
Starting from the indicated boundary conditions, I shall calculate what the initial state of the universe must be.  It is, as Kelvin and Maxwell conjectured at the end of the nineteenth century, a state of zero entropy.  This unique quantum state is consistent with the Standard Model only if the only field present is the $SU(2)_L$ field of the SM.  I shall compute the solution to the Yang-Mills-Einstein equations for this unique state, and show that it naturally yields, via electroweak tunneling, more matter than antimatter, and also the correct baryon to photon ratio $\eta$.  The baryons thus generated are the source of the perturbations from which all the structure of the universe is generated, and I shall show that observed scale free Harrison-Zel'dovich spectrum arises naturally from the generated baryons.  The flatness, horizon and isotropy problems are automatically resolved given the required unique initial state.  In particular, the observed flatness of the universe is a result of the familiar quantum mechanical wave packet spreading.

\medskip	 
There remain the dark matter and the dark energy problems.  I point out that these problems have a solution if the initial $SU(2)_L$ gauge field managed to avoid thermalization in the early universe.  If it did, then necessarily this field is the Cosmic Microwave Background Radiation (CMBR), and the dark matter would be a manifestation of an interchange of energy between the SM Higgs field, and the CMBR.  The dark energy would then be the manifestation of the residual positive cosmological constant which must exist if the SM is to be consistent with general relativity.  

\medskip
It is an undisputed fact that the CMBR is Planckian; in fact, the observations show that the CMBR fits a Planck distribution with temperature $T=2.723$ K with an accuracy so great that the error bars are smaller than the width of the line in most diagrams of the data.  To my mind this level of fitness is too close to be due to a mere statistical process like thermalization.  We would expect such precision to be forced by physical law. 

\medskip
I show that this is exactly the case.  Specifically, I show that {\it any} massless classical gauge field in a Friedmann-Robertson-Walker universe necessarily obeys the Wien Displacement Law, and a quantized massless gauge field necessarily has a Planckian distribution, whatever its actual temperature, with the reciprocal of the scale factor $R$ playing the role of the temperature.  In other word, the fact the the CMBR follows the Planck distribution may indicate not that it is thermalized radiation, but instead that this radiation field is in a universe that is homogeneous and isotropic, which in fact it is.  And remember that I shall also demonstrate that quantum field theory requires the very early universe to be exactly homogeneous and isotropic.
	 
\medskip
I point out that having the CMBR be a pure $SU(2)_L$ gauge field would solve one of  the outstanding problems of cosmic ray physics, namely the existence of ultra high energy (UHE) cosmic rays.  How such particle could exist has been a mystery ever since the discover of the CMBR.  Such cosmic rays should not be able to propagate in the CMBR.  And indeed they should not --- if the CMBR were an electromagnetic field.  I show that if the CMBR were a pure $SU(2)_L$ gauge field, then UHE protons could travel through the CMBR for cosmological distances.  The CMBR {\it could} be a pure $SU(2)_L$; according to the Standard Model, the electromagnetic field is not fundamental, but composite: a mixture of the $SU(2)_L$ gauge field and a $U(1)$ gauge field.  In effect, I  am proposing that the CMBR is ``missing'' half of its field.
		
\medskip
A CMBR that is $SU(2)_L$ gauge field should have manifested itself in the Sunyaev-Zel'dovich effect.  The effect of such a CMBR on the SZE would be most prominent in the determination of the Hubble constant using the SZE.  I shall point out that there is an overlooked discrepancy in the published analyses which use the SZE observations to determine the  Hubble constant, and this discrepancy is just what one would expect if the CMBR were a $SU(2)_L$ gauge field .

\medskip
Finally, I shall describe several simple experiments to test the idea that the CMBR is a pure $SU(2)_L$ gauge field.  In particular, I shall show that observing the CMBR through a filter of 290 \AA$\,$ of graphite would yield a 39\% greater flux if the CMBR were a $SU(2)_L$ gauge field than if the CMBR is an electromagnetic field.

\section{Physical Reality as Pure Number: The Platonic-Pythag-orean Ideal}

Is the nature of the physical universe uniquely determined by the nature of pure mathematics?  Plato and Pythagoras thought so.  Plato believed that nature reflected, imperfectly, the precise and austere beauty of Euclidean geometry.  The stars moved on a perfect sphere.  The planets moved in perfect circles.

	Unfortunately for Plato, Nature did not follow his simple model.  The Sun, not the Earth, is the center of the Solar System.  The planets did not even move alone perfect ellipses around the Sun, but in chaotic orbits controlled not only by the Sun, but also by the planetsÕ mutual gravity.

	But the Platonic ideal has never lost its fascination.  Physicists have continued to deduce a Theory of Everything from considerations of mathematical beauty.  Sometimes this approach works.  Dirac derived his equation from the purely mathematical requirements of linearity, correspondence with Schr\"odinger's equation, and sameness between space and time in the order of the highest derivative.  But more often, the Platonic idea misleads physicists.  Eddington's Fundamental Theory is a notorious example.  Eddington's inference that the number of protons was a definite and unchanging integer was refuted by the discovery of antimatter --- pair creation meant that the number of protons was not constant --- and by the discovery that the reciprocal of the fine structure constant is not exactly 137, even if measured at low energy, but rather a number that depends on the energy, and possibly on the cosmological epoch.

	The Platonic ideal was a prime motivation of the development of string theory.  The hope was that the there would be a unique mathematically consistent string equation, which would have a single unique solution.  Alas, this hope has been dashed.  String perturbation theory is term by term finite, but the entire perturbation series is as divergent as the corresponding quantum field theory perturbation theory.  And as string theory was extended to M-theory, the number of solutions was also realized to be infinite.

	But in spite of these failures, there have been some notable successes in inferring the nature of reality from the structure of mathematics, or more simply, from the simple requirement of mathematical consistency.  A classic example is found in Einstein's 1905 paper ``On the Electrodynamics of Moving Bodies.''  Einstein referenced no experiments in this paper.  Instead, he pointed out that the two fundamental theories of the day, Maxwell's equations for the electromagnetic field, and Newton's equations for the motion of charged particles in an electromagnetic field, were mutually inconsistent: the former were invariant under the Lorentz group, whereas the latter were invariant under the Galilean group.  Einstein, in his Autobiography, gave a simple way to see that the Galilean group was inconsistent with Maxwell's equations.  Imagine, wrote Einstein, a plane electromagnetic wave.  Use a Galilean transformation to move to the rest frame of this wave.  In this frame, the wave should appear as a stationary sinusoidal field.  But there are no such solutions to Maxwell's equations.  

	Einstein realized that this inconsistency could not removed by modifying Maxwell's equations to make them Galilean invariant, without the resulting theory being inconsistent with experiment.  But he realized that the same was not true of Newtonian mechanics, because a modification would involve terms of the order v/c, which would be tiny in the experiments conducted up to 1905.  The rest, as they say, is history.  Imposing mutual mathematical consistency on the theories of the day was to use mathematics to understand reality.

	Another example is relativistic quantum field theory.  In the 1940's, the aging revolutionaries Einstein, Schr\"odinger, Dirac and Heisenberg claimed that only a revolutionary new way of looking at reality could combine relativity and quantum mechanics.  All four of these great physicists (even Einstein!) attempted to construct a Òfinal theoryÓ using their mathematical intuition.  All four attempts failed.  Instead, Feynman and Schwinger developed QED by reformulating quantum mechanics in a language that was consistent with special relativity.  Dyson showed that the Schwinger operator language was equivalent to the Feynman path integral language, and the latter could yield a renormalization algorithm that could yield a finite value for the S-matrix at each order in perturbation theory.  Both Weinberg (1995, p. 38), and Dyson (2004) have emphasized the conservative nature of the Feynman-Schwinger construction of QED.  Of course, Dyson's hope that QED perturbation theory could form the basis for a Final Theory was dashed by none other than Dyson himself when he showed that the perturbation series, though term by term finite (after the ultraviolet divergences were swept under the run by charge and mass renormalization) was nevertheless a divergent series.

	According to Aristotle, the Pythagoreans
\begin{quote}
É devoted themselves to mathematics, they were the first to advance this study, and having been brought up in it they thought its principles were the principles of all things.  Since of these principles, numbers are by nature the first, and in numbers they seemed to see many resemblances to the things that exist and come into being --- more than in fire and earth and water (such and such a modification of numbers being justice, another being soul and reason, another being opportunity --- and similarly almost all other things being numerically expressible); since, again, they saw that the attributes and the ratios of the musical scales were expressible in numbers; since, then, all other things seemed in their whole nature to be modeled after numbers, and the numbers seemed to be the first things in the whole of nature, they supposed the elements of numbers to be the elements of all things, and the whole heaven to be a musical scale and a number.  And all the properties of numbers and scales which they could show to agree with the attributes and parts and the whole arrangement of the heavens, they collected and fitted into their scheme; and if there was a gap anywhere, they readily made additions so as to make their whole theory coherent.  (Metaphysics A5, 985b, 24-33, 986a, 1-7, (Barnes 1984), p. 1559) 
\end{quote}

This ideal of the primacy of number is what I wish to develop in this section.  As we shall see, by ``number'' the Greeks   probably meant ``real number'' and not ``natural number'' (positive integer), but in this section I shall follow modern (post 18th century) opinion and assume that ``number'' means ``integer''.

	The physical evidence that integers are fundamental comes from quantum mechanics.  James Clerk Maxwell, in a little known article for {\it Encyclopedia Britannica,} was the first to state that energy of certain systems was quantized: that is, the energy of these systems could not be a continuous variable, but instead would have to be discrete.  In the system analyzed by Maxwell, the system energy could have only two possible values (see Tipler 1994, pp. 230-231 for a discussion of Maxwell's discovery of quantized energy).  Max Planck in 1900 and Albert Einstein in 1905 established that the energy of the electromagnetic field was quantized.

	The most general expression of the discreteness of physical reality is the information Bound established by Jacob Bekenstein (1981, 1984, 1988, 1989) and by Bekenstein and Schiffer (1989)

\begin{equation}
    I \leq \frac{2\pi ER}{\hbar c\ln2} = 2.57 \times 10^{45}\left( \frac{M}{ 1\, {\rm kilogram}}\right)\left( \frac{R}{ 1\, {\rm meter}}\right) \,\,\,{\rm bits}
\label{eq:Bekenstein}.
\end{equation}

\noindent
where $E$ is the energy enclosed in a sphere of radius $R$, and $I$ is the information contained in the quantum states in the sphere.  The $\ln2$ factor comes from defining the information as the logarithm to the base 2 of the number of quantum states.  It is assumed that the vacuum state is unique, and hence carries no information.  As we shall see, this means that we can apply the Bekenstein Bound formula even in $S^3$ closed universes {\it provided} we do not count the energy of the cosmological vacuum when we add all the contributions to the energy.  Indeed, Penrose (1982) and Tod (1983) have shown that the total energy of a closed universe is zero!  It is the effect of the cosmological vacuum that is the physical source for the total energy summing to zero: the positive energy contributions must be perfectly balanced by negative gravitational energy.  As we shall see, the vacuum energy modes are the forms of energy that are perfectly isotropic and homogeneous.  In a universe that was always perfectly homogeneous and isotropic --- Friedmann-Robertson-Walker (FRW) for its entire history --- the information content of the universe would be zero for its entire history. 
	
\medskip
An upper bound to the information content of the universe can be obtained if we assume all the non-gravitational energy in the universe is in the form of baryons, assume that the universe is at the critical density, and ignore the gravitational energy. Penrose pointed out in 1973 that putting these assumptions into the Bekenstein Bound, and choosing $R$ to be the radius of the visible universe ($\sim 10^{10}$ lyrs), one obtains $10^{123}$ bits as the upper bound to the amount of information in the visible universe at the present time.  A better estimate of the upper bound to the information content of the universe would have been obtained if just the baryonic content of the universe, 4\% of the critical density, were inserted into the Bekenstein inequality.  This would have given a number some two orders of magnitude lower than the Penrose Number, but as Penrose himself noted, it is still much too high.  We shall see why in later sections of this paper.
	
\medskip
Two years before Penrose obtained his upper bound to the amount of information in the visible universe, Carl Friedrich von Weizs\"acker argued, independently of Bekenstein and Penrose, that the universe at the present time had to have an upper bound to its information content, namely $10^{120}$ bits of informtation (Weizs\"acker 1971, p. 259; Weizs\"acker 1980, p. 218).   Weizs\"acker's number is closer to the actual upper bound required by the baryon content than is Penrose's.  Thomas G\"ornitz in a series of papers (G\"ornitz 1986, 1988a, 1988b and G\"ornitz and Ruhnau 1989) have connected the Weizs\"acker and Bekenstein cosmological bounds, and used it to obtain a new solution for the interior of black holes which explicitly eliminates the horizons inside black holes.
	
\medskip
A few points about the Bekenstein Bound should be emphasized.  First, the Bound applies in strong gravitational fields, so indeed it can be applied to the entire universe, as Penrose claimed.  But care needs to be used in the cosmological case.  If there are particle or event horizons (in the sense these terms are defined by Hawking and Ellis (1973)), then the factor $R$ is the radius of the horizon measured from the initial or final singularity respectively.  If there are no event horizons, then $R$ is the radius of the entire universe.  Second, the Bound is a non-local bound in the sense that it has an effective size tacitly built in.  It will imply, for instance, that the entire universe, if is closed and possesses horizons, can contain no information when said universe has the Planck radius.  This was first pointed out by Bekenstein (1989), who considered it an argument against the existence of the initial singularity.  However, Bekenstein has since accepted (2000, 2003, 2004) a reformulation of his Bound due to R. Bousso (1999, 2000, 2002, and 2003), which does not yield the unexpected result of zero information (and entropy) near a singularity with horizons.  I think that Bekenstein's original conclusion was correct, and that Bousso's reformulation is incorrect.  The initial singularity did indeed possess zero entropy, and further, singularities in the future must be of a very special nature, quite different from the singularities one would expect to see inside black holes in asymptotically flat spacetimes.  I shall show at length in this paper that Bekenstein original calculaltion was correct, and has some remarkable testable implications.  One should beware of reformulating a physical law so that it will yield {\it a priori} conclusions.  The most interesting implications of physical laws are precisely those which run counter to our intuitions.
	
\medskip
If the universe is closed --- I shall argue in later sections that quantum mechanical consistency requires it to be not only spatially compact but a three-sphere $S^3$ --- then the Bekenstein Bound shows that the complexity of the universe at any time to be finite.  Or more precisely, the Bound requires {\it a} universe of the {\it multiverse} to be finite at any given time.  As we shall see, there are an uncountable number of universes in the multiverse, but there are only a finite number of physically distinguishable universes in the multiverse of a given size and non-vacuum energy content.  So fixing the size and non-vacuum energy content, there must be an uncountable number of identical copies of each universe with a given information content.  For example, a universe the size of the visible universe and with the non-vacuum energy content assumed by Penrose could be in any one of $10^{10^{123}}$ possible quantum states.  (In double exponentiation, it doesn't matter if one uses 2 or 10 as the lowest base: $10^{10^{123}} \simeq 2^{10^{123}}$.)  There will be an uncountable number of identical copies of each of these $10^{10^{123}}$ universes.  As time increases, these identical copies will differentiate, but at any time there will be an uncountable number of identical copies of {\it each} possible quantum state allowed by the laws of physics.

	The Second Law of Thermodynamics tells us that the complexity of the universe increases with time.  Or, alternatively, we can use the Second Law as the definition of time direction:  time is said to increase in the direction in which entropy increases.  It is well known to physicists (e.g. Feynman 1963, pp. 46-8 and 46-9; Feynman 1995, pp. 18Ñ21) that the Second Law is not entirely a statistical phenomenon but instead the Second Law arises from the cosmological boundary conditions.  The most natural initial condition to impose on the initial state is that the entropy of the universe is zero.  This proposal goes back to Kelvin and Maxwell, but I shall show in a later section that quantum field theory requires zero entropy to occur at least at one singularity if there is an initial and a final singularity.  Zero entropy means that the information content of the universe is zero: the state of the universe is entirely determined by the laws of physics.  Since zero entropy will hold in {\it all} the universes of the multiverse, the entire multiverse will have zero entropy initially, and since unitarity implies that the von Neumann entropy is conserved, the entropy of the entire multiverse will remain zero: the multiverse is determined entirely by the laws of physics (where we regrad the universal boundary conditions as physical laws).

	But the entropy of a single universe is not required to remain zero.  Indeed, the entropy of the universe in which we find ourselves is far above zero and increasing.  If the entropy of our universe had a least upper bound, this number would be a universal constant, and its value would require an explanation.  Conversely, no explanation is required if the ``constantÕÕ takes on all of its possible values.  Since the entropy $S \geq 0$, the value of the entropy of the universe will require no explanation if $S \rightarrow +\infty$ as our universe approaches its final state, since $S$ is now increasing from its initial value of zero.

	Let me outline an argument based on the Bekenstein Bound that the entropy of our universe {\it must} diverge to infinity as its final state is approached.  Hawking showed in 1974 that if a Black Hole were to evaporate to completion, then unitarity would be violated.  (See Wald 1994 pp. 182--185 for a detailed defense of Hawking's argument.  Hawking's argument is valid even if he himself no longer believes it, as has been reported in the popular press)!)  Black holes have been seen, and if the universe were to expand forever --- either because the universe is open, or because it accelerates forever --- these observed black holes would certainly have time to evaporate to completion.  But unitarity cannot be violated, hence the universe cannot exist long enough for the observed black holes to evaporate to completion.  The universe, in other words, can exist for only a finite time in the future.  The universe must end in a final singularity after a finite proper time.  Unitarity also forces this final singularity to be all-encompassing:  all future-directed timelike curves must hit the final singularity in a finite time. The Second Law of Thermodynamics says the amount of entropy in the universe cannot decrease, but I shall show in a later section can be that the amount of entropy already in our universe will eventually contradict the Bekenstein Bound near the final singularity unless there are no event horizons, since in the presence of horizons the Bekenstein Bound implies the universal entropy $S \leq constant\times R^2$, where $R$ is the radius of the universe, and general relativity requires $R \rightarrow 0$ at the final singularity.  The absence of event horizons by definition means that the universe's future c-boundary (see Hawking and Ellis 1973 for a detailed discussion of the concept of c-boundary) is a single point, call it the {\it Omega Point}.  MacCallum has shown that an $S^3$ closed universe with a single point future c-boundary is of measure zero in initial data space.  Barrow has shown that the evolution of an $S^3$ closed universe into its final singularity is chaotic.  Yorke has shown that a chaotic physical system is likely to evolve into a measure zero state if and only if its control parameters are intelligently manipulated.  Thus life ($\equiv$ intelligent computers) almost certainly must be present {\it arbitrarily close} to the final singularity in order for the known laws of physics to be mutually consistent at all times.  Misner has shown in effect that event horizon elimination requires an infinite number of distinct manipulations, so an infinite amount of information must be processed between now and the final singularity.  Each manipulation will generate at least one bit of entropy, since each manipulation will require first observing the universe, and each (irreversible) observation will require increasing the entropy by at least one bit.  This yields $S\rightarrow +\infty$ as the final singularity is approached.  Furthermore, the amount of information stored at any time diverges to infinity as the Omega Point is approached, since the divergence of the universeÕs entropy implies the divergence of the complexity of the system that must be understood to be controlled.

	So we have obtained two divergences for the price of one!  Not only must the entropy of the universe diverge, but so must the information coded in the biosphere.  The laws of physics require progress and life to continue to the very end of time, and improve to infinity.  If the laws of physics be for us, who can be against us?

	One interesting consequence of this divergence of information coded in the biosphere I have developed at length in my book {\it The Physics of Immortality}.  The finiteness of information coded in our universe at the present time means that the entire universe at the present time could be emulated --- simulated exactly --- in a computer of sufficient power.  The divergence of information coded in the biosphere means that a computer of sufficient power will eventually exist in the far future.  I have argued in my book that life's drive to total knowledge in the far future will cause our far future descendants to carry out this emulation of their distant ancestors.  After all, we are now attempting to reproduce our ultimate biological ancestor, the first living cell from which all life on Earth is descended.  We would be the first rational beings from which all rational beings in the far future would be descended, so in reproducing us in these far future computers, life in the far future would just be learning about their history.   So the laws of physics will not only be for us in the sense of requiring the biosphere to survive, they are for us in the sense that they will eventually allow every human who has ever lived have a second chance at life.

	Notice that this ``life goes on forever'' picture really makes use only of the integers.  At any one time, the complexity of the universe is finite.  In fact, we could now be an emulation in a digital computer!  But since we have no way of reaching the computer from inside the emulation, we could just regard the emulation as fundamental.  This would mean regarding physical reality as a subset of mathematical reality.  This is the Platonic universe: physical reality is not ``real'' ultimately; only number --- the integers comprising the true ultimate reality --- is actually real.  What does mathematics tell us about this ultimate integer reality?

	To answer this question, let us first remind ourselves of a few basic notions from logic (See Jech 2003, pp. 155--157 or more details).  A {\it language} is roughly a set of symbols (which includes symbols for relations, functions and constants).  A {\it model} for a given language is a pair (A, I), where A is the collection of symbols and I is the interpretation function which maps the symbols into the relations, functions, and constants of the language.  A formula without free variables is called a {\it sentence}.  A set S of sentences is said to be {\it consistent} if a formal proof of contradiction does not exist.  The Propositional Calculus can be proven to be consistent (See chapter 5 of Nagel and Newman 1958).  A set of sentences S is called {\it complete} if for every sentence T in S, either the sentence T or its negation not-T is a theorem in the set, where as usual, {\it theorem} means that the sentence follows from a subset of ``fundamental'' sentences called {\it axioms}.  Roughly speaking, S is said to be {\it decidable} if there is an effective procedure of deciding whether any given sentence is a theorem of the set.  A (axiomatizable) complete theory is decidable (Boolos and Jeffery 1974, p. 180).  The converse is not true; a decidable (axiomatiable) theory can be incomplete (Boolos and Jeffery 1974, p. 180).

	Set theory is usually based on the Zermelo-Fraenkel or ZF axioms (see Jech 2004 for a technical description, Cohen and Hersh 1967, p. 114 for a popular introduction).  Of the ZF axioms, there are three axioms that can be regarded as controversial by physicists: first, the Axiom of Infinity, which asserts the existence of an infinite set.  Bryce DeWitt, in a famous 1973 {\it Science} review of Hawking and Ellis' {\it The Large-Scale Structure of Space-Time}, accepted the Axiom of Infinity only with great reluctance, because physicists have never observed an infinity.  The second controversial axiom, the Power Set Axiom, says that the collection of all subsets of a given set is itself a set, the power set.  The third controversial axiom, the Axiom of Choice, which asserts that it is possible to form a set made up of exactly one element from each of an infinite number of sets, DeWitt put absolutely beyond the pale.  Hawking and Ellis' book was ``deeply flawed'' according to DeWitt, because they had presented the Geroch-Choqut-Bruhat theorem that there exists a unique maximal Cauchy development from a given set of initial data.  The proof used Zorn's lemma, a form of the Axiom of Choice.  As we shall see, the Axiom of Choice has some consequences that are difficult for physicists (and even a few great mathematicians) to accept.  One problem with the Axiom of Choice is its equivalence to Zermelo's Theorem:  Every set can be well-ordered, which means that every non-empty subset of a given set has a first element (a set with its well-ordering is called an {\it ordinal}).  So in particular the set of the real numbers must have a first element.  (With the usual ordering, the real numbers have no first element.)  Set theory without the Axiom of Choice is denoted ZF, with Choice ZFC, without the Power Set Axiom ZF$^-$ and with Choice but without Power Set ZFC$^-$.

G\"odel proved three theorems of central importance in mathematical logic.  First, G\"odel's Completeness Theorem says that every consistent set of sentences has a model.  G\"odel's First Incompleteness Theorem states that Peano Arithmetic (basically the arithmetic with which we are all familiar, with addition, subtraction, multiplication, and division) and any of its consistent extensions (like ZF or ZFC), is undecidable.  That is, there are sentences in the theory that cannot be proven true or false in the theory. Peano Arithmetic is both incomplete and undecidable. G\"odel's Second Incompleteness Theorem asserts that Peano Arithmetic or ZF cannot prove its own consistency in any finite number of logical steps.

	If Peano Arithmetic cannot be proved consistent, then we have to consider the possibility that it is inconsistent.  Since it is easy to show that  {\it any} statement can be deduced from a contradiction, for instance the statement 2 = 1, an inconsistent Peano Arithmetic would have to be abandoned.  One might wonder why, if Peano Arithmetic were inconsistent, this fact has not been discovered to date.  The reason could that in physics and in every day life we really make use of only the part of Peano Arithmetic which is consistent.

	Presburger Arithmetic, which is arithmetic with addition and substraction only, is complete, decidable and consistent. See Boolos and Jeffrey Chapter 21 for a proof of decidability, and Mendelson 1964, p. 116 for a proof of completeness.  A proof of the consistency of Presburger Arithmetic seems to be unavailable in English, but in German, a proof has been given by Hilbert and Bernays (1934, pp. 359--366).  So if it is later discovered that Peano Arithmetic or the ZF axioms are inconsistent, then physicists need not lose faith in the physical laws.  It might just mean that the laws of physics would have to be reformulated in a language that uses only addition and subtraction.  Admittedly this would mean giving up differential equations and replacing them with difference equations, but such a replacement has already been done in practice in computational physics.  In many cases, computers effectively use algorithms for multiplication and division that reduce these operations to addition and subtraction.  In general, multiplication and division are introduced because Presburger Arithmetic is super-exponentially hard: if the generic problem statement requires n symbols to express, then exp(exp(n)) operations will be required to generate an answer (Machtey and Young 1981).  So if standard arithmetic --- and standard mathematics, including the familiar calculus of infinitesimals --- is found to be inconsistent, we will come to regard multiplication and division as necessary illogical evils introduced to simplify calculations, evils which will not lead to contradictions if backed up by an addition and subtraction model.

	On the other hand, the G\"odel theorems do not prove that no proof of consistency of Peano Arithmetic is possible.  The theorems merely show that a valid proof cannot be mapped into arithmetic in which sentences must be of finite length.  It might be the case, for example, that a valid proof of consistency can be obtained if we allow proof of infinite length.  To this possibility we now turn.

\section{The Empiricist Dissent}
	The Turing machine is an ideal machine representation of a human mathematician's working out of a mathematical proof via pencil and paper.  Turing believed that his machine reflected the fundamental nature of mathematical proof. However, for certain problems, like the factorization of a number into its prime factors, a classical Turing machine will require (assuming NP is not equal to P) exp(n) steps.  A quantum computer, in contrast, can find the prime factors in only n steps.  Reflecting on Turing's computer and the quantum computer, Feynman remarked, ``Turing's mistake was to think he understood pencils and paper''.  This example shows that in formulating the foundations of mathematics, human mathematicians have made tacit assumptions about the physical universe in which they do mathematics, assumptions which are build into the foundations of mathematics, but which are not an accurate reflection of ultimate physical reality.
\medskip
	Physical reality is ultimately quantum mechanical, and quantum mechanics is fundamentally a theory of linear superposition, based on the continuum.  The ``natural'' numbers, which are tacitly in the mental background when the ZF axioms are formulated (think {\it finite} number of symbols, {\it finite} number of steps allowed in an acceptable proof), are not a natural foundation at all.  Rather, it is the continuum that is the basic entity, and the positive integers a derivative quantity.  Specifically, the integers we see in the world around us --- five coins, six birds, the distinct lines of the Balmer series --- are expressions of the Exclusion Principle and the discrete eigenfunctions of the Schr\"odinger equation applied to atoms.  But the Schr\"odinger equation also has plane wave solutions, and these solutions have a continuous spectrum.   Ultimate reality is continuous, not discrete.  Discreteness --- the integers --- arises from boundary conditions imposed on an underlying continuum.
	
\medskip
	The axioms of set theory, in contrast, are tacitly based on the integers as fundamental.  The ZFC axioms consist of 9 axioms, rather than $\aleph_0$ axioms ($\aleph_0$ being the cardinality of the entire set of integers, or $2^{\aleph_0}$ axioms, $2^{\aleph_0}$  being the cardinality of the continuum.  Peano arithmetic (based on five axioms) is deduced from the ZF axioms, and Peano Arithmetic itself starts from the integers, and derives the reals, by such techniques as Dedekind cuts (a real number such as $\pi$ is the {\it set} of all rationals less than $\pi$).  As the 19th century mathematician Leopold Kronecker (1823-1891) put it ``God made the integers, all the rest is the work of man.ÕÕ

\medskip
	This idea of the integers as fundamental seems to have first arisen in the 19th century.  The ancient Greek mathematicians did not regard the integers as fundamental.  Instead, they regarded all numbers as lengths --- actually, straight line segments --- areas, or volumes, which are entities of the continuum in one, two, or three dimensions respectively.  For Euclid, a ``rationalÕÕ number was not merely a number which can be expressed as a ratio $m/n$ of two integers $m$, $n$, as we moderns would have it, but also any number of the form $(m/n)\rho$, where $\rho$ is the length of any given straight line segment, whether this length is rational or not (see Heath 1981, p. 403 for a short discussion, or Euclid, {\it Elements} Book X).  A prime number was called by the Greeks a {\it rectilinear} or {\it linear} number, meaning that it can be thought of as a straight line segment only, instead of an area like the composite number $5\times3$  (Heath 1981, pp. 72--73).  Euclid defined a prime to be a number that can be measured (is commensurate with) a unit alone (Heath 1981, Euclid {\it Elements} Book VII, definition 11 (Heath 1956, p. 278) the ``unitÕÕ being some standard but from our perspective arbitrary line segment of length $\rho$.  Euclid's development of number theory (in Books VII through X) had many diagrams of the numbers, always picturing numbers as line segments whose length were referred to a basic ``unit'' segment.  Multiplication was the formation of areas or volumes from the line segment outlining these geometrical objects.  Since there was, for Euclid, no physical and hence no mathematical meaning to a four or higher dimensional geometric object, he allowed no multiplication of more than 3 numbers.  For an example of this, see Euclid's celebrated proof of the Prime Number Theorem (that there are an infinite number of primes) in {\it Elements} Book IX, Proposition 20 (Heath 1956, p. 412).

\medskip
	For Euclid and the other Greek mathematicians, the continuum was the fundamental mathematical entity, from which other mathematical entities were derived.  Contra Plato, in the mental background there was always the idea that the physical world should form the model of the mathematical universe, as witness the later discomfort with the Parallel Postulate.  There were also some of Euclid's contemporaries who challenged his Postulates on physical grounds, for example the Atomists.  But the Atomists proposed no mathematical alternative of equal usefulness to Euclidean geometry, and in any case their atoms moved in continuous Euclidean space.  (Other examples of physical challenges by other Greek mathematicians to the Postulates of Euclid can be found in Heath 1981.)  In summary, for the Greeks, the continuum, not the integers, was fundamental, and was based on physics, even though the continuum is unobservable by humans (a point made by the Atomists, and to a lesser extent, by Zeno of the Paradoxes).  Modern physics, specifically quantum mechanics, operationally takes the same view.

\medskip
	However, in one sense, the integers were fundamental for Euclid as well as for contemporary mathematicians.  Euclid, Hilbert, and G\"{o}del allowed only a finite number of steps in a valid mathematical proof.  But we should consider whether this constraint is merely a consequence of the human inability to check a proof with an infinite number of steps rather than a constraint coming from mathematical or physical reality.  If the constraint comes from human limitations, is there then any difference between an actual infinity of steps, and a huge, but still finite, number of steps in a proof?

\medskip
	This last question came to the painful attention of the mathematical community when Thomas Hales announced he had proven Kepler's Sphere Packing Conjecture that the face-centered cubic lattice is the most efficient way to pack spheres (gives the greatest number density). Hales submitted his proof, which was of gigantic length because computers had been used in many of the steps, to the most prestigious of all mathematics journals,  {\it Annals of Mathematics}, whose editor assembled a team of 12 referees, headed by Fejes Toth.  In early 2004, Toth delivered a report to the editor that although he and the rest of the team were 99\% certain that HalesÕ proof was correct, they were not completely certain, and after five years of effort they had become convinced that they would never be certain.  The length of the proof was such that a single human could never check the proof in a ``reasonable'' amount of time.  A computer could check the proof, and certify the proof as correct, but a correct proof of what?  Perhaps indeed a proof of the Kepler Conjecture, but perhaps in reality a proof of some other theorem.  No human could be sure (Devlin 2003).  So {\it Annals of Mathematics} accepted the part of the proof that the human mathematicians had certified as valid, while the remaining part of the proof was accepted by a computer journal.  HalesÕ proof may or may not be a proof of the Kepler Conjecture, but it is a proof that human mathematicians have reached the stage where there is no practical difference between ``huge'' and infinite.  Physicists will recall George Gamow's ``One, Two, Three, --- Infinity.''
	
\medskip
	If we allow an infinite number of steps in a mathematical proof, then a proof of the consistency of Peano Arithmetic is possible.  Gerhard Gentzen provided just such a proof in 1936 (he overcame the G\"odel barrier by using transfinite induction up to a sufficiently great ordinal; see Kleene 1950 pp. 440 -- 479).   A computer cannot mechanically go through the series of steps, but a human can ``see'' the validity of the steps, provided the human accepts the necessary generalization of logic.  In Cohen's (1966, p. 41) proof of G\"odel's First Incompleteness Theorem, an indecidable statement is actually constructed, and then shown --- by an argument that cannot be mapped into arithmetic --- to be false.  So mathematicians accept arguments that cannot be arithmetized.  Nevertheless, the general reaction of most human mathematicians is that assuming the validity of transfinite induction is more dubious than simply assuming the consistency of Peano Arithmetic.
	
\medskip
	A major theoretical reason for thinking there is no fundamental difference between a finite number of postulates and a (countable) infinite number of postulates is the {\bf L\"owenheim-Skolem Theorem}:  Let M be a model for a collection T of constant and relation symbols.  Then there exists an elementary sub-model of M whose cardinality does not exceed that of T if T is infinite and is at most countable if T is finite (Cohen 1966 p. 18).  The proof of this theorem uses a weak version of the Axiom of Choice (hereafter AC); see Boolos and Jeffrey 1974, p. 133 and p. 158).  Skolem regarded this theorem as an argument that ZFC cannot form a ``reasonable'' foundation for mathematics because it implies there is a countable sub-model for the uncountable set of real numbers (Yandell 2002, p. 64).  If we want an axiom system for the real numbers that yields only the uncountable real numbers as a unique model, we will have to have an uncountable number of axioms.  If we regard the continuum as the fundamental entity in reality, and if we want the postulates giving the real numbers to yield only the real numbers, then the continuum must be governed by an uncountable number of postulates.  A finite axiom system will yield a countable number of consequences, but so will a countable number of postulates.  Our preference for a finite number of axioms may just reflect our human finiteness.  I shall argue below that a countable infinity of axioms in the form of having a countable infinity of terms in the Lagrangian (all the invariants that can be formed from the Riemann tensor and all of its covariant derivatives) allow unitarity to force the finiteness of quantum gravity coupled to the Standard Model of particle physics.  It has been known for decades that even if you start the Hilbert action for gravity, the path integral will give you the entire countable infinity of terms, and these additional terms, if artificially suppressed, will yield a quantum theory of gravity that is either non-renomalizable, or not unitary.  Conversely, if we accept quantum field theory, the fact that gravity is curvature (and recall that Cartan showed even Newtonian gravity is curvature --- see (Misner, Thorne and Wheeler 1973), and locally special relativistic, then we have to accept the countable infinity of terms in the fundamental Lagranian.  Physicists have always hoped that when the equations of the Theory of Everything were found, it could be shown that there was only one model for this system of equations, namely the actual universe.  What the L\"owenheim-Skolem Theorem demonstrates is that this hope cannot be fulfilled with a finite set of equations, or a finite set of constants, if the actual universe is actually infinite.  If we regard the boundary conditions on the universal wave function as an ``axiom'', then the boundary conditions on a continuous function will be in effect a set of axioms whose cardinality is that of the continuum.  A system with a countable number of terms in the gravitational Lagrangian and an uncountable number of ``axioms'' in the boundary conditions may, by the L\"owenheim-Skolem Theorem, have a unique (uncountable) model.

\medskip
	The cardinality of the integers is $\aleph_0$ and the continuum has cardinality $2^{\aleph_0}$.  Cantor, using his famous diagonal argument, first proved that $\aleph_0 < 2^{\aleph_0}$.  Recall that two sets are said to have the same cardinality if they can be put in one-to-one correspondence.  The {\it cardinal number} of a set is the least ordinal number that be placed in one-to-one correspondence with it (Devlin 1977, p. 8).  The ordinals thus give a infinite sequence of cardinals, represented by $\aleph_n$.  If we accept AC, then the cardinality of the continuum --- more generally, every infinite set, since, if AC holds, all sets are ordinals --- is an aleph, and further, $2^{\aleph_0} \geq \aleph_1$ (Jech 2003, pp. 47--48).

\medskip
	How many cardinals are there?  The Power Set Axiom generates an infinite hierarchy $2^{\aleph_n}$.  What is the relationship between the alephs and this hierarchy?  Cantor conjectured that $2^{\aleph_0}= \aleph_1$; this is called the {\it Continuum Hypothesis} (CH).  More generally, the {\it Generalized Continuum Hypothesis} (GCH) asserts that $2^{\aleph_n}= \aleph_{n+1}$.  G\"odel in the 1930Õs showed that ZF was consistent with CH, GCH, and AC by showing that if one restricted attention to sets that were, roughly speaking, generated by the ZF axioms, then one could prove CH, GCH, and AC.  Cohen in the 1960Õs, by inventing a remarkable new technique called ``forcing,ÕÕ constructed a model for ZF in which $2^{\aleph_0} = \aleph_2$, in contradiction to CH (Cohen 1966; Jech 2003, pp. 219--220).  Together, the G\"odel-Cohen theorems showed that both CH and GCH were independent of the ZF axioms; that is, accepting ZF (or ZFC) allowed one to either accept CH and GCH or deny either or both.  Mathematicians have been divided on the CH ever since.

\medskip
	It is important to note that the two greatest mathematical logicians of the 20th century, Kurt G\"odel and Paul Cohen, disbelieved in CH.  G\"odel wrote `` $\dots$one may on good reasons suspect that the role of the continuum problem in set theory will be this, that it will finally lead to the discovery of new axioms which will make it possible to disprove Cantor's conjectureÕÕ (G\"odel 1947, p. 524).  Cohen agreed with G\"odel in his 1966 book: ``A point of view which the author feels may eventually come to be accepted is that CH is {\it obviously} false [Cohen's emphasis].  The main reason one accepts the Axiom of Infinity is probably that we feel it absurd to think that the process of adding only one set at a time can exhaust the entire universe.  Similarly with the higher axioms of infinity.   Now $\aleph_1$ is the set of countable ordinals and this is merely a special and the simplest way of generating a higher cardinal.  The [continuum] is, in contrast, generated by a totally new and more powerful principle, namely the Power Set Axiom.  It is unreasonable to expect that any description of a larger cardinal which attempts to build up that cardinal from ideas deriving from the Replacement Axiom can ever reach [the continuum].  Thus [the cardinality of the continuum] is greater than $\aleph_n$, $\aleph_\omega$, $\aleph_\alpha$, where $\alpha = \aleph_\omega$ etc.  This point of view regards [the continuum] as given to us by one bold new axiom, which can never be approached by any piecemeal process of constructionÕÕ (Cohen 1966, p. 151).  Cohen expressed the same opinion about CH and the cardinality of the continuum in (Cohen 2002, p. 1099).

\medskip
	Mathematicians who accept the GCH often argue for the {\it Axiom of Constructability}: the only sets mathematics really needs to use --- and hence according to the Constructability Axiom are the only sets that exist --- are the sets which G\"odel generated from a restricted set of the ZF axioms: recall from the above discussion that if these generated sets are the only sets allowed in mathematics, then one can prove that CH, GCH, and AC hold.   One problem with the Axiom of Constructability is that the Axiom of Choice implies that there are subsets of the real line that are not Lebesque measurable:  The Banach-Kuratowski Theorem (Jech 2003, p. 133) states that if there is a measure on the continuum, then $2^{\aleph_0} > \aleph_1$.  Physicists routinely assume that all subsets of the continuum that appear in calculations are measurable.  This is one of the reasons why DeWitt was not willing to accept the Axiom of Choice.  Myclelski (2003) has written a recent review giving reasons contra DeWitt why AC and GCH should be accepted.

\medskip
	For mathematicians disinclined to accept CH, GCH, and AC, there are the {\it Large Cardinal Axioms} (Kanamori 1994; Woodin 1994), which, as the name implies, assert the existence of ``very large'' infinite sets.  Perhaps the simplest example of a Large Cardinal Axiom is the Axiom that an uncountable ``strongly inaccessible'' cardinal exists.  Roughly, an ordinal number is {\it inaccessible} if it is not the successor of some other ordinal.  For example, the ordinal zero is strongly inaccessible, because it is the first non-negative integer in the usual ordering of the non-negative integers.  The least infinite ordinal $\omega$, the ordinal of the set of non-negative integers, is also inaccessible (the cardinal number of $\omega$ is $\aleph_0$.)  The implications of quite a few Large Cardinal Axioms have been studied.  

\medskip
	There is even a connection between the Large Cardinal Axioms and the Theory of Games.  Consider a game $G_C$ with two players Mr. A and Ms. B.  First, A chooses a natural number $a_0$, then B chooses a natural number $b_0$, then A chooses another natural number $a_1$, followed by B's choice of $b_1$ and so on until $\omega$ steps have been played.  If the sequence $(a_0,b_0,a_1, \ldots)$ is in a pre-selected subset $C \subset \omega^\omega$, Mr. A wins, otherwise Ms. B wins.  A rule telling a given player what move to make depending on the moves previously made is called a ``strategy''.  A ``winning strategy'' for a given player is a strategy that makes a win certain, whatever the other player does.  The game $G_C$ is said to be {\it determined} if one of the players has a winning strategy.  The {\it Axiom of Determinacy} (AD) asserts that for every set $C \subset \omega^\omega$, the game $G_C$ is determined. (Jech 2003, p. 627).

\medskip
	Is AD a ``reasonable'' axiom?  It is inconsistent with AC (though it does imply a weak version of AC). AD does imply that all subsets of the reals are Lebesgue measurable, a desirable feature for physics.  

\medskip
A physicist is left with the impression that all of this recent set theory work lacks any sort of an anchor in physical reality.  Indeed, many set theorists boast of their disconnect from physics; ``{\it mathematical truth is what we have come to make of it.} As for knowledge, description ultimately provides no insights beyond description. $\ldots$ it may be that we cannot say anything other than that {\it the acquisition of mathematical knowledge may be just what happens''} (Kanamori 1994, p. 480; his emphasis).  What an unfortunate change from the relationship between mathematics and physics for most of human history!  Prior to the mid 19th century, there was no distinction between theoretical physicists and mathematicians.  The change began with the application of Fourier analysis to construct functions with no possible application in physical reality with its use of piecewise differentiable functions.  Weierstrass, using Fourier series, constructed a function on $[0,1]$ that was everywhere continuous but nowhere differentiable.  It was no coincidence that an expert on Fourier analysis, Georg Cantor, created set theory and the infinite hierarchy of infinite cardinal numbers.

\medskip
	I have argued that physics assumes the continuum, but is any higher infinity required?  The Power Set Axiom generates an infinite hierarchy, but {\it physically} only the continuum appears.  Physics deals with the set of all functions defined on n-dimensional space, and in general this set would be the set of all subsets of n-dimensional space.  But if we impose the condition that the only allowed functions are continuous --- if all the curvature derivatives are present, as required by quantum theory, the only allowed functions are only continuous but $C^\infty$ --- then the cardinality of this restricted set of functions is the cardinality of the continuum, since a continuous function is completely determined by its values at the rational points in its domain.  This will not change in the path integral.  So we have no justification from physics for any set of higher cardinality than the continuum.  If we accept physics as our guide to mathematical reality, the Power Set Axiom must be abandoned.  Physics begins and ends with the continuum.  It is possible that the continuum should not be thought of as a set, but as a class.  But even in ZF there are interesting classes that are not sets.  The class of all ordinal numbers is not a set, for example (Dugundji 1970, p. 43).  Or it may be that we should reformulate the Power Set Axiom to yield the continuum, but not the infinite hierarchy beyond the continuum.  My intuition is the same as Paul Cohen's: that the continuum is larger than any of the alephs.  For me, the continuum plays the same role as Cantor's Absolute Infinity.  I shall argue that physical reality of the continuum is manifested in several ways.  I shall show that physics requires the cosmological singularity to exist and have the cardinality of the continuum.  The cardinality of the universes in the multiverse is the cardinality of the continuum.  And all 25 constants in the (extended) Standard Model of particle physics assume all possible values of the continuum somewhere in the multiverse.

\medskip
	If reality is based on the integers, we would expect to find the integers 1 or 2 at the base of reality.  On the other hand, if the continuum is basic, we would expect to find irrational, transcendental but fundamental continuum numbers like $\pi$ and $e$ at the foundation of nature.  In the next section, I shall suggest a way in which we might determine which set of numbers Nature has decided to use as its starting point.

\section{The 25 ``Constants'' of the Particle Physics (Extended) Standard Model}

The main objections to the Standard Model (SM) as the ultimate theory of nature as follows:
\begin{enumerate}
\item{The Standard Model does not include gravitational forces}
\item{The Gauge Hierarchy Problem:  The SM Higgs mass must be less than a few hundred GeV, and the largest Yukawa coupling constant is that of the top quark, with a mass of about 180 GeV.  These masses are very much less than the Planck mass of $10^{19}$ GeV}
\item{The Triviality Problem (Callaway 1988): The mutual interaction of the coupling constants, which should cause the values to be pulled up to the Planck mass or beyond}
\item{The presence of 25 parameters, which are not given by theory, but which must be determined by experiment.  These parameters are}
\begin{enumerate}
\item{6 lepton masses (Yukawa coupling constants)}
\item{6 quark masses (Yukawa coupling constants)}
\item{4 lepton mixing angles}
\item{4 quark mixing angles}
\item{ 2 Higgs fields constants ($v = 246.22$ GeV, and the Higgs mass $m_H$, say)}
\item{3 gauge coupling constants.}
\end{enumerate}

\end{enumerate}

Before the discovery of non-zero neutrino masses and neutrino mixing, the 3 neutrino masses and the 4 lepton mixing angles were set equal to zero (this restriction gives the (un-extended) Standard Model, so this would give 18 parameters to be discovered by experiment.  But if the Standard Model is correct, and there is nothing beyond the (extended) Standard Model, then one of the quark masses must actually be zero, most likely the up quark mass $m_u$.  For only if at least one of the quark mass is zero will there be a SM mechanism to eliminate the Strong CP Problem.  A zero up quark mass has been proposed as the solution to the Strong CP Problem for over 30 years, and refutations of this resolution have been published almost as fact as new variations of the proposal.  However, all the refutations have been based on approximations to the SU(3) color force, and not on exact calculations based on quantum chromodynamics itself.  In view of the recent acknowledgement of the difficulty in calculating the correct value of the QCD correction to the muon g--2 factor, I think we cannot rule out a zero up quark mass.  As I said, a zero quark mass is the only solution to the Strong CP Problem if the SM is correct.  (I shall discuss this point in more detail later.)  The only quark mass which could be zero is the up quark mass.  By symmetry between the lepton and quark sectors, I would expect the electron neutrino, the analogue of the up quark, to have zero mass also.  If these two masses are zero, the SM would have 23 ``free'' parameters.

\medskip
	The word ``free'' is in quotes because as is well-known, the values of the 23 parameters are not completely arbitrary.  If the parameters were set at too large a value, the SM would become inconsistent.  For example, if the value of the Higgs mass (given the top quark mass at about 180 GeV) were larger than about 200 GeV, the Higgs vacuum would become unstable.  However, if the universe were to end in a final singularity, and if we were sufficient close to the final singularity, then this instability would not matter because the universe would end before the Higgs vacuum instability could manifest itself.

\medskip
	And the closer we are to the final singularity, the larger the Higgs mass --- and indeed all the coupling constants can be before the instability can be realized.  There is no limit to this:  for any possible values of the coupling constants, there will be a time sufficiently close to the final singularity so that the SM will be consistent for all times before the universe ends in the final singularity.  If we imagine that the parameters do indeed get larger with cosmic time, and there is indeed a final singularity near which the coupling constants become arbitrarily large, then notice that the Gauge Hierarchy Problem disappears.  There will be a time in the far future at which the particle masses are comparable in magnitude to the Planck mass, and beyond the Planck mass.  This suggests that perhaps SM problems 2 and 3 are connected.

\medskip
	One of the cosmological parameters recently determined by the WMAP observations is the Hubble constant $H_0$.  But of course the Hubble parameter is not a constant.  It varies with cosmological epoch, having the value $+\infty$ at the initial singularity, and the value $-\infty$ at the final singularity.  The value $H_0$ = 71 km/sec-Mpc  (Spergel {\em et al} 2003) is a measure of our epoch of cosmological history rather than a constant of nature.  The Hubble parameter H takes on all possible values at some point in universal history. Perhaps the ÒconstantsÕ of the SM are the same: not constants, but parameters taking on all possible values allowed by mathematical consistency over the whole of universal time and over the entire multiverse.  In this case, the solution to SM model problem 4 would be the same as the solution to problems 2 and 3.

\section{Finite Quantum Gravity}

Most high energy physicists and general relativists firmly believe that there is no quantum theory of gravity.  In fact, Feynman (1995) in effect discovered in 1962 that quantum field theory yields a qualitatively unique theory of quantum gravity.  I shall show in this section that with the appropriate boundary conditions, this unique theory of quantum gravity has a perturbation series that is not only term by term finite, but further the same mechanism that forces each term in the series to be finite also forces the entire series to converge.

\medskip
	All attempts to quantize gravity within quantum field theory treat gravity as a spin 2 field $h_{\mu\nu}$ and generally start with the Hilbert action

\begin{equation}
S = \int d^4x\, \sqrt{-g}(\Lambda + \frac{1}{8\pi G}R)
\label{eq:HILBERT},
\end{equation}

Anyone who starts with this action immediately soon sees that it does not yield a consistent quantum field theory.  The fact that the coupling constant $1/G$ has dimensions of mass$^{-2} = 1/M_{Pk}^{-2}$ implies that the theory is non-renomalizable.  Cancellation of divergences requires the presence of derivatives of $h_{\mu\nu}$ higher than second order in the Lagrangian.  One also realizes that path integral quantization --- a sum over {\it all} histories --- tends to require all terms that are consistent with the symmetries of the theory.  In general relativity, the symmetries are the group of general linear transformations, and not only is the Ricci scalar $R$ invariant under this group, but so is $R^2$, $R^2$ $\dots$ $R^n$ $\ldots$, and so are all invariants formed from the Riemann tensor, and powers of these invariants,and all invariants formed from all covariant derivatives of the Riemann tensor.

\medskip
	So basic quantum field theory quickly forces upon us the general invariant action
	
 \begin{equation}
S = \int d^4x\, \sqrt{-g}(\Lambda + \frac{1}{8\pi G}R + c^2_1R^2 +c^3_1R^3 \ldots + c^2_2R_{\mu\nu}R^{\mu\nu} + \ldots + c^3_1R_{\mu\nu;\alpha}R^{\mu\nu;\alpha} + \ldots)
\label{eq:gravS},
\end{equation}

This is the qualitatively unique gravitational Lagrangian picked out by quantum mechanics.  Physicists don't like it because (1) it has an infinite number of (renormalizable) constants $c^i_j$, all of which must be determined by experiment and (2) it will {it not} yield second order differential equations which all physicists know and love.  But the countable number of constants are in effect axioms of the theory, and I pointed out in an earlier section that the L\"owenhein-Skolum theorem suggests there is no real difference between a theory with a countable number of axioms and a theory with a finite number of axioms.  The finite case is just easlier for humans to deal with, provided the ``finite'' number is a small number.  Further, as Weinberg (1995, p. 499, pp. 518--519) has emphasized, this Lagranian generates a quantum theory of gravity that is just as renormalizable as QED and the Standard Model.

\medskip
	Since quanum field theory itself is forcing the Lagrangian (\ref{eq:gravS}) on us, I propose that we accept the judgment of quantum mechanics and accept (\ref{eq:gravS}) (and the countable number of additional terms involving the non-gravitational fields interacting with the $h_{\mu\nu}$ as the actual Lagangian of reality.

\medskip
	Donoghue (1994, 1996) has shown that Lagrangian (\ref{eq:gravS}) will not contradict experiment provide the (renormalized) values of the infinite number of new coupling constants are sufficiently small.  For a given interaction, a Feynman diagram of order n will behave at large energy like $\int E^{A - nd}\,dE$, where $A$ depends on the interaction, but not on $n$, and $d$ is the dimensionality of the interaction (Weinberg 1979).  For the $R$ term of gravity, $ d= -2$, so the Feynman diagrams for this process will grow with the cut-off $\Lambda_{cut}$ as $\Lambda_{cut}^{A +2n+1}$.  The key question is, is there a physical mechanism to fix  $\Lambda_{cut}$ at some value less than $+\infty$, and in a relativistically invariant way?

\medskip
	I propose the following mechanism to fix $\Lambda_{cut}$.  Hawking showed in the early 1970's that {\it if} trapped surfaces were to form and event horizons were to develop around them, then in the semi-classical approximation, these black holes would violate unitarity if they were to evaporate to completion.  And they would so evaporate if the trapped surfaces were sufficiently small.  A sufficient condition (though perhaps not a necessary condition) to avoid such a violation of unitarity is to impose boundary conditions on the universe that prevents ``small'' trapped surfaces from forming, except near the final singularity, where the universe will end before a black hole will have time to evaporate completely.  Consider what is implied by the integral over a loop in a Feynman diagram.  As the energy in the loop is increased to infinity, the spatial region in which the energy is concentrated decreases to zero.  At the Planck energy, the energy would be concentrated in sphere with the Planck radius, thus forming a Planck-size black hole.  By assumption, this cannot happen, so there is a natural upper bound to $\Lambda_{cut}$, namely $\Lambda_{cut} < M_{Pk}$.  Notice that since a {\bf No-Mini-Trapped-Surface (NMTS) condition} is relativistically invariant, the value of $\Lambda_{cut(t)}$ will depend on the coordinate system in a way that will leave the $U$ matrix invariant.  Also, notice that this condition is time dependent.  Sufficiently close to the final singularity, $\Lambda_{cut(t)}$ is allowed to increase to $+\infty$.  As I have said, Hawking's calculation is only at the semi-classical level, ignoring the higher order terms in (\ref{eq:gravS}).  So we cannot fix $\Lambda_{cut(t)}$ even at the present time.  But  some consequences are easy to calculate.  For instance, a $\Lambda_{cut(t)}$ that depends on the cosmological time $t$ means that the renormalized particle masses and coupling constants change with cosmological time.
	
	For example, recall that the running value at one-loop of the fine structure constant $\alpha(\mu^2)$ measured at energy $\mu$ depends on $\Lambda_{cut}(t)$ as	
	
\begin{equation}
\alpha(\mu^2,t) =   \frac{\alpha_0}{1 + {\frac{\alpha_0}{3\pi}}\ln\left({\frac{\Lambda^2(t)}{\mu^2}}\right)}
\label{eq:A}
\end{equation}	

This gives the values of $\alpha(\mu^2,t)$ at two different times $t_{then}, t_{now}$ as

\begin{equation}
\alpha(\mu^2,t_{then}) =   \frac{\alpha(\mu^2, t_{now})}{1 + {\frac{\alpha(\mu^2, t_{now})}{3\pi}}\ln\left({\frac{\Lambda^2(t_{then})}{\Lambda^2(t_{now})}}\right)}
\label{eq:B}
\end{equation}	

	 Let us assume that some power of $\Lambda_{cut}(t)$ is proportional to the proper time difference between the current cosmological epoch $t_{now}$ and the the total length of time $t_{total}$ which our particular universe of the multiverse will exist:
	 
\begin{equation}
t_{total} - t_{now} = A{\left({\frac{\Lambda_{cut}(t_{then})}{M_{Pk}}}\right)^n}
\label{eq:C}
\end{equation}	

For example, the lifetime of a black hole $t_{BH} \sim t_{Pk}(M/M_{Pk})^3$, so we might expect the cut-off mass arising from the No-Mini-Trapped-Surface Condition to scale the same way.  Then, assuming that $t_{now}/t_{total} \ll 1$ and $t_{then}/t_{total} \ll 1$, we obtain

\begin{equation}
\frac{\alpha(\mu^2,t_{then})}{{\alpha(\mu^2, t_{now})}} =   \frac{1}{1 + {\frac{\alpha(\mu^2, t_{now})}{3\pi}}\left[\left({\frac{2}{n}}\right)\left({\frac{t_{now} - t_{then}}{t_{total}}}\right)\right]}
\label{eq:D}
\end{equation}

This gives for the change $\Delta\alpha$ of the fine structure constant:

\begin{equation}
\frac{\Delta\alpha}{\alpha} \equiv \frac{\alpha(\mu^2,t_{then}) - \alpha(\mu^2, t_{now})}{{\alpha(\mu^2, t_{now})}} \simeq  -  {\frac{\alpha(\mu^2, t_{now})}{3\pi}}\left[\left({\frac{2}{n}}\right)\left({\frac{t_{now} - t_{then}}{t_{total}}}\right)\right]
\label{eq:E}
\end{equation}	

Barrow {\em et al.} (2001) (but see (Uzan (2003) for a recent review) have apparently measured $\Delta\alpha = - 0.72 \pm \times 10^{-5}$ for $\Delta t \sim 10^{10}$ years.  Putting this into (\ref{eq:E}) and assuming that $n\sim3$ would give a total life time of the the universe of $t_{total} \sim 10^{12}$ years, which is consistent with the measured degree of flatness,  but quite lower than the predictions I made in (Tipler 1994) based on the divergence of entropy near the final singularity.  Recall that a divergence of entropy (or for that matter the validity of the Second Law of Thermodynamics) is consistent with quantum field theory (the Bekenstein Bound) only if horizons disappear near the final singularity, and this in turn is likely only if life continues to exist all the way into the final singularity.  But life can survive only if its information is transferred to a medium that can survive arbitrarily high temperature.  I argued in (Tipler 1994)  that such a transfer requires the universe to exist for much longer than $10^{12}$ years.  Indeed, I predicted (Tipler 1994, p.402) for our universe

\begin{equation}
10^{17}\, {\rm years} < t_{total} < 10^{19} \,{\rm years}
\label{eq:TipUnivage}
\end{equation}

Damour (2003) has recently argued that the Barrow {\em et al} value for $\Delta\alpha/\alpha$ is much too high.  It is are especially too high if the theory described here for the origin of the constant variation is correct, because in this theory, {\it all} the constants vary with cosmic time.  (Recall that only by having the constants take on all their possible values can one solve the problem of why the 25 constants have the values they do in the context of the SM.)  In particular, the Oklo natural nuclear reactor (Damour and Dyson 1996; Shlyakhter 1982) places a very strong constraint on the variation on the strong force, which I shall assume is essentially the same as the variation of the color force coupling constant $\alpha_3$.  The variation of $\alpha_3$ can be obtained from (\ref{eq:B}) via the replacement

\begin{equation}
\frac{\alpha(\mu^2)}{3\pi} \rightarrow  \frac{\alpha_3}{4\pi}\left[\frac{2}{3}n_f - 11\right]
\label{eq:alpha3}
\end{equation}	

\noindent
where $n_f$ is the number of quark flavors the the energy $\mu$ ($n_f = 2$ for $\mu < 100\,{\rm MeV}$), and $\alpha_3 \sim 1/7$ for $\mu < 1\,{\rm GeV}$.  The Oklo data gives $\Delta\alpha_S/\alpha_S < 2 \times 10^{-10}$ for $\Delta t \sim 2\times 10^9$ years.  Setting $\Delta\alpha_S/\alpha_S = \Delta\alpha_3/\alpha_3$ yields, when we combine  (\ref{eq:E}) and  (\ref{eq:alpha3}):

\begin{equation}
t_{total} > 10^{18}\, {\rm years}
\label{eq:Univage}
\end{equation}	

Inequality (\ref{eq:Univage}) is consistent with my earlier estimate (\ref{eq:TipUnivage}), and these two inequalities together show that a slight improvement in the precision of measurement of $\Delta\alpha_3/\alpha_3$ should show some variation of $\Delta\alpha_3/\alpha_3$, if the theory which I have developed above is correct.

\medskip
Notice that the running coupling constant equation (\ref{eq:A}) contains the bare coupling constant $\alpha_0$, which is not directly measurable, but which must be non-zero.  This constant is the true fundamental constant, independent of both cosmic time, and universe of the multiverse.  If reality is ultimately based on integers, we would expect this true fundamental constant to be either 1 or 2.  If on the other hand the continuumis fundamental, we would expect it to be either $\pi$ or $e$.  In principle, this bare coupling constant could be measured, and we would then know whether reality is digital or continuous.  

	Having a finite cut-off makes the standard perturbation series for the $U$ matrix term by term finite.  (``$U$ matrix'' since the S-matrix will not exist in a closed universe.  However, it has been known for more than 50 years that the S-matrix series diverges.  What happens is that when the  S-matrix is expanded in power of $\alpha$, the number of Feynman diagrams increases with the order $n$ very rapidly.  Hurst (1952a,b) and Riddle (1953) estimated that the number of Feynman diagrams in QED increases as $(n!)^{3n/2}$ for large $n$. Thirring (1953) and later Decalan and Rivasseau (1982) analyzed a similar enormous growth rate in the scalar field.   Dyson (1949) used the approximation $n =137$ for the order at which the QED series terms started getting larger.  Rosensteel {\em et al} (1976) estimate that the number $N$ of nth order diagrams for a process with only 2 external lines satisfies
	
\begin{equation}
(2n+1)!! < N < 4(2n+1)!!
\label{eq:Rosensteel}
\end{equation}	

All of these estimates imply that the S-matrix series must diverge.  If we ask when $\alpha^n(2n+1)!! = 1$, the answer is at $n = 180$, at which order there will be about $10^{384}$ irreducilbe Feynman diagrams.   With Dyson's estimate of $n = 137$, there would be $(2(137)+1)!! \sim 10^{277}$ irreducible diagrams.  When one considers that for QED vertex diagrams there are 7 diagrams at 2 loops, 72 diagrams at 3 loops, 891 diagrams at 4 loops, and 12,672 diagrams at 5 loops, and that the evaluation of the two loop diagrams was not completed until 1957 Kinoshita 2003), the three loop diagrams took until 1996, and we have not yet completely evaluated the four loop diagrams (Hayes 2004), there is no immediate worry that we will reach the point at which the S-matrix series begins to diverge.  In fact, the Bekenstein Bound of $10^{123}$ bits allowed in the visible universe today shows that it is impossible in principle to evaluate Feynman diagrams to the order required to see the S-matrix sum begin to grow, at least in the current epoch of universal history.  It is only near the final state of the universe, where the disappearance of event horizons will allow an unlimited amount of information to be stored, that all the Feynman diagrams of the 180th order and above could be evaluated. Nevertheless, the QED S-matrix perturbation series could converge only if the magnitude of the typical nth order Feynman diagram dropped more rapidly than than $\alpha^n(2n+1)!! $, which is impossible in the usual renormalization scheme.  It is also impossible in string perturbation theory: Gross and Vipul (1988) have shown that the number of irreducible string loops increases faster than $n!$.

\medskip
But it {\it might} be possible using the mechanism I have described for obtaining the ultraviolet cut-off.  In my proposal, the NMTS condition restricts the amount of energy in a small volume.  At n-loops, {\it all} diagrams at n-order will superimpose, necessitating a reduction of the cut-off at each order.  Since there are $\sim(2n+1)!!$ loops at nth order, the cut-off should be reduced so that each diagram would be reduced by a factor of $\sim[(2n+1)!!]^{-1}$, cancelling the increase in the number of diagrams, with the result that the series expansion in the fine structure constant $\alpha$ becomes a convergent power series.  Of course, this means that there are an infinite number of new constants --- one at each order --- but I have already argued above that quantum gravity will itself require an infinite number of new constants.  (The usual infinite number of equations for unitarity will give some constraints on the cut-offs.  Conversely, the infinite number of new constants guarantee that unitarity can be satisfies for the U-matrix.)  The crucial point, however, in both cases is that the new constants will introduce new measurable physics only near the final singularity.  We physicists must keep in mind that the only constraints we may reasonably impose on a physical theory is (1) it must agree with all experiments we are capable of carrying out, and (2) the theory must be consistent.  

\medskip
It is instructive to revisit Dyson's (1952) analysis of the physical origin of the QED perturbation series divergence.  Dyson pointed out that if the series in powers of $\alpha$ converged, then a series formed by the replacement $\alpha \rightarrow -\alpha$ should also converge, since the new series is just an alternating series formed from a convergent positive terms power series.  Dyson then considered the physics of the alternate universe in which $-\alpha$ was negative.  He showed that the vacuum of such a world was unstable (see Gibbons and Rasheed 1996 for a more recent discussion in which gravitational effect are considered in the Dyson alternate universe).  Dyson therefore conjectured that the divergence of the QED S-matrix power series was a manifestation of a problem with the choice of the QED vacuum.  Gross and Vipul (1988) conjectured that it was the ambiguity in the choice of the string vacuum that was the ultimate cause of the divergence of string perturbation theory.

\medskip
I think Dyson, and Gross and Vipul are correct.  Notice first that Dyson's alternate world cannot physically exist in the sense that my proposed cut-off mechanism will not work in his alternate universe.  The replacement $\alpha \rightarrow -\alpha$ would result in either naked singularities or mini-black holes (Gibbons and Rasheed 1996), either of which would violate the NMTS condition.  Furthermore, the full quantum gravity Lagrangian {\it assumes} tacitly that the Hilbert term is the dominant term, and that the higher order terms will not come into play until very high energies are reached near the final singularity.  This effectively means that the classical universe of the Hilbert action will be the ground state of physical reality, which is to say, the true vacuum state of the full Lagrangian is the classical closed universe.  The universe of Minkowski space is a good approximation to the classical closed universe locally, but at high energies --- or when high order terms become important --- the true vacuum state must be used.  I remind the reader again that the Penrose quasi-local energy naturally yields a value of zero for the total energy of a closed classical universe.  Actually, the true classical closed universe will be a multiverse since all quantum objects are multiverses.  Let me now construct the unique classical closed universe multiverse.

\section{The Quantized FRW Universe}

\subsection{Quantizing the FRW Universe in Conformal Time} 

Since I have argued above that the classical closed universe multiverse requires only the Hilbert action, I am allowed to quantize the universe using the Wheeler-DeWitt equation.  (This refutes the anti-Wheeler-DeWitt equation argument expressed in DeWitt (1999, 2003).   The Wheeler-DeWitt equation is the equation for the vacuum state of the universe.) I shall
construct a quantized FRW universe in which the
only field is a gauge field (actually a perfect fluid for
which $p = \rho/3$) and show that imposing the
boundary condition that classical physics hold exactly
at ``late times" (any time after the first minute)
implies that classical physics is good all the way into
the initial singularity.

\medskip
Recall that the Wheeler-DeWitt equation is 

\begin{equation}
\hat {\cal H}\Psi = 0
\label{eq:WheeDeW}
\end{equation}	

\noindent
where $\hat{\cal H}$ is the super-Hamiltonian
operator.  This operator contains the equivalent of
time derivatives in the Schr\"odinger equation.  I say
``the equivalent'' because the Wheeler DeWitt equation does not
contain time as an independent variable.  Rather, other
variables --- matter or the spatial metric --- are used
as time markers.  In other words, the variation of the
physical quantities {\it is} time.  Depending on the
variable chosen to measure time, the time interval
between the present and the initial or final
singularity can be finite or infinite --- but this is
already familiar from classical general relativity.  As I shall so later, in
the very early universe, conformal time measures the
rate at which particles are being created by instanton
tunnelling, that is it measures the rate at which new
information is being created.  Therefore, the most
appropriate physical time variable is conformal time,
and thus we shall select an appropriate combination of
matter and spatial variables that will in effect result
in conformal time being used as the fundamental time
parameter in the Wheeler-DeWitt equation.  Conformal
time is also the most natural physical time to use for
another reason:  I shall argue below that the matter in the early universe
consists entirely of the SM $SU(2)_L$ gauge field, and the
Yang-Mills equation is conformally invariant; a gauge
field's most natural time variable is conformal time.

	Since the Bekenstein Bound tells us that the
information content of the very early universe is zero, this
means that the only physical variable we have to
take into account is the scale factor $R$ of the
universe, and the density and pressure of the gauge
field.  So we only have to quantize the FRW universe for 
a radiation field, or equivalently, a perfect fluid for
which $p = \rho/3$.

	If matter is in the form of a perfect fluid, the Hilbert action
$S$ in the ADM formalism can be written

\begin{equation}
S = \int (R + p)\sqrt{-g}\, d^4x = \int L_{ADM}\,dt
\label{eq:ADMaction}
\end{equation}	

\noindent
where $p$ is the fluid pressure and $R$ is the
Ricci scalar as before.  If the spacetime is assumed
to be a Friedmann universe containing isentropic perfect
fluids, Lapchinskii and Rubakov (1977) have shown the
canonical variables can be chosen $(R,\phi,s)$, where
$R$ is now the scale factor of the universe, and $\phi,s$
are particular parameterizations of the fluid variables
called {\it Schutz potentials} (Schutz 1971).  (I am using the same symbol for the Ricci scalar and the scale factor of the universe, but hopefully this will not lead to confusion.)  The momenta
conjugate to these canonical variables will be written
$(p_R,p_\phi, p_s)$.

	The ADM Lagrangian in these variables can be shown to
be

\begin{equation}
L_{ADM} = p_RR' + p_\phi \phi' + p_ss' - N(H_g + H_m)
\label{eq:ADML}
\end{equation}

\noindent
where the prime denotes the time derivative,

\begin{equation}
H_g = - \frac{p^2_R}{24R} - 6R
\label{eq:ADMHG}
\end{equation}	

\noindent
 is the purely gravitational super-Hamiltonian, and
 
 \begin{equation}
H_m = N^2R^3[(\rho + p)(u^0)^2 +pg^{00}] = p^\gamma
_\phi R^{3(1-\gamma)} e^s
\label{eq:ADMHM}
\end{equation}

\noindent
is both the coordinate energy density measured by a
comoving observer and the super-Hamiltonian of the
matter.  The momentum conjugate to $R$, the scale
factor of the universe, is
 
 \begin{equation}
p_R = -\frac{12RR'}{ N}
 \label{eq:pR}
\end{equation}	

The constraint equation for the Friedmann universe is 
obtained by substituting (\ref{eq:ADML}) through (\ref{eq:pR}) into (\ref{eq:ADMaction})
and varying the lapse $N$.  The result is the
super-Hamiltonian constraint:

 \begin{equation}
0 = {\cal H} = H_g + H_m = -\frac{p^2_R}{ 24R} -6R +
p^\gamma_\phi R^{3(1-\gamma)}e^s 
 \label{eq:Hconstra}
\end{equation}

When the perfect fluid is radiation the last term is
$H_m = p_\phi^{4/3}e^s/R$, and so if we choose the
momentum conjugate to the {\it true} time $\tau$ to be

\begin{equation}
p_\tau = p^{4/3}_\phi e^s 
 \label{eq:ptau}
\end{equation}

\noindent
then the super-Hamiltonian constraint becomes

\begin{equation}
0 = {\cal H} = -\frac{p_R^2}{ 24R} -6R + \frac{p_\tau}{ R}
 \label{eq:tauH}
\end{equation}

The ADM Hamiltonian is obtained from $H_{ADM} =
p_\tau$, or 
 
 \begin{equation}
 H_{ADM} = \frac{p_R^2}{ 24} + 6R^2
 \label{eq:SHOH}
\end{equation}	

\noindent
which is just the Hamiltonian for a simple harmonic
oscillator.

The lapse $N$ is fixed by solving Hamilton's equation

 \begin{equation}
 \tau' = 1 = \frac{\partial (N[H_g + H_m])}{ \partial
p_\tau} = \frac{N}{ R}
 \label{eq:lapse}
\end{equation}

\noindent
which says that $N = R$; that is, {\it true} time is
just conformal time, which is why I have called it
$\tau$.

	If we quantize by the replacement $p_\tau \rightarrow
\hat p_\tau = -i \partial /\partial \tau$, and $p_R
\rightarrow \hat p_R = -i\partial /\partial R$, together
with a reversal of the direction of time
$\tau\rightarrow -\tau$ in the super-Hamiltonian
constraint (\ref{eq:tauH}), the Wheeler-DeWitt equation
(\ref{eq:WheeDeW}) will then become (if we ignore factor ordering
problems) Schr\"odinger's equation for the simple
harmonic oscillator with mass $m = 12$, spring
constant $k=12$ and angular frequency $\omega = 1$:
  
 \begin{equation}
i{\partial\Psi \over\partial \tau} = -{1\over
24}{\partial^2\Psi\over\partial R^2} + 6R^2\Psi
 \label{eq:SHOQM}
\end{equation}

\subsection{Consistency between Copenhagen and Many-Worlds {\bf Requires} a Delta Function Initial Boundary Condition}

\bigskip
We need to find what boundary conditions to impose on
equation (\ref{eq:SHOQM}).  The boundary condition that I
propose is the unique boundary condition that will
allow the classical Einstein equations to hold
exactly in the present epoch: that is, I shall require
that on the largest possible scales in the present
epoch, classical mechanics holds exactly.  To see how
to impose such a boundary condition, let us consider
the general one-particle Schr\"odinger equation
 
\begin{equation}
i\hbar{\partial\psi\over \partial t} = -{\hbar^2\over
2m}\nabla^2\psi + V(\vec x)\psi
 \label{eq:QM}
\end{equation}

If we substitute (Jammer 1974 p. 280; Landau and Lifshitz 1977, p. 51--52; Bohm 1952, 1953)
 
\begin{equation}
\psi = {\cal R}\exp(i{\varphi}/h)
\label{eq:Pilotwave}
\end{equation}

\noindent
into (\ref{eq:QM}), where the functions ${\cal R} = {\cal
R}(\vec x,t)$ and $\varphi = \varphi(\vec x,t)$ are real,
we obtain
 
 \begin{equation}
 {\partial {\cal R}\over\partial t} = -{1\over2m}\left[
{\cal R}\nabla^2\varphi + 2\vec\nabla
{\cal R}\cdot\vec\nabla \varphi\right] 
\label{eq:Bohm1}
\end{equation}

 \begin{equation}
 {\partial \varphi\over\partial t} = - {(\vec\nabla
\varphi)^2 \over 2m} - V  +\left(\hbar^2\over
2m\right){\nabla^2{\cal R}\over {\cal R}}
\label{eq:Bohm2}
\end{equation}

Equation (\ref{eq:Bohm2}) is just the classical Hamilton-Jacobi
equation for a single particle moving in the potential

 \begin{equation}
U = V  -\left(\hbar^2\over
2m\right){\nabla^2{\cal R}\over {\cal R}}
 \label{eq:quanpot}
\end{equation}

Equations (\ref{eq:Bohm1}) and (\ref{eq:Bohm2}) are fully equivalent to
Schr\"odinger's equation (\ref{eq:QM}),  and this way of
expressing  Schr\"odinger's equation, which I shall call the
Bohm--Landau Picture, is the most convenient formulation of QM when one wishes to
compare QM with classical mechanics.  

\medskip
There is, however, a crucial difference between Schr\"odinger's equation (\ref{eq:QM}) and the classical Hamilton-Jacobi equation (equation (\ref{eq:Bohm2}) with the quantum potential set equal to zero).  Since the H-J equation is non-linear, it will almost always develop cusp singularities where the wave field $\varphi$ ceases to be $C^1$.  For a simple example, imagine a plane wave incident on an attractive spherically symmetric central potential like the familiar $V = -e^2/r$.  The wave will be focussed into a point a finite distance from the center of the potential on the side opposite the incoming wave.  This focus point will be the cusp.   For all times after the formation of the cusp, the H-J equation will cease to define a global $C^1$ solution: the H-J equation will cease to be deterministic.  There will be a $\varphi$ field after the cusp develops, so this cusp will constitute a singularity in the laboratory.  In other words, classical mechanics in its most powerful form is not deterministic, contrary to what Laplace (and most physicists today) believed.  Adding the quantum potential saves the day, for the coupled equations (\ref{eq:Bohm1}) and(\ref{eq:Bohm2}) are equivalent to the linear Schr\"odinger's equation, which has no singularities.  A global $C^1$ solution is guaranteed by $C^1$ initial data.  Quantum mechanics is {\it more} deterministic than classical mechanics!  Alternately, we can regard the transition from classical mechanics to quantum mechanics --- the addition of the quantum potential to the classical H-J equation --- as being forced by the requirement of global determinism.  Quantum mechanics can be regarded as a specialization of the most general classical mechanics, because the quantum equation (\ref{eq:Bohm2}) does not allow an arbitrary potential $V$; it requires all acceptable potentials to be of the form (\ref{eq:quanpot}).  Once again, this specialization is required by determinism.

\medskip
The normals to
surfaces of constant phase, given by $\varphi(\vec x,t)
= \rm\, constant$, define trajectories: those curves
with tangents
 
\begin{equation}
\vec\nabla \varphi = {\hbar\over 2im}\ln\left(\psi
\over \psi^\ast\right) = Re\left[\left(\hbar\over
i\right)\ln\psi\right]
\label{eq:traj}
\end{equation}

The density of the trajectories is conserved, since this
density is given by $\rho = \psi \psi^\ast = {\cal R}^2$,
satisfying 

\begin{equation}
{\partial\rho\over\partial t} + \vec\nabla\cdot
\left(\rho{\vec\nabla \varphi\over m}\right)=0
\label{eq:trajden}
\end{equation}

\noindent
which is just (\ref{eq:Bohm1}) rewritten.

\medskip
The surfaces of constant phase guide an infinite
ensemble of particles, each with momentum $\vec p =
m\vec\nabla \varphi$:  {\it all} the trajectories
defined by (\ref{eq:traj}) are real in quantum mechanics.  In all
quantum systems, Many Worlds are present, though if
we make a measurement, we will see only one particle.
But we must keep in mind that in actuality, there are an uncountably
infinite number of particles --- infinitely many histories
--- physically present.  The same will be true in
quantum cosmology.

\medskip
But we will be aware of only one universe in quantum
cosmology, so the requirement that classical
mechanics hold exactly in the large in the present
epoch can only mean that this single universe of which
we are aware must obey exactly the classical
Hamilton-Jacobi equation: that is, we must require
that

\begin{equation}
{\nabla^2{\cal R}\over {\cal R}} = 0
\label{eq:classicalcon}
\end{equation}	

By requiring (\ref{eq:classicalcon}) to be imposed on the wave
function of the universe in the present epoch, I have in
effect unified the Many-Worlds Interpretation of
quantum mechancs with Bohr's version of the
Copenhagen Interpretation.  In what is universally
regarded as Bohr's definitive article on the
Copenhagen interpretation (Bohr 1959), Bohr never once claims
that the wave function must be ``reduced''; i.e., undergo
non-unitary evolution, nor does he ever claim that
macroscopic systems such as human beings are not
subject to the unitary time evolution of atomic
systems.  Bohr instead asserts: ``$\ldots$ it is
decisive to recognize that, {\it however far the
phenomena transcend the scope of classical physical
explanation, the account of all evidence must be
expressed in classical terms.}'' (Bohr 1959, p. 209, Bohr's
italics) ``$\ldots$ This recognition, however, in no
way points to any limitation of the scope of the
quantum-mechanical description $\ldots$'' (Bohr 1959, p.
211).  

\medskip
But quantum mechanics has unlimited validity
only if it applies equally to human-size objects as
well as to atoms, and thus the requirement that
accounts of phenomena expressed at the human size
level and larger must be in classical mechanical terms
can only mean that the larger objects must obey
classical and quantum laws simultaneously.  And this
is possible only if the human-size objects and larger
obey Schr\"odinger's equation and the classical H-J
equation simultaneously, which requires that the
boundary condition (\ref{eq:classicalcon}) hold.

\medskip
But it is only required to hold at the present epoch
(more precisely, after the first minute), and only on
the largest possible scale, that of the universe as a
whole.  In other words, the Copenhagen Interpretation is
to be regarded as something like the Second Law of
Thermodynamics: it applies only to large scale
systems, and it holds exactly only on the largest
possible scale, the scale of the universe as a whole. 
The Many-Worlds Interpretation holds always, just as
statistical mechanics holds always.  But boundary
conditions must be imposed on the MWI to yield exactly
the Copenhagen Interpretation in the large, just
boundary conditions must be imposed in statistical
mechanics to yield exactly the Second Law in the
Thermodynamic Limit.

However, as we shall see, imposing  (\ref{eq:classicalcon}) today will
yield a classical evolution from the initial singularity
to the present day, thus justifying the use of classical
field equations in the very early universe arbitrarily
close to the initial singularity, as I shall do in subsequent sections.

\medskip
If ${\cal R}$ is bounded above --- as it would be if
$\psi$ were a function in a Hilbert space --- equation(\ref{eq:classicalcon}) requires $\nabla^2{\cal R}=0$.  This in turn
implies (since ${\cal R}$ is bounded above) that ${\cal
R}=\rm\, constant$.  But the only allowed way in
quantum mechanics to obtain ${\cal R}=\rm\,
constant$ is to extend the Hilbert space to a Rigged
Hilbert space (Gel'fand triple) (B\"ohm 1978) that includes delta
functions.  For example, when $V=0$, a delta function
in momentum space yields ${\cal R}=\rm\,constant$,
and the plane wave, which indeed satisfies the
classical H-J equation, and indeed the trajectories
which are everywhere normal to the constant phase
surfaces are the straight lines with tangents
proportional to the momentum vector.

\medskip
It is important to emphasize that ${\cal R}=\rm\,
constant$ is going to yield a non-normalizable wave
function, and that the only allowed non-normalizable
wave function are indeed delta functions.  For,
as B\"ohm (1978) has pointed out, the most fundamental
expression of the wave function, the Dirac kets, are
themselves delta functions, so delta functions are
physically real states that are actually physically
more fundamental than the Hilbert space states.  Thus we
should not be surprised to find that the initial state of
the universe is one of the most fundamental states.

\medskip
The wave function of the universe $\Psi(R,\tau)$ in the
mini-superspace described above is a function of two
variables, the scale factor of the universe $R$ and the
conformal time $\tau$.

\medskip
If the initial boundary condition

\begin{equation}
\Psi(0,0) = \delta(R)
\label{eq:halfline1}
\end{equation}	

\begin{equation}
{\left[\partial\Psi(R,\tau)\over
\partial R\right]_{R=0}} = 0
\label{eq:halfline2}
\end{equation}

\noindent
is imposed, then the resulting wave function will have
classical H-J trajectories for $\tau >0$.  (Boundary
condition (\ref{eq:halfline2}) is imposed in addition to the delta
function condition (\ref{eq:halfline1}) for the following reason.  The
wave function is assumed to have no support for $R <
0$.  However, we cannot get this by imposng the DeWitt
boundary condition $\Psi(0,\tau) = 0$, because it
contradicts (\ref{eq:halfline1}).  But (\ref{eq:halfline2}) is sufficient for
self-adjointness of the SHO Hamiltonian on the
half-line $R \in [0,+\infty)$; see (Tipler 1986) for a discussion.) 
The wave function satisfying boundary conditions
(\ref{eq:halfline1}) and (\ref{eq:halfline2}) is just the Green's function $G(R, \tilde R,\tau)$ defined on the entire real line for the
simple harmonic oscillator, with $\tilde R$ set equal
to zero.  The wave function of the universe is thus

\begin{equation}
\Psi(R,\tau) = \biggl[{6\over \pi L_{Pk}\sin \tau}
\biggr]^{1/2}\exp \biggl[{i6 R^2\cot \tau\over
L_{Pk}^2}- {i\pi\over 4} \biggr]
\label{eq:ufunction}
\end{equation}

\noindent
where $L_{Pk}$ is the Planck length.  This wave
function is defined only for a finite conformal time: $0
\leq \tau \leq \pi$.  (The initial and final singularities
{\it are} in the domain of the wave function!)

\begin{figure}
\includegraphics[width=4.0in]{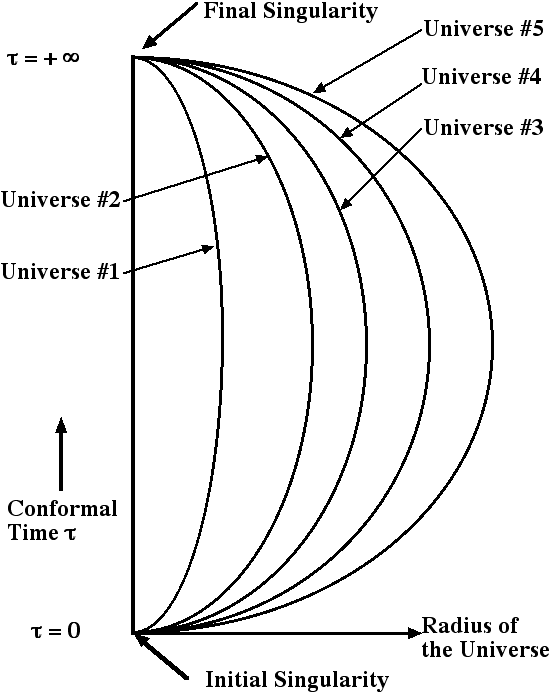}
\caption{\label{fig:multiverse}The Mulitverse: an infinite number of classical closed universe, all beginning at the same initial singularity, and ending at the same final singularity.  The probability is 1 that the particular universe we happen to find ourselves in is extremely large.  Thus the resolution of the Flatness Problem is wave packet spreading from an initial delta function.  Notice that this resolution is kinematic: it does not depend on any assumed dynamics.  The future conformal time limit is labeled $+\infty$, even though (\ref{eq:ctraj}) obviously has $+\pi$ as its upper bound, because the actual universe has a single point as its future c-boundary --- equivalently, it has no event horizons --- which requires infinite future conformal time.}
\end{figure}

\medskip
Notice that the magnitude of the wave function (\ref{eq:ufunction}) is independent of the scale factor of the universe $R$.  Since the scale factor plays the role of
``spatial position'' in the simple harmonic oscillator
equation (\ref{eq:SHOQM}), we have $\nabla^2R = 0$, and hence
from the discussion on phase trajectories above,
we see that the phase trajectories for the wave
function (\ref{eq:ufunction}) are all the classical trajectories for
a simple harmonic oscillator.  That is, the phase
trajectories are all of the form

\begin{equation}
R(\tau) = R_{max}\sin \tau
\label{eq:ctraj}
\end{equation}	

\noindent
which are also all the classical solutions to the
Einstein field equations for a radiation-dominated
Friedmann universe.  The phase trajectories of the multiverse are pictured in Figure \ref{fig:multiverse}.

\medskip
We can also show that the phase trajectories are given
by (\ref{eq:ctraj}) by direct calculation.  Since in the natural
units $L_{Pk}=1$, the phase $\varphi$ is $\varphi =
6R^2\cot\tau - {\pi\over4}$, we have $\nabla\varphi =
\partial\varphi/\partial R =12R\cot\tau$.  The tangents
are defined by $p_R =  \nabla\varphi$, which implies 

\begin{equation}
{1\over R}{dR\over d\tau} = \cot\tau
\label{eq:nphi}
\end{equation}

\noindent
using (\ref{eq:pR}), $N=R$, and $\tau\rightarrow -\tau$. 
The solutions to (\ref{eq:nphi}) are (\ref{eq:ctraj}).  

\medskip
With the boundary condition (\ref{eq:halfline1}), {\it all} radii at
maximum expansion, $R_{max}$, are present; all
classical paths are present in this wave function.  We
thus see that, with the boundary condition (\ref{eq:halfline1}),
both the phase trajectories and the wave function
begin with an initial singularity and end in a final
singularity.  In other words, with this wave function,
the universe behaves quantum mechanically just as it
does classically.  The singularities are just as real in
both cases.  Conversely, we can run the calculation in
reverse and conclude that in a SHO potential with 
${\cal R}=\rm\, constant$, we see that the universal
wave function must have been initially a delta
function.  This shows that each universe of the multiverse is classical all the way into the initial singularity.

\subsection{Singularity Structure of the Multiverse}

The muliverse (\ref{eq:ufunction}), the phase trajectories of which are pictured in Figure \ref{fig:multiverse}, has a simple singularity structure.  Let us call the different parts of the boundary of a spacetime or multiverse a {\it hypostasis} (plural: hypostases) which is the Greek word for ``foundation''.  The word ``different'' will mean ``usually constructed by different techniques.'' For example, Minkowski spacetime has five hypostases: $\imath^-$ (past timelike infinity), $\Im^-$ (past null infinity), $\imath^0$ (spacelike infinity), $\Im^+$ (future null infinity), and $\imath^+$ future (timelike infinity).  Future timelike infinity is the ``point'' outside of spacetime that timelike geodesics end up as time goes to future infinity, future null infinity is the infinite future ``endpoint'' (actually $R\times S^2$ in the Penrose topology) of null geodesics, and so forth.    Universal anti-de Sitter space has three hypostases, $\Im$, $\imath^-$ and $\imath^+$.  Maximally extended de Sitter spacetime has two hypostases, $\Im^-$ and $\Im^+$.  A classical Friedmann dust or radiation filled closed universe has two hypostases, $\Im^-$ and $\Im^+$, which are the initial and final singularities respectively.  See Hawking and Ellis (1973) and Hawking, King, and McCarthy (1976) for a more detailed discussion of the distinct boundaries to spacetime.

\begin{figure}
\includegraphics[width=4.0in]{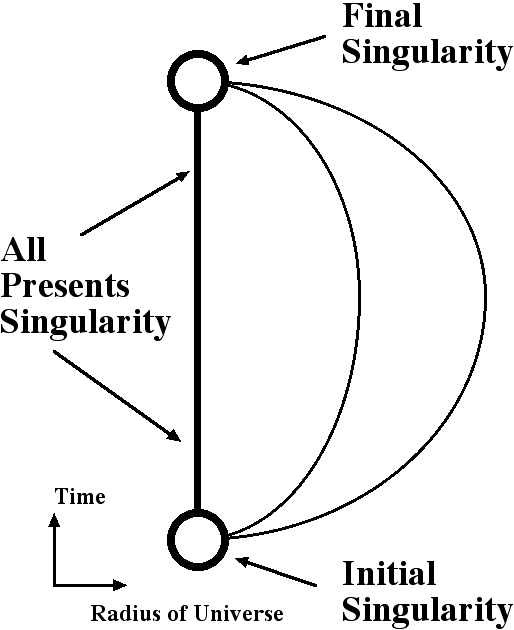}
\caption{\label{fig:threehypostases}The singularity structure of the mulitverse: three hypostases, consisting of the initial singularity, the final singularity, and the ``all presents singularity'' which connects the initial and final singularity.}
\end{figure}

\medskip
The multiverse has three hypostases, pictured in Figure \ref{fig:threehypostases}.  Each classical universe begins in an initial singularity, and ends in a final singularity.  These two singularities are constructed by moving inside each classical universe, by Cauchy completion with the natural positive definite metric $g_{\mu\nu} + 2u_\mu u_\nu$, where $u_\mu$ is a globally defined unit timelike vector field (Tipler, Ellis, and Clarke 1980, p. 154).  The unit normals to any foliation of the closed Friedmann universe by Cauchy hypersurfaces will give the same structure in the particular situation we consider in this paper:  The Cauchy completion will be a single point for the past singularity, and a single point for the future singularity.

\medskip
The third hypostasis is visible only from the multiverse. Since the multiverse pictured here is really the ground state of the universal wave function, we can use the natural metric on superspace given by the dynamics of this ground state, which satisfies the Wheeler DeWitt equation.  This natural metric is a six dimensional hyperbolic metric with signature $- +++++$.  As 
Figure \ref{fig:multiverse} indicates, all geodesics in this six dimensional superspace are incomplete (see DeWitt 1967 for a proof), and we can pick any of the natural timelike geodesic families in equation (5.8) of (DeWitt 1967) to define a positive definite metric $G_{AB} + 2U_A U_B$ and form the Cauchy completion.  This will yield the third hypostasis.  I have pictured the three hypostases in Figures  \ref{fig:multiverse} and  \ref{fig:threehypostases} using conformal time to spread out the third hypostasis.  In the positive definite metric constructed from the DeWitt superspace metric, the three hypostases are collapsed into a single point.

\subsection{Solution to Flatness Problem in Cosmology}

Since $\rho(R(\tau)) = \psi \psi^\ast = {\cal R}^2$
measures the density of universes with radius $R$,
for normalizable wave functions, it implies the Born
Interpretation: the probability that we will find
ourselves in a universe with size $R$ is proportional to ${\cal
R}^2$.   Similarly, if ${\cal R}=\rm\, constant$, we are
equally likely to find ourselves in a universe with any
given radius.  However, since $R>0$, if we ask for the
relative probability that we will find ourselves in
a universe with radius larger than any given radius
$R_{given}$ or instead find ourselves in a universe with
radius smaller than $R_{given}$, we see that the
relative probability is one that we will find ourselves
in a universe with radius larger than $R_{given}$, since
$\int^\infty_{R_{given}}{\cal R}^2\,d {\cal R} = +\infty$
while $\int^{R_{given}}_0{\cal R}^2\,d {\cal R}$ is
finite.  Thus with probability one we should expect to
find ourselves in a universe which if closed is
nevertheless arbitrarily close to being flat.  This
resolves the Flatness Problem in cosmology, and we
see that we live in a near flat universe because (1) the
Copenhagen Interpretation applies in the large,
or equivalently, because (2) the quantum universe began
as a delta function at the initial singularity, or
equivalently, because (3) classical physics applies on
macroscopic scales.

	Notice a remarable fact:  although the above
calculation was done using the Wheeler-DeWitt
equation, the same result would have been obtained if I
had done it in classical GR (in its Hamilton-Jacobi
form), or even done it in Newtonian gravity (in its
Hamilton-Jacobi form).  Just as one can do FRW
cosmology in Newtonian gravity, so one can also do
FRW cosmology in quantum gravity.  The conclusion is
the same in all theories: the universe must be flat. 
This conclusion does not, in other words, depend on the
value of the speed of light, or on the value of Planck's
constant.  In short, the flatness conclusion is robust!

\subsection{ Solution to the Standard Cosmological Problems: Homogeneity, Isotropy, and Horizon, and a Proof that the Universe Must Be Spatially $S^3$}

	One might think that quantum fluctuations would
smear out the classical nature of spacetime near the
initial singualrity.  Let me now prove that this is
false; that in fact the consistency of quantum
mechanics with general relativity requires these
fluctuations to be suppressed.  It is not the quantum
fluctuations at the instant of their formation that
gives rise to an inconsistency, but rather how such
fluctuations would evolve in the far future. 
Fluctuations will in general yield mini--black holes,
and it is the evolution of black holes, once formed,
that give rise to inconsistencies.

\medskip
Astrophysical black holes almost certainly exist, but Hawking has shown (1976; Wald 1994, Section 7.3) that if black holes are allowed to exist for unlimited proper time, then they will completely evaporate, and unitarity will be violated.  Thus unitarity requires that the universe must cease to exist after finite proper time, which implies that the universe has spatial topology $S^3$. 
The Second Law of Thermodynamics says the amount of entropy in the universe cannot decrease, but Ellis and Coule 1994 and I (Tipler 1994) have shown that the amount of entropy already in the CMBR will eventually contradict the Bekenstein Bound near the final singularity unless there are no event horizons, since in the presence of horizons the Bekenstein Bound implies the universal entropy $S \leq constant\times R^2$, where $R$ is the radius of the universe, and general relativity requires $R \rightarrow 0$ at the final singularity.  If there are no horizons, then the (shear) energy density can grow as $R^{-6}$, which means that the total available energy grows as $(R^{-6})R^3 \sim R^{-3}$, and so the Bekenstein Bound yields $ER \sim (R^{-3})R \sim R^{-2}$ which diverges as $R^{-2}$ as $R \rightarrow 0$ at the final singularity.  The absence of event horizons by definition means that the universe's future c-boundary is a single point, call it the {\it Omega Point}.  MacCallum (1971) has shown that an $S^3$ closed universe with a single point future c-boundary is of measure zero in initial data space.  Barrow (1982, 1998) Cornish \& Levin (1997) and Motter (2003) have shown that the evolution of an $S^3$ closed universe into its final singularity is chaotic.  Yorke {\em et al} (1990, 1992) have shown that a chaotic physical system is likely
to evolve into a measure zero state if and only if its control parameters are intelligently manipulated.  Thus life ($\equiv$ intelligent computers) almost certainly must be present {\it arbitrarily close} to the final singularity in order for the known laws of physics to
be mutually consistent at all times.  Misner (1968, 1969a,b) has
shown in effect that event horizon elimination requires
an infinite number of distinct manipulations, so an
infinite amount of information must be processed
between now and the final singularity.  The amount of
information stored at any time diverges to
infinity as the Omega Point is approached, since
$S\rightarrow +\infty$ there, implying divergence of
the complexity of the system that must be understood
to be controlled.

\medskip
Let me now expand out the argument in the
preceeding paragraph.  First, let me show in more
detail that unitarity (combined with the Hawking effect
and the Bekenstein Bound) implies that the spatial
topology of the universe must be $S^3$.  The argument I
shall give is independent of the dynamics; it only
depends on the basic structure of quantum mechanics
and Riemannian geometry.  A dynamical argument
would be sticky if one does not want to make any a
priori assumptions about the cosmological constant: a
deterministic (globally hyperbolic) universe with a
negative cosmological constant {\it always} will exist
for only a finite proper time, whatever the spatial
topology (Tipler 1976).  A dynamical proof for $S^3$ can be
found in (Barrow {\em et al} 1986) .

\medskip
I have pointed out in an earlier section that the Bekenstein Bound,
in the presence of particle horizons, implies that each
region of space inside the paricle horizon must have
less than one bit of information when this spatial
region becomes less than a Planck length in radius. 
Since less than one bit physically means that there is
no information whatsoever in the region --- that is, the
laws of physics alone determine the structure of space
--- this region must be isotropic and homogeneous,
because information must be given to specifiy
non-FRW degrees of freedom.  Now the Bekenstein
Bound is not merely a local bound, but a global
constraint, in the sense that it constrains a region
with radius less than the Planck length to have zero
information, rather merely some open ball of with no
{\it a priori} minimum size.  But we can overlap these balls
of Planck length in radius, to conclude that there is no
information anywhere in the spatial universe at the
Planck time.  

\medskip
Now a non-compact FRW universe at the Planck time
would still be of infinite volume, and thus would
eventually create an infinite number of protons and
neutrons.  Zel'dovich has shown that the lifetime
of a proton to decay via the Hawking process is
$10^{122}$ years (the actual value doesn't matter; it
just has to be finite for my argument).  If the universe
held an infinite number of protons and neutrons, the
probability is one --- a virtual certainty --- that at
least one proton or neutron would decay via the
Hawking process in the next instant after the Planck
time, so the probability is one that unitarity would be
violated.  But unitarity cannot be violated, so the
probability is one that the universe is spatially
compact.
	
\medskip
An alternative argument for spatial compactness, not using the Hawking argument for unitarity violation in complete black hole evaporation, is as follows.  It is well-known that a non-compact universe necessarily has event horizons (Hawking and Ellis 1973, p. 263).  The mechanism described earlier in this paper to stabilize quantum gravity and the SM requires a final singularity, and I have just argued that near a final singularity the mutual validity of the Second Law of Thermodynamics and QFT ( the Bekenstein Bound) requires event horizons to be absent.  Hence the mutual consistency of QFT and the Second Law of Thermodynamics requires the universe to be spatially compact.  Furthermore, the absence of event horizons automatically solves the Black Hole Information Problem. 

\medskip
We can now apply the Bekenstein Bound to this
compact universe, and note once again that the
Bekenstein Bound is a global bound; in particular, 
it implies that the amount of information is zero when
the volume of the universe is the Planck volume.  But
if the universe were not simply connected, the
topology itself would contain information, which is
not allowed.  Hence the universe must be spatially
compact and simply connected.  The homogeneity and
isotropy of the universe, inferred above, implies that
the spatial sectional curvatures are constant. 
Compactness implies (Cheeger and Ebin 1975, p. 11) that spatially, the
universe is complete.  It is well-known (e.g., (Cheeger and Ebin 1975, p.
40) that the only complete, simply connected compact
three-manifold with constant sectional curvature is
the three-sphere.

\subsection{Solution to Standard Model Hierarchy Problem Revisited}

\bigskip
Since the validity of the Standard Model of Particle
Physics --- especially of the SM electroweak physics
--- is the essential assumption in this paper, I shall
now further justify this assumption by pointing out
that the finite quantum gravity theory developed above, combined with the
Hawking black hole evaporation effect and the
requirement of unitarity as discussed above,
automatically resolves the Heirarchy Priblem.

\begin{figure}
\includegraphics[width=4.0in]{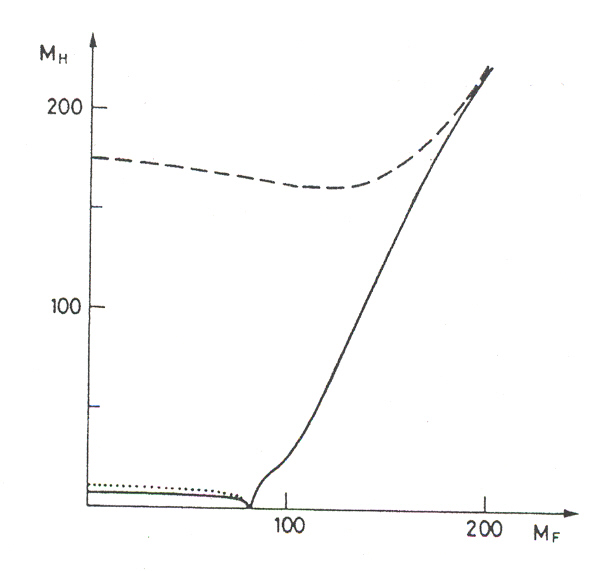}
\caption{\label{fig:Cabibbo}Stability bounds on the Higgs Boson mass $M_H$ vs. the top quark mass $M_F$.  Dashed line is the upper bound and the solid line is the lower bound,  Figure is reproduced from Cabibbo {\em et al} (1979), Figure 1, p. 301.  Cabibbo {\em et al} assume three quark generations and $\sin^2\theta_W = 0.2$  The curves do not change significantly if the more accurate value for $\sin^2\theta_W$ is used.}
\end{figure}

\medskip
Recall that the Heirarchy Problem is explaining why
the Higgs mass --- and hence all the particle masses
--- are not dragged up to the the Planck mass (or
higher!) by the self-interactions as expressed by the
renormalization group equation.  Let us first note that
the measurement of the top quark mass at around $178.0 \pm 4.3\,
{\rm GeV}$ (Abazov {\em et al} 2004) forces the SM Higgs boson mass to be around 200
GeV, because otherwise the SM Higgs potential would
become unstable due to the higher order quantum
corrections: the highest order term in the Higgs
potential when the quantum corrections are taken into
account is no longer $\lambda\phi^4$, but rather
$C\phi^4\ln(\phi^2/M^2)$ (to one loop order), and the
constant $C$ becomes negative, if the top quark mass
and the Higgs mass becomes greater than about 190
GeV and 200 GeV respectively. (I have argued above that this one-loop renormalization
group calculation is valid to all orders because the gravitational regularization
mechanism described above prevents the one
and higher loop self-energy corrections to the mass of
the Higgs boson alone from dragging the Higgs mass to
the Planck mass and beyond.)  The one-loop renormalization calculation of (Cabibbo {\em et al} 1979) is particularly constraining:  the most recent top quark mass data puts the universe so close to Cabibbo's instability line that the Higgs mass is forced to be $m_H = 196\pm4\, {\rm GeV}$ (see Figure \ref{fig:Cabibbo}).  (Sher (1989) and Ellis {\em et al} 1990) give a somewhat different stability curve.  See (Froggatt {\em et al} 2001) for another discussion of how one can obtain the Higgs boson mass from the stability of the electroweak vacuum.))

\medskip
The experimental fact that the SM Higgs vacuum
potential is, given the observed top quark mass, only
marginally stable is of fundamental significance:
when combined with the Hawking effect, it provides
the mechanism that solves the Hierarchy Problem.  

\medskip
Suppose on the contrary that the one and
higher loop corrections to the mass of the Higgs
boson increased the Higgs mass to greater than the
allowed $\sim 200$ GeV.  Then the mass of the Higgs
would be pulled over the vacuum stability bound, and
the mass of the Higgs would grow at least to the
Planck mass, and the mass of the down quark would
thus also increase to within an order of magnitude of
the Planck mass.  But this would mean that a neutron,
with two valence down quarks, would become unstable
via the Zel'dovich effect discussed above to the
formation of a mini-black hole of near Planck mass,
which would then evaporate via the Hawking process,
violating unitarity.  Hence, the one-loop and higher
self-energy terms cannot act to increase the mass of
the Higgs beyond the 200 GeV upper bound allowed by
vacuum stability, since this would violate unitarity.

\medskip
This also shows that the one and higher loop
corrections, which are integrals over the energy in the
loops, necessary have a cut-off at an energy less than
the Planck mass, a cut-off arising from quantum
gravity.  As I argued above, the cut-off is given by the requirement that
the energy in the loop cannot increase to the level that
would result in the formation of a mini-black hole
even virtually.  This cut-off is Lorentz and gauge invariant.  To see this,
ignore for a moment all effects except for
self-energy of a single particle.  Then the upper bound
to the value of the energy would be Planck energy,
defined by the condition that no trapped surface of
Planck energy is allowed to be formed, since such
would give rise to a violation of unitarity.  But the
trapped surface condition is a Lorentz and gauge
invariant condition.

\medskip
With the Hawking effect and unitarity, we see once again that no breakdown of the finite quantum gravity theory occurs.  Instead, the higher order
curvature terms generate a more intense gravitational
field than the first order Einstein Lagrangian, and thus
would force a mini-black hole at a lower cut-off than
the Einstein term.  This means that in the current
epoch of universal history, these higher order terms
must be suppressed by unitarity.  They will
be important near the final singularity --- when the
time before the final singularity is less than the
Planck time --- but they are essentially suppressed
at earlier times, in particular in the present epoch and
near the initial singularity.   So we can ignore these
terms today and in the past, and thus the fact that
adding an infinite number of terms to the Lagrangian
necessarily involves an infinite number of constants
that must be determined by experiment.  The
experiments need not be conducted, indeed cannot be
conducted until the universe is within a Planck time
of the final singularity.  Measuring the values of these
constants are among the infinity of measurements
that must be carried out by life as the universe moves
into the final singularity.  At all times, however, the
``effective'' quantum gravity theory will be term by
term finite, where I have placed the quotes because I
claim that this standard quantum gravity theory can
in fact be regarded as the true theory of quantum
gravity. 

\medskip
Recognizing that the Hawking effect plus unitarity
requires a Lorentz and gauge invariant upper bound to
the energy in a loop integral --- in other words, yields
a Lorentz invariant ultraviolet cut-off --- also solves
the well-known problem of infinite particle
production by time dependent gravitational fields. 
Expressing a quantized field as

$$\phi({\vec x}, t) = (2\pi)^{-3/2}\int
d^3\,{\vec k}[A_{\vec k}\phi_{\vec k}(t)e^{i{\vec
k}\cdot{\vec x}} + A_{\vec k}^\dagger\phi^\ast_{\vec
k}(t)e^{-i{\vec k}\cdot{\vec x}}]$$

The operators $\phi_{\vec k}(t)$ and $\phi^\ast_{\vec
k}(t)$ define what we mean by particles at time $t$. 
Given this definition, the corresponding definition at
time $t_0$ is given by 

$$\phi_{\vec k}(t_0) = \alpha_{\vec k}(t_0)\phi_{\vec
k}(t) + \beta_{\vec k}(t_0)\phi^\ast_{\vec
k}(t)$$

It was established by Parker more than thirty years
ago that the necesary and sufficient condition for the
particle number density to be finite is (Fulling 1978)

$$\int|\beta_{\vec k}(t_0)|^2d^3\,{\vec k} < \infty$$

Since in many cases of physical interest,
$|\beta_{\vec k}(t_0)|^2$ drops off only as $k^{-2}$,
this integral will diverge if the upper limit of the
energy is infinity.  However, the integral is a loop
integral, and thus having the upper bound extend past
the Planck energy would cause the spontaneous
formation of a mini-black hole, which would
immediately evapoate, violating unitarity.  Once again,
this ultraviolet cut-off does not violate Lorentz
invariance, because what is giving  the upper bound is
the non-formation of a trapped surface, and whether a
given 2-sphere is a trapped surface is a Lorentz
invariant condition.  So the definition of particle via
the Hamiltonian diagonalizaton procedure (which is
the definition used above) makes perfect sense given
unitarity and the Hawking effect, so I must disagree
with Fulling who opined in 1979 that no one
should ever again take the idea of Hamiltonian
diagonalization seriously (Fulling 1979, p. 824). 

\medskip
In the above I have made reference only to the down
quarks in the Zel'dovich--Hawking effect.  There
is a reason for omitting the up quarks.  Recall that the
Zel'dovich upper bound is the average time required for
two massive quarks to come within a Schwarzschild
radius of each other, the Schwarzschild mass being
assumed to be the Higgs quark mass.  A particle with
zero Yukawa coupling to the Higgs field would thus
have zero Schwarzschild radius, and thus two such
particles would have an infinite time before coming
within a Schwarzschild radius of each other.   Thus any
massless quark would not be subject to the Ze'dovich
mechanism.  The evidence suggests that the mass of the up quark is zero.

\medskip
Recall that the other outstanding theoretical problem
with the Standard Model of particle physics is the 
strong CP problem.  Now that the B factories have seen
CP violation, the solution of spontaneous CP violation
is now ruled out, at least in the sense that all such
models proposed to date predict that CP violation in B
decay should be too small to be observed in the
experiments where it was observed  (I am grateful
to Paul Frampton for a discussion on this point).  The
axion solution is generally considered to be ruled out
by the required fine tuning in the early universe
--- though I would rule it out because the axion has not
been detected.  The only remaining solution to be
strong CP problem is for the mass of up quark to be
zero.

\medskip
Standard current algebra analysis (e.g. Weinberg 1996, p. 231)
giving the ratios of quark masses in terms of the
masses of various mesons indicate that the up quark
has a non-zero mass, but Weinberg (1996, p. 458) points
out that inclusion of terms second order in the strange
quark mass might allow the up quark mass to vanish.
Kaplan and Manohar (1986) for example claim that $m_u =
0$ is allowed provided 30\% of the squares of the
meson masses arise from operators second order in the
chiral symmetry breaking, and also that ``The most
striking feature of our result is that a massless up
quark is not in contradiction with the past success of
$SU(3) \times SU(3)$ chiral perturbation theory." (p. 2006).

\medskip
Setting $m_u =0$ solves the Strong CP Problem, and
including standard quantum gravity effects in the
Standard Model solves the Hierarchy Problem.  Since
these were the main theoretical problems with the
Standard Model, we can be confident in the application
of the Standard Model to the early universe --- and
also confident in the Standard Model's claim that
electromagnetism is not a fundamental field but
instead is a composite of a $U(1)$ and an $SU(2)_L$
field.

\medskip
In summary, the one and higher self-energy
corrections to the Higgs boson indeed pull the Higgs
boson mass up to a higher value --- the loop integral
pulls the Higgs (and top quark) mass up to the
maximum value it can have consistent with vacuum
stability.  It cannot pull it up further than this,
because a further value would violate unitarity via the
Hawking effect.  The Hawking effect, by imposing an
upper bound to the energy (ultraviolet cut-off), an
upper bound coming from the requirement that this
stability bound be not exceeded, makes the Standard
Model fully consistent.

\section{The Spectra of Gauge Fields in a FRW Background}

I shall now consider the question of what spectral distribution is allowed for massless gauge fields in a FRW universe.  I shall show that a classical (massless) gauge field must obey a Wien Displacement Law, and a quantized (massless) gauge field must have a Planckian distribution, and that in both cases, this is true irrespective of the temperature of the fields.  That is, isotropy and homogeneity {\it alone} are sufficient to force the CMBR to have its observed Planckian spectrum.  So the observations showing that the CMBR has a Planckian spectrum says nothing whatsoever about its actual temperature.  It could even now have zero temperature!  But a zero temperature field is also a zero entropy field, and only zero entropy is consistent with the requirement coming from the Bekenstein Bound that the information content of the universe is zero when the radius of the universe is less than the Planck length.

\medskip
It has long been known (e.g, Weinberg 1972, p. 515; Weinberg 1977, p. 72) that the Rayleigh-Jeans long wavelength limit of the Planck distribution can be produced non-thermally, since this limit is simply an expression of the number of wave modes allowed in a given volume.  But I shall demonstrate that {\it any} radiation field (defined as matter whose energy density is inversely proportional to the fourth power of the scale factor in a FRW cosmology) will necessarily obey the Wien displacement law, irrespective of whether it is quantized or what statistics the field obeys.

\medskip
My derivation of the Planck distribution without assuming thermal equilibrium is analogous to
Hawking's deriviation of a Planckian distribution for the emission of radiation from a black hole.  In Hawking's original calculation, no assumption of thermal equilibrium was made initially, but he discovered that the black hole radiation emission was Planckian, with the black hole surface gravity playing the role of the temperature.  I shall show that in a FRW cosmology, a quantized gauge boson field must also have a Planck spectrum, with the quantity $\hbar c/R$, where $R$ is the radius of the universe, playing the role of temperature.  However, because of the isotropy and homogeneity of the FRW cosmology, there is no possibility of interpreting this quantity as a temperature.
	
\subsection{Proof that {\it all} classical Massless Gauge Fields Necessarily Obey a Wien Displacement Law in a FRW Universe}

I shall first show that the spectral distribution of radiation --- that is, any field whose energy density
is inversely proportional to the fourth power of the radius of the universe --- in any FRW cosmology necessarily obeys the Wien displacement law in the form

\begin{equation}
 I(\lambda, R) = {f(\lambda/R)\over R^5} =
{\phi(\lambda/R)\over \lambda^5}
 \label{eq:Wien}
\end{equation}

\noindent
where $R = R(t)$ is the FRW scale factor at any given
time $t$, $\lambda$ is the wavelength, $I(\lambda, R)$
is the monochromatic emissive power of the radiation,
and $f(\lambda/R)$ and $\phi(\lambda/R)$ are unknown
functions of the single variable $\lambda/R$.  Notice
that in the form (\ref{eq:Wien}), the reciprocal of the scale
factor has replaced the temperature $T$.  The
temperature does not appear in the form (\ref{eq:Wien}), and
no assumption of thermodyamic equilibrium will be
used in the derivation of (\ref{eq:Wien}).  Once again I emphasize that the
spectral distribution (\ref{eq:Wien}) will apply no matter what
the thermodynamic state of the radiation is; it will
even be consistent with the radiation being at the temperature of
absolute zero.

\medskip
Recall that in the standard derivation of the Wien
displacement law, the first step is to establish
Kirchhoff's law, which yields the fact that the
intensity $I$ of the radiation at a given wavelength
$\lambda$ depends only on $\lambda$ and the absolute
temperature.  In the standard derivation of Kirchhoff's
law, the assumption of thermal equilibrium is required
to establish this.  If we have a single radiation field in
a FRW cosmology, then $I = I(\lambda, R)$ --- the
intensity at a given wavelength depends only on the
wavelength and the scale factor --- because there are
no other variables in the cosmology upon which the
intensity of radiation could depend.

\medskip
Second, we recall that in a closed universe, the number
of modes $\cal N$ is constant under the expansion:

\begin{equation}
{\cal N} = {R\over \lambda} = {R'\over \lambda'}
 \label{eq:NmodesA}
\end{equation}

\noindent
where the primes denote the quantities at some other time.  Equation (\ref{eq:NmodesA}) can be re-written

\begin{equation}
{R'\over R} = {\lambda'\over \lambda}
 \label{eq:NmodesB}
\end{equation}

An alternative calculation following (Misner et al 1973, pp. 777--778) shows that in addition the same relation
between the wavelengths and expansion factors also hold infinitesimally: $ {d\lambda/ R)} =
{d\lambda'/ R')}$, or

\begin{equation}
{d\lambda'\over d\lambda} = {R'\over R}
 \label{eq:NmodesC}
\end{equation}

During the expansion, the energy density $U$ of a radiation dominated universe also changes.  We have

\begin{equation}
dU = \left({4\over c}\right) I(\lambda, R) d\lambda
 \label{eq:EdensityA}
\end{equation}
 
The energy density of any gauge field satisfies

\begin{equation}
 {dU\over dU'} = {\left({R'\over
R}\right)}^4
 \label{eq:EdensityB}
\end{equation}

Thus\ combining (\ref{eq:EdensityA}) and (\ref{eq:EdensityB}) gives
\begin{equation}
{dU\over dU'} = {{\left({4\over c}\right) I(\lambda, R)
d\lambda}\over \left({4\over c}\right) I(\lambda', R')
d\lambda'} = {\left({R'\over R}\right)}^4 
 \label{eq:EdensityC}
\end{equation}

\noindent
which gives upon solving for $I(\lambda, R)$ while
using (\ref{eq:NmodesA}) and (\ref{eq:NmodesB}):

\begin{equation}
I(\lambda, R) = {\left({R'\over R}\right)}^4
{d\lambda'\over d\lambda}I(\lambda', R') = {\left({R'\over
R}\right)}^5I({\lambda R'\over R}, R') =
{\left({\lambda'\over \lambda}\right)}^5I({\lambda
R'\over R}, R')
 \label{eq:RWien}
\end{equation}

As in the usual derivation of the Wien displacement law, we note that since the LHS of equation (\ref{eq:RWien}) does not contain the variables $R'$ or $\lambda'$,
neither can the RHS.  Thus (\ref{eq:RWien}) can be written

\begin{equation}
I(\lambda, R) = {f(\lambda/R)\over R^5} =
{\phi(\lambda/R)\over \lambda^5}
 \label{eq:WienF}
\end{equation}

\noindent
which is the Wien displacement law.  Notice that if there were several non-interacting radiation fields present, then each would satisfy the Wien dispalcement law, but possibily with different
functions of $R$, since we are not assuming thermal equilibrium.

\medskip
The maximum of the distribution (\ref{eq:WienF}) is obtained by
differentiating the first form of (\ref{eq:WienF}) with respect
to $\lambda$:

$${dI(\lambda, R)\over d\lambda}\bigg\vert_{\lambda
= \lambda_m} = {d\over d\lambda}{f(\lambda/R)\over
R^5}\bigg\vert_{\lambda
= \lambda_m} = {1\over R^6}f'(\lambda_m/R) = 0$$

\noindent
which tells us that the wavelength $\lambda_m$ of the maximum of the distribution satisfies 

\begin{equation}
{\lambda_m\over R} = {\rm constant}
 \label{eq:Wienmax}
\end{equation}

\medskip
Of course, the above calculation is meaningful only
if there {\it is} a maximum. The Rayleigh-Jeans law
obeys the Wien dispalcement law, and has no
maximum.  But the Rayleigh-Jeans law also suffers
from the ultraviolet divergence, and so is unphysical. 
The actual distribtion $I(\lambda, R)$ must either have
a maximum, or asymptotically approach a limiting
value as $\lambda \rightarrow 0$.

\subsection{Proof that {\it All} Quantized Massless Gauge Fields Necessarily Have a Planckian Specturm in a FRW Universe}

I shall now show that if the radiation field of the previous subsection is in addition a quantized massless gauge boson gas, the spectral distribution will follow the Planck distribution irrespective of whether the gas is in thermal equilibrium.  The key idea will be to follow Planck's original derivation of his Law (Planck 1959, Jammer 1966), which remarkably did NOT assume that the gas was at a maximum of the entropy (i.e., did not assume equilibrium), though, as is well-known, he did assume in effect that the energies of the gas particles were quantized, and that these particles obeyed Bose statistics.  As in the derivation of the Wien displacement law, the reciprocal of the scale factor of the FRW cosmology will replace the temperature in the Planck distribution law.

\medskip
The first part of the derivation will be the same as the standard derivation of the Planck distribution (e.g. Leighton 1959).

\medskip
Let us define the following quantities:

\medskip
$g_s\, \equiv$ number of modes in the $s$ energy
level;

\medskip
$n_s\, \equiv$ number of particles in the $s$ energy
level;

\medskip
$\epsilon\, \equiv$ the energy of the $s$ energy level;
and 

\medskip
$Q \, \equiv n_s/g_s\, =$ the occupation index.

\medskip
For a boson gas, the number of distinct arrangements is

\begin{equation}
P_s = {{(n_s + g_s
-1)!}\over{n_s!(g_s-1)!}}
 \label{eq:Narrage}
\end{equation}

The number of distinct arrangements $P$ for all energy
levels is thus

\begin{equation}
P = \prod_{s=1}^{s_{max}}P_s
= \prod_{s=1}^{s_{max}}{{(n_s + g_s -1)!}\over {n_s!(g_s
-1)!}}
 \label{eq:NarrageE}
\end{equation}

The information in this collection of
possible arrangements is

\begin{equation}
I \equiv \log P = \sum_{s=1}^{s_{max}}\lbrack
\log(n_s + g_s - 1)! - \log n_s! - \log(g_s -1)!\rbrack
 \label{eq:Info}
\end{equation}

If we assume $g_s \gg 1$, and  use Stirling's formula, (\ref{eq:Info}) becomes

$$
I = \sum_{s=1}^{s_{max}} \lbrack(n_s + g_s)\log(n_s
+ g_s) - n_s\log n_s - g_s\log g_s\rbrack = $$

\begin{equation}
I = \sum_{s=1}^{s_{max}} g_s\left[\left(1+ {n_s\over
g_s}\right)\log\left(1 + {n_s\over g_s}\right) -
{n_s\over g_s}\log{n_s\over g_s}\right]
 \label{eq:InfoB}
\end{equation}

\noindent
where each term in the sum will be denoted $I_s$, the
information in each energy level.
 
\medskip
Now I have already shown in my derivation of the Wien Displacement law that the number of modes ${\cal N} \propto ER$.  In the situation of perfect isotropy and homogeneity
this must apply for each state $s$ independently, and
for each mode.  Since the information per mode can
depend only on the scale factor $R$ --- in the FRW
universe there is no other possible function for the
information per mode to depend on --- and since the
argument in the Wien Displacement derivation gives a linear dependence on $R$ for
all values of the scale factor, we have:
 
 \begin{equation}
 d(I_s/g_s) = {\cal T}R\epsilon_sd(n_s/g_s )\
= {{\partial(I_s/g_s)}\over{\partial(n_s/g_s)}}
d(n_s/g_s)
 \label{eq:diffInfo}
\end{equation}

\noindent
where $\cal T$ is a constant to be determined. 
Equation (\ref{eq:diffInfo}) can be written
 
  \begin{equation}
 {{\partial(I_s/g_s)}\over{\partial(n_s/g_s)}} =
{\cal T}\epsilon_sR
 \label{eq:diffInfoB}
\end{equation}
 
From equation (\ref{eq:InfoB}) we have 

$${I_s\over g_s} = \left(1+ {n_s\over
g_s}\right)\log\left(1 + {n_s\over g_s}\right) -
{n_s\over g_s}\log{n_s\over g_s}$$

\noindent
and so substituting for simplicity $Q \equiv n_s/g_s$ 
we can evaluate the partial derivative in (\ref{eq:diffInfoB}):

$${{\partial(I_s/g_s)}\over{\partial(n_s/g_s)}} =
{d\over dQ}\left[(1+Q)\log(1 +Q) - Q\log Q\right] =
\log\left({{1+Q}\over Q}\right) = \epsilon_s{\cal T}R$$

\noindent
which can be solved for 
 
 \begin{equation}
n_s = {g_s\over{\exp(\epsilon_s{\cal T}R) -1}}
 \label{eq:NS}
\end{equation}

As is well-known, the infinitesimal number of modes
$d{\cal N}$ in a volume $V$ in the frequency interval
$d\omega$ is

$$d{\cal N} = {{V\omega^2d\omega}\over{\pi^2c^3}}$$

\noindent
so the energy per mode is

$$dE = \hbar\omega n_s/g_s = \hbar\omega d{\cal N}/g_s = 
{{\hbar\omega V\omega^2d\omega}\over{\pi^2c^3
(\exp(\hbar\omega{\cal T}R) -1)}}$$

\noindent
which yields a spectral energy density $dU = dE/V$ of

 \begin{equation}
dU = {{\hbar\omega^3d\omega}\over{\pi^2c^3
(\exp(\hbar\omega{\cal T}R) -1)}}
 \label{eq:dU}
\end{equation}

Using $I(\lambda,R) = (4/c)dU$ we obtain

 \begin{equation}
I(\lambda,R) = {{2\pi c^2h}\over{\lambda^5
(\exp({\cal T}chR/\lambda) - 1)}}
 \label{eq:PlanckI}
\end{equation}

Equation (\ref{eq:dU}) can be integrated over all frequencies
from 0 to $+\infty$  to give a total energy density

\begin{equation}
U = {{\pi^2}\over{15\hbar^3c^3}}\left({1\over{{\cal
T}R}}\right)^4
 \label{eq:TotalU}
\end{equation}

In a radiation dominated closed FRW universe, we have
(e.g. (Misner et al 1973, p. 735)

\begin{equation}
U = {{3R^2_{max}c^4}\over{8\pi GR^4}}
 \label{eq:FRWU}
\end{equation}

\noindent
where $R_{max}$ is the scale factor at maximum
expansion.

\medskip
Equating  (\ref{eq:TotalU}) and  (\ref{eq:FRWU})  yields the constant $\cal
T$ (for a radiation universe):

\begin{equation}
{\cal T} = \left({8\pi^3\over{45}}\right)^{1/4}
\left(L_{Pk}\over
R_{max}\right)^{1/2}\left(1\over{\hbar
c}\right)
 \label{eq:T}
\end{equation}

If we integrate equation (\ref{eq:PlanckI}) over all $\lambda$ to
obtain the total information in the universe, we obtain 

\begin{equation}
I_{Total} = {2\pi^4\over 15}\left({45\over 8\pi^3}
\right)^{3/4}\left[{R_{max}\over L_{Pk}}\right]^{3/2}
\approx 4\left[{R_{max}\over
L_{Pk}}\right]^{3/2}
 \label{eq:Tinfo}
\end{equation}

This is independent of the scale factor of the universe
$R$ --- so the information in the gauge field does not
change with time  (a Planck distribution is unchanged
by the universal expansion), but nevertheless it is
non-zero, which may be contrary to expectation; one
might expect that the information in the universe is
zero in the absence of thermalization.

\medskip
Before addressing the origin of this information, let me
first point out that the number (\ref{eq:Tinfo}) is completely
consistent with the Bekenstein Bound, which is 

\begin{equation}
I \leq  \left({2\pi\over \hbar
c}\right)\left(ER\right)
 \label{eq:Bek}
\end{equation}

Let me replace the constant $2\pi/\hbar c$ with the
constant $\cal T$, which will give an upper bound to the actual information content:

\begin{equation}
I = {\cal T}\left(ER\right)
 \label{eq:eqBek}
\end{equation}

\noindent
where $\cal T$ can be written

\begin{equation}
{\cal T} = {1\over(90\pi)^{1/4}}\left({L_{Pk}\over
R_{max}}\right)^{1/2}\left({2\pi\over \hbar c}\right)
 \label{eq:TBek}
\end{equation}

So, provided 
\begin{equation}
R_{max} \geq L_{Pk}
 \label{eq:RgLPK}
\end{equation}

\noindent
we will have 

\begin{equation}
{\cal T} < {2\pi\over \hbar c} 
 \label{eq:Tlhbar}
\end{equation}

\noindent
and thus the Bekenstein Bound will hold.

\medskip
If we happened to be in a universe in which $R_{max}
< L_{Pk}$, then the crucial approximation $g_s \gg 1$
and the use of Stirling's formula,which allowed me to
re-write (\ref{eq:Info}) as (\ref{eq:InfoB}), would no longer be
valid, and we would obtain a different $\cal T$, but still
consistent with the Bekenstein Bound in this universe.

\medskip
 Thus the origin of the information in the universe is
the particular value for $R_{max}$, which is just one
of the possible values; as we have seen earlier,
there are an infinity of possible values for $R_{max}$, the radius of the universe at maximum expansion, and our
``selection'' of the particular scale factor at maximum
expansion in our particular universe of the multiverse, generates the information.

\medskip
In fact, it is clear from (\ref{eq:TBek}) that had we simply
imposed the requirement that the information be of
order 1 at the Planck radius, say by setting ${\cal T} =
1$, then $R_{max} \sim L_{Pk}$.  Alternatively,  let us
try to eliminate all reference to $R_{max}$, by the
dimensionally allowed replacement $kT \rightarrow
\hbar c/R$.  Then, using the standard expression
above for the energy density $U$ of radiation with
termperature $T$, we get

$$U = {\pi^2(kT)^4\over 15 (\hbar c)^3} = {\pi^2\over 15
(\hbar c)^3}\left({\hbar c\over R}\right)^4 =
{\pi^2\hbar c\over 15 R^4} = {3c^4\over
8\pi GR^2_{max} \sin^4\tau} = {3R^2_{max}c^4\over
8\pi GR^4}$$

\noindent
or, 

$${3R^2_{max}\over 8\pi G} = {\pi^2 \hbar c\over 15}$$

\noindent
which yields

\begin{equation}
R_{max} = {2\pi\over\sqrt15}L_{Pk}
\approx (1.6)L_{Pk}
 \label{eq:Pkuniv}
\end{equation}

Which is to say, the scale factor at maximum
expansion is of the order of the Planck length, if we try to force the elimination of the constant $R_{max}$.

\medskip
Another (incorrect) way we could try to set the constant $\cal T$
is to simply require that the information in the gauge
field be of the order of one bit.  (Since the expansion
is adiabatic, the radius will cancel out of the
calculation.)   Recall that the entropy of radiation is
given by

\begin{equation}
S = {4\pi^2 kV(kT)^3\over45(c\hbar)^3}
 \label{eq:Srad}
\end{equation}

Setting $kT = 1/{\cal T}R$ and using the volume of a
closed universe $V = 2\pi^2R^3$, we get
 
\begin{equation}
 {\cal T} = {2\pi\over\hbar c}\left({\pi\over
45}\right)^{1/3}\left({k\over S}\right)^{1/3}
 \label{eq:SradT}
\end{equation} 

Setting $S/k = 1$ gives
 
 \begin{equation}
 {\cal T} = {2\pi\over\hbar c}\left({\pi\over
45}\right)^{1/3} \approx {2.6\over \hbar
c}
 \label{eq:SradTB}
\end{equation} 

\noindent
which, as we have already seen, gives $R_{max}
\approx L_{Pk}$.

\medskip
A third (incorrect) alternative we could try would be to set the
number of quanta in the universe to be equal to one. 
The total number $N_{\gamma}$ of quanta in the entire closed universe is

  \begin{equation}
 {N_{\gamma} = {2\zeta(3)\over \pi^2c^3\hbar^3}\left(kT\right)^3
V = {2\zeta(3)\over
\pi^2c^3\hbar^3}\left({1\over  {\cal
T}R}\right)(2\pi^2R^3) = {4\zeta(3)\over (c\hbar {\cal
T})^3}}
 \label{eq:Nphotons}
\end{equation} 

Setting $N_{\gamma}=1$ gives 

$${\cal T}c\hbar =
(4\zeta(3))^{1/3} \approx 1.7$$

\noindent
which once again gives $R_{max} \approx L_{Pk}$.

\medskip
As Bekenstein has often pointed out, when horizons
are present, the correct ``size'' $R$ that really should
be put into the Bekenstein Bound is the horizon radius;
in the case of the early universe, this is the particle
horizon radius.  Let be show now that with this value
for ``size'', indeed the choice (\ref{eq:T}) gives less than
one bit of information inside the horizon when the
particle horizon is the Planck Length.

\medskip
The equation for the particle horizon radius is

$$ds^2 = -dt^2 + R^2(t)d\chi^2 \equiv -dt^2
+(dR_{Particle})^2 = 0$$

\noindent
which when integrated (setting $R(0) = 0$) yields
 
  \begin{equation}
R_{Particle} = t
 \label{eq:Rparticle}
\end{equation} 

It is well known that for a radiation dominated FRW
closed universe the scale factor can be expressed in
closed form in terms of the proper time $t$:
 
 \begin{equation}
 R(t) = R_{max}\left( {2t\over R_{max}} - {t^2\over
R_{max}^2}\right)^{1/2}
 \label{eq:radRpropt}
\end{equation} 

\noindent
which can be solved for the proper time $t$:
 
 \begin{equation}
  t = R_{max}\left(1- \sqrt{1-\left({R\over R_{max}}
\right)^2}\right)
 \label{eq:sqrtRpropt}
\end{equation} 

\noindent
valid for $0 < t \leq R_{max}$.  For $R\ll R_{max}$,
this is approximately
 
 \begin{equation}
 R_{Particle} \approx
{R^2(t)\over2R_{max}}
 \label{eq:Rpartapprox}
\end{equation} 

The information inside the particle horizon in the early
universe is thus

$$I = {\cal T}(UR^3_{Particle})(R_{Particle}) =
{\cal T}UR^4_{Particle} 
={\cal T}\left({3R^2_{max}c^4\over 8\pi
GR^4}\right)\left({R^8\over 2R^4_{max}}\right) =$$
 
 \begin{equation}
 I =\left({8\pi^3\over
45}\right)^{1/4}\left({L_{Pk}\over
R_{max}}\right)^{1/2}\left({3R^4\over 8\pi
R_{max}^2L^2_{Pk}}\right)
 \label{eq:Rpart}
\end{equation} 
 
Putting (\ref{eq:Rpartapprox}) into (\ref{eq:Rpart}) gives 
 
 \begin{equation}
 I = \left({8\pi^3\over
45}\right)^{1/4}\left({L_{Pk}\over
R_{max}}\right)^{1/2}\left({12R^2_{Particle}\over
L^2_{Pk}}\right)
 \label{eq:infoRpart}
\end{equation} 

which is much, much less than one for
$R_{Particle}\leq L_{Pk}$.

\medskip
Heretofore, I have considered the universe to be radiation dominated, which is true in the early universe, but of course not true today, nor is it likely to be true in the future.  So the constant $\cal T$ will be slightly  different in a universe that contains both dark matter and dark energy.  Equation (\ref{eq:FRWU}) will need to be generalized.  The necessary equation is obtained from the Friedmann equation

 \begin{equation}
\left(\dot R\over R\right)^2 = {8\pi G\over 3}\left[\rho_{rad} + \rho_{matter} + \rho_{vac}\right] - {1\over R^2}
 \label{eq:FRWeq}
\end{equation} 

\noindent
in the closed universe case, where $\rho_{rad}$, $\rho_{matter}$, $ \rho_{vac}$ are the mass densities of the radiation, matter, and vacuum respectively.  At maximum expansion, we have $\dot R(t = t_{max}) = 0$, so we have for the mass density $\rho^{max}_{rad}$of the radiation at maximum expansion:

$$\rho^{max}_{rad} = {3\over 8\pi GR^2_{max}}\left[1 + \left(\rho^{max}_{matter} \over\rho^{max}_{rad}\right) + \left(\rho^{max}_{vac}\over \rho^{max}_{rad}\right)\right]^{-1}$$

Since $\rho_{rad} = (R_{max}/R(t))^4\rho^{max}_{rad}$, we have for the energy density of a massless gauge field at all times:

 \begin{equation}
U = {3R_{max}^2c^4\over 8\pi GR^4} \left[1 + \left(\rho^{max}_{matter} \over\rho^{max}_{rad}\right) + \left(\rho^{max}_{vac}\over \rho^{max}_{rad}\right)\right]^{-1} 
 \label{eq:gaugeden}
\end{equation} 

Equating (\ref{eq:gaugeden}) and (\ref{eq:TotalU}) we obtain the constant $\cal T$ in the more general case:

 \begin{equation}
{\cal T} = \left({8\pi^3\over{45}}\right)^{1/4}
\left(L_{Pk}\over
R_{max}\right)^{1/2}\left(1\over{\hbar
c}\right) \left[1 + \left(\rho^{max}_{matter} \over\rho^{max}_{rad}\right) + \left(\rho^{max}_{vac}\over \rho^{max}_{rad}\right)\right]^{1/2}
 \label{eq:genT}
\end{equation}

Since if the occurrence of a maximum expansion requires that $\rho^{max}_{matter} >> \rho^{max}_{vac}$, and we expect that $\rho^{max}_{matter} >> \rho^{max}_{rad}$, we can ignore the first and third factors in the bracket in (\ref{eq:genT}).  Comparing the energy density $U$ of the thermalized radiation in the usual Planck distribution with $U$ in equation (\ref{eq:genT}) yields the effective replacement:
\bigskip

\def\boxit#1{\vbox{\hrule\hbox{\vrule\kern3pt \vbox{\kern3pt#1\kern3pt}\kern3pt\vrule}\hrule}}

\boxit{$$kT \rightarrow \left(\hbar c\over R\right)\left(45\over8\pi^3\right)^{1/4} \left(R_{max}\over L_{Planck}\right)^{1/2} \left(\rho^{max}_{rad}\over \rho^{max}_{matter}\right)^{1/2}$$}

The question I address next is what kinds of gauge fields are allowed in the very early universe.

\subsection{Exact Solution of EYM Equations with Constant SU(2) Curvature}

\bigskip

The Yang-Mills field (curvature) is

$$W^{\mu\nu}_a = \partial^\mu W^\nu_a - \partial^\nu
W^\mu_a + gf_{abc}W^\mu_bW^\nu_c$$

\noindent
where $f_{abc}$ are the structure constants of the Lie
group defining the Yang-Mills field, the Latin index
is the group index, and we sum over all repeated
Latin indicies.  In the absence of all other fields
except gravity, the YM fields satisfy the equations (Atiyah 1979
p. 13)

$$\nabla\wedge W = 0$$

\noindent
and

$$\nabla\wedge ^*W = 0$$

\noindent
where $\nabla$ is the gauge and spacetime covariant
derivative.  The first equation is the Bianchi identity
for the YM fields, while the second is the Yang-Mills
equation.  It is obvious that a self-dual ($W=^*W$) or
anti-self-dual ($W=-^*W$ will automatically satisfiy
both equations if it satisfies one.

\medskip
In more conventional notation, the Bianchi identity is

$$D_\mu W^a_{\nu\lambda} + D_\nu W^a_{\lambda\mu}
+ D_\lambda W^a_{\mu\nu} =0$$

\noindent
where

$$D_\lambda W^a_{\mu\nu} = W^a_{\mu\nu;\lambda} -
f^a_{bc}A^c_\lambda W^b_{\mu\nu}$$

\noindent
with the semicolon denoting the spacetime
covariant derivative and $A^c_\lambda$ being the gauge
potential, in terms of which the gauge field
$W^a_{\mu\nu}$ can be expressed as

$$ W^a_{\mu\nu} = A^a_{\nu;\mu} - A^a_{\mu;\nu} +
f^a_{bc}A^b_\mu A^c_\nu$$

$$= A^a_{\nu,\mu} - A^a_{\mu,\nu} + f^a_{bc}A^b_\mu
A^c_\nu$$

\noindent
where the last equality is valid in any coordinate
basis; that is, if the spacetime covariant derivative is
expressed in a coordinate basis, the spacetime
covariant derivatives can be replaced by partial
deiviatives.  The same is true in the Binachi identify
for the gauge fields $W$.

\medskip
The Yang-Mills equation in more conventional notation
is (Weinberg 1996, p. 13):
$$D^\nu W^a\,_{\mu\nu} = 0 =
W^a\,_\mu\,^\nu\,_{;\nu} - f^a_{bc}A^c\,_\nu
W^b\,_\mu\,^\nu$$

 \medskip
The Lagrangian for the YM field is $L =
-(1/16\pi)W^{\mu\nu}_aW^a_{\mu\nu}$, and the
standard expression for the stress energy tensor
$T_{\mu\nu} = -2\delta L/\delta g^{\mu\nu} +
g_{\mu\nu}L$ yields

\begin{equation}
T^{YM}_{\mu\nu} =
{\hbar c\over4\pi}\left[W^a_{\mu\beta}W_{a\nu}^\beta -
{1\over4}g_{\mu\nu}W^a_{\alpha\beta}
W^{\alpha\beta}_a\right] \label{eq:stressenergyYM}
\end{equation} 

\noindent
where I have made the units explicit: the dimensions of $W^a_{\mu\nu}$ is $({\rm length})^{-4}$, and the energy in the W field is scaled in terms of $\hbar$.

\medskip
For any $T^{YM}_{\mu\nu}$, we have $T^\mu_\mu =
0$, and so for an isotropic and homogeneous universe,
any YM field must have $T_{\hat i \hat i} \equiv p =
{1\over3}T_{\hat t \hat t}$, where $T_{\hat t \hat t} 
\equiv \rho$ in any local orthonormal frame.  In other
words, any YM field satisfies in a FRW universe a
perfect fluid equation of state with adiabatic
index $\gamma =4/3$.

\medskip
However, very few Yang-Mills fields are consistent
with isotropy and homogeneity.  It is well-known that
a non-zero electromagnetic field --- a U(1) YM field ---
cannot exist in a FRW universe.  (Proof: eq. (5.23)
of (Misner et al 1973, p. 141), gives the stress energy tensor for the
EM field in an orthnormal basis, in particular
$T^{{\hat0}{\hat j}} = (\vec{E} \times
\vec{B})^{\hat j}/4\pi$, which equals zero since in FRW
there can be no momentum flow.  Thus $\vec{B}$ must
be a multiple of $\vec{E}$, so set $\vec{B} = a\vec{E} =
aE\hat{x}$.  Computing the diagonal components of
$T^{\mu\nu}$ gives $T^{{\hat0}{\hat0}} =
E^2(1+a^2)/8\pi \equiv \rho$, and $T^{\hat{x}\hat{x}} =
-\rho = -T^{\hat{y}\hat{y}} = -T^{\hat{z}\hat{z}}$.  But
for FRW isotropy requires $T^{\hat{x}\hat{x}} =
T^{\hat{y}\hat{y}} =T^{\hat{z}\hat{z}}$, so $\rho =
({\vec E}^2 + {\vec B}^2/8\pi) = 0$, which implies ${\vec
E} ={\vec B} = 0$).  However, any non-Abelian YM field
with an SU(2) normal subgroup {\it can} be non-zero in
a closed FRW, basically because SU(2) {\it is} a
homogeneous and isotropic 3-sphere.

\medskip
If the YM connection is a left invariant 1-form, that is,
if the connection is a Maurer-Cartan form, then the
resulting YM curvature will be given spatially by the
structure constants of the Lie group.  The YM curvature
will be

$$W^{\mu\nu}_a = gf_{abc}W^\mu_bW^\nu_c$$

\noindent
where
$$W^\mu_a = R^{-1}(t)\delta^\mu_a$$

\noindent
with $a = 1,\, 2,\, 3$ being the spatial indices
and $R(t)$ being the FRW scale factor.  It is easily
checked that the above expression for
$T^{YM}_{\mu\nu}$ gives $T^{\hat{j}\hat{j}} =
(1/3)T^{\hat{0}\hat{0}} \propto R^{-4}$ and all other
components zero, provided that $f_{abc}
=\epsilon_{abc}$, the structure constants for SU(2).

\medskip
It is clear that the above expression for the gauge
field is consistent only if $R(t) =$ constant.  The true
time dependent SU(2) gauge field is

$$ W^{\mu\nu}_a = \left[\epsilon_{abc} \delta^\mu_b
\delta^\nu_c \pm
{1\over2}\epsilon^{\mu\nu\alpha\beta}\epsilon_{abc}
\delta^b_\alpha\delta^c_\beta\right]{A\over R^2(t)}$$

\noindent
where $A$ is a dimensionless constant, which I shall compute below.  The plus sign gives a self-dual
gauge field ($W^{\mu\nu}_a = +^*W^{\mu\nu}_a$), and
the minus sign gives an anti-self-dual field 
($W^{\mu\nu}_a =-^*W^{\mu\nu}_a$).
 
This solution to the Einstein-Yang-Mills equations for the case of an $SU(2)$ fields has been independently discovered (Galt'sov and Volkov 1991); see (Gibbons and Steif 1994) for a detailed list.  The point I want to make that this solution is the {\it only} possible gauge field allowed by the Bekenstein Bound in it original form.  As I pointed out above, a $U(1)$ gauge field like electromagnetism is not allowed by the symmetries.

\subsection{Particle Production by Instanton Tunnelling in a FRW Universe}

Since the Bekenstein Bound requires a unique initial
state, and since the only allowed non-zero initial field
is the isotropic and homogeneous SU(2) field, the initial
baryon number is necessarily zero; the Bekenstein
Bound thus {\it requires} a mechanism to produce a net
baryon number.  Baryogenesis requires satisfying the
three Sakharov conditions: 

\medskip\noindent 
(1) violation of baryon number B and
lepton number L conservation;

\medskip\noindent
(2) violation of C and CP invariance; and
 
\medskip\noindent
(3) absence of thermal equilibrium

\medskip
The SM has a natural method of baryogensis via the
triangle anomaly, which generates both baryon and
lepton number (B + L is not conserved but B - L is
conserved), and since the self-dual gauge field
generates fermions rather than anti-fermions, it
violates C.  The anomaly function
$^*W^a_{\mu\nu}W_a^{\mu\nu}$ can be written ${\bf
E}^a\cdot{\bf B}_a$.  Since ${\bf E}^a$ is odd under
parity while ${\bf B}_a$ is even, the anomaly function
will violate CP.  At zero temperature, all processes
will be effectively non-equilibrium process.  So
baryogenesis via the triangle anomaly at zero
temperature is the natural method of SM baryogenesis.

\medskip
The Standard Model violates CP perturbatively via the
complex phase in CKM matrix.  In the early universe,
this perturbative mechanism fixes whether fermions or
anti-fermions will be created via the triangle anomaly;
that is, it fixes the SU(2) gravitational sphaleron to be
a self-dual rather than an anti-self-dual solution to the
EYM equations.  At high temperatures, the triangle
anomaly will on average generate almost as many
anti-fermions as fermions, because in thermal
equilibrium the SU(2) gauge fields will be
anti-self-dual as often as self-dual.  Thus, in
equilibrium, the CP violating CKM complex phase acts
to suppress SM baryogeneis; the excess of fermions
over anti-fermions is suppressed by the Jarlskog
determinant factor.  As is well-known, this
suppression can wash out at 100 GeV any fermion
excess generated at higher temperature by other
mechanisms.  

\medskip
In the usual high temperature SM baryogenesis
calculation (e.g. Rubakov and Shaposhnikov 1996) the baryon to photon ratio in
dimensionaless units, known (e.g. Spergel et al 2003) from
nucleosynthesis to be $\eta = 6.1\pm 0.3 \times 10^{-10}
$, is too small by a factor of about $10^{-8}$,
because the net creation rate is suppressed by the
smallness of the CP violation in the CKM matrix
described above, even when the problem of washing out
any net baryon number is ignored.  These problems are
avoided in my proposed mechanism, for two reasons:
first, the only role played by the CP violation is to
select a self-dual rather than an anti-self dual field
as the unique field (in the absence of CP violation
there would be no unique SU(2) field; the actual
magnitude of the CP violation is irrelevant.  Second,
the baryon number generation is always at zero
temperature, so there will never be any anti-fermions
generated, and never any washout of fermions created. 
In fact, my model has the opposite problem from
the usual electroweak high temperature model: my
model tends to produce too many baryons relative to
the number of SU(2) pseudo-photons.

\medskip
The net baryon plus lepton number ($B+L$) generated by the $SU(2)_L$
sphaleron is given by (Rubakov and Shaposhnikov 1996); Weinberg 1996, p. 454] ($g_2$ is the $SU(2)_L$ gauge coupling constant; Note Weinberg's remarks on the normalization):

\begin{equation}
N = {-g^2_2\over{32\pi^2\hbar c}}\int \sqrt{-g} d^4x
\,[{1\over2}\epsilon
^{\alpha\beta\mu\nu}W^a\,_{\alpha\beta}
W^b\,_{\mu\nu}(tr \,t_at_b)]
 \label{eq:baryonN}
\end{equation} 
 
\noindent
The net total baryon number of the entire universe $N_B$ will be half of $N$, since half of the particles generated will be leptons.

\medskip
I have set the $SU(3)$ gauge field to zero.  Once
again, this is required by uniqueness.  There are
uncountably many $SU(2)$ subgroups in $SU(3)$, but
thery are all in the same congugacy class.  A simple
proof of this is as follows (this simple proof was
pointed out to me by J. Bryan).

\medskip
Suppose $G$ is a subgroup of $SU(3)$ and $G$ is
isomorphic to $SU(2)$. Then the action of $SU(3)$ on
$C^3$ (complex Euclidean 3-space) induces a three
dimensional representation of $G$.  Since any
representation is the direct sum of irreducible
representations, this representation must be (1) the
unique irreducible representation of dimension three
(spin one representation), or (2) a direct sum of the
unique representation of dimension two (spin one half
representation) plus the one dimensional (trivial,
spin zero) representation, or (3) a direct sum of three
one dimensional (trivial) representations. (I use the
fact that $SU(2)$ has a unique irreducible
representation in each dimension). It cannot be (3)
since $G$ is a subgroup and so acts non-trivially. It
cannot be (1) since this representation is isomorphic
to the adjoint representation of $SU(2)$ on its Lie
algebra and so the negative of the identity element
acts trivially and so $G$ would not be a subgroup. 
Therefore the representation must be a direct sum of
the standard two dimensional representation and the
trivial representation.  Choose a unitary basis of $C^3$
so that the last factor is the trivial representation and
first two factors are the standard representation and
this change of basis will conjugate $G$ into the
standard embedding of $SU(2)$ into $SU(3)$. QED.  (As
an aside, note that we have the double cover $SU(2)
\rightarrow SO(3) \subset SU(3)$. The induced
representation on $SU(2)$ in this case will in fact be
the irreducible three dimensional one, but in this case
the subgroup is $SO(3)$, not $SU(2)$.)

\medskip
However, even though all $SU(2)$ subgroups are in the
same congugacy class, they are not physically
equivalent.  Each different $SU(2)$ subgroup is
generated by a different Lie subalgebra corresponding
to a different linear superpostion of gluons.  Each
such linear superposition is physically different.  Thus
there are uncountably many physically distinct $SU(2)$
subgroups of $SU(3)$, each capable of generating a
homogeneous and isotropic metric (since isomorphic
to the electroweak $SU(2)_L$ used above).  This means
the only way to have a unique $SU(3)$ field over a FRW
spacetime is to set {\it all} the gluons fields to zero.

\medskip
I have exhibited the two essentially unique vacuum
solutions to the ETM equations in the previous subsection; since, as I
have argued above, the self-dual is the unique solution
required by the SM, we take the plus sign:

 \begin{equation}
W^{\mu\nu}_a = \left[\epsilon_{abc} \delta^\mu_b
\delta^\nu_c +
{1\over2}\epsilon^{\mu\nu\alpha\beta}\epsilon_{abc}
\delta^b_\alpha\delta^c_\beta\right]{A\over R^2(t)}
 \label{eq:selfW}
\end{equation}

Putting (\ref{eq:selfW}) into (\ref{eq:baryonN}), using $2\pi^2R^3(t)$ for the total volume of an $S^3$ closed universe, using $\alpha_W \equiv g^2_2/4\pi\hbar c$, and finally remembering that $N_B = (1/2)N$ gives

\begin{equation}
N_B = {\alpha_W\over{16\pi}}\int {6A^2\over R^4} \sqrt{-g}
d^4x = {3\pi\alpha_W\over 4}A^2\int^t_0{c\,dt\over R}
 \label{eq:Nself}
\end{equation}

The last integral is of course conformal time; net
fermion production is proportional to conformal time,
and this suggests that the most appropriate time
variable for quantum gravity is conformal time, since
conformal time and only conformal time measures the
rate at which something new is occurring to break
perfect symmetry: fermions are appearing in the pure
$SU(2)_L$ gauge field.  This fact of the nature of the natural
time variable justifies my use of conformal time to quantize the FRW universe.

The constant $A$ in equation (\ref{eq:Nself}) can be evaluated by first using equations (\ref{eq:stressenergyYM}) and (\ref{eq:selfW}) to calculate $T^{YM}_{\hat0\hat0} = (\hbar c/8\pi)(3A^2/R^4)$, and then equating this energy density as measured in a local Lorentz frame to the energy density of radiation $U$ given in equations (\ref{eq:gaugeden}) or equation (\ref{eq:TotalU}) with constant $\cal T$ given by equation (\ref{eq:genT}).  Noting that

\begin{equation}
\eta \equiv \frac{n_B}{n_\gamma} = \frac{N_B}{N_\gamma}
 \label{eq:etaNB}
\end{equation} 
 
\noindent
where $\eta$ is the baryon to photon ratio, $n_B$ and $n_\gamma$ are the baryon and photon number densities respectively, and $N_\gamma$ is the total number of ``photons'' (here pure $SU(2)_L$ gauge particle) in the entire universe given by equation (\ref{eq:Nphotons}).  Simply assuming a universe that is radiation dominated over its entire history, and using the relationship between conformal time and the maximal size of such a universe, the correct value of $\eta$ is obtained when the average pseudo-temperature is $10^5$ GeV.  (The energy will be even higher if the more realistic value of the constant $\cal T$ is used.)  Thus as I pointed out, the problem to be solved if the initial field is a pure $SU(2)_L$ field is how to avoid too many baryons rather than too few.  It is possible that the pure $SU(2)_L$ gauge field were to be thermalized into an electromagnetic field at this energy, thus cutting off baryogenesis.  As we shall see, I shall suggest that the $SU(2)_L$ field has survived to the present time, and thus I shall propose that there is some other mechanism cutting off baryogenesis.  (Possibly an interaction of the $SU(2)_L$ with the Higgs field of the Standard Model.)

\subsection{The Coherent  $SU(2)_L$ Initial State Required by QFT and the resulting baryogenesis yields a Harrision-Zel'dovich Spectrum of the Correct Magnitude}

The argument originally used by Harrison and Zel'dovich was roughly as follows.  Define the spectrum of perturbations as they cross the particle horizon (B\"orner 1992, p. 352):

\begin{equation}
\delta_H = C\left(\frac{M_H}{M_*}\right)^{-\alpha}
 \label{eq:HZspectrum}
\end{equation} 

\noindent
where $C$ and $M_*$ are to be fixed by observation.  The flat spectrum $\alpha$ was recommended by Harrison and Zel'dovich because if $\alpha<0$, the perturbations would diverge on large scales, and if $\alpha > 0$ they would diverge on small scales (Brandenberger {\em et al} 1992).  However, if the perturbations are regarded as a field in a flat universe, then necessarily $\alpha = 0$, because only in this case would the perturbations be scale invariant, which flat space is: multiplication of the scale factor $R$ by a constant has no effect on the physics, whereas the physics of the perturbations would be effected by such a multiplication if $\alpha\not=0$.  One of the main reasons for the continued interest in the inflation mechanism is that it provides a natural way of generating a scale free perturbation spectrum on superhorizon scales.  But in the pure $SU(2)_L$ model discussed in this paper, the initial $SU(2)_L$ field is forced to be coherent on superhorizon scales by the Bekenstein Bound, that is, by the requirements of quantum field theory.  Once again, this coherence arises from kinematics; it is independent of any dynamics.  

The baryons generated by quantum tunneling are the natural source of the perturbations breaking the exact isotropy and homogeneity.  Like all coherent quantum perturbations we would expect the perturbation spectrum to be Gaussian, and the magnitude of the perturbations to be proportional to the square root of the measure of the baryons, that is to the baryon to photon ratiio $\eta$.  But since the baryons arise from a coherent quantum state we would expect the perturbations to be adiabatic.  Thus we would expect

\begin{equation}
\frac{\Delta T}{T} = \frac{1}{3}\frac{\Delta n_B}{n_B} = F\sqrt\eta
 \label{eq:tempperturb}
\end{equation} 

\noindent
where $F$ is a constant, and $n_B$ is the number density of baryons.  Setting the constant $F=1$, its natural value, accounts for the observations $\eta = 6.1\pm 0.3 \times 10^{-10}$ and $\Delta T/T \sim 10^{-5}$ (Bennett {\em et al} 2003).

\section{Could the $SU(2)_L$ Gauge Field Have Survived to the Present Day? Could the CMBR Be a Pure $SU(2)_L$ Gauge Field?}

\bigskip

The Bekenstein Bound requires the Higgs field to
initially have the value $\phi =0$, as has been
discussed in the previous section.  Since the present
day value is $<\phi> = 246\, {\rm GeV}$, the Higgs
field must have rolled down the incline of its
potential.  It is usual to assume that this Higgs energy
has long since been thermalized into the CMBR, but as I
have shown, the CMBR gauge field component would have a
Planck distribution in a FRW background whether it is
thermal or non-thermal, I shall instead investigate the
possibility that the Higgs energy has never been
thermalized, but instead has been partly absorbed by
the $SU(2)_L$ gauge field, and partly decreased in density
by the expansion of the universe.  The Bekenstein Bound requires the SM Higgs field to have been zero at the Planck radius and earlier, but of course now the Higgs field must be very near its minimum value, so that the SM will be observationally correct.  If the Higgs field has never been thermalized, then what must have happened is that the field ``rolled'' down its potential well to its minimum, with the expansion of the universe providing the damping to near the minimum of the potential.  But the sloshing of the Higgs field in its potential would not have been entirely damped by the expansion of the universe, and I shall argue that this incomplete damping is responsible for the observations of the dark matter.

\medskip
Let us recall what the expansion of the universe must
have been like so that the nucleosyntheis calculations
are still vaild, and so that the structure formation 
computer simulations are also still valid.  During the
nucleosynthesis era, the expansion rate must have
been $R(t)\sim t^{1/2}$ corresponding to radiation
domination --- as discussed earlier, any massless
gauge field will generate this behaviour if it is the
dominant energy density in the universe.  After
decoupling, the expansion rate must change from the
radiation domination rate to the matter dominated rate
of $R(t) \sim t^{2/3}$ with a slight mixture of a
$\Lambda$ term, the change being required by
structure formation simulations, which agree best
with the $\Lambda{\rm CDM}$ model (actually, the
best results are obtained by assuming that the CDM is
slightly ``warm'', but as we shall see, this will be
allowed in my proposal for the dark matter).  The
nucleosynthesis data indicate that baryons are only a
small fraction of the dark matter --- which by
definition is that ``substance'' which is responsible for
the $R(t) \sim t^{2/3}$ expansion rate.

\subsection{Solution to the the ``Dark Matter Problem'': What it is and How it Has Eluded Detection}

\medskip
In the early universe in the theory developed here, there are only two fields outside of the
baryons and leptons: the $SU(2)_L$ gauge field and the Standard Model
Higgs field.  I shall now argue --- but not prove ---
that it is {\it possible} for these two fields
interacting together to produce the observation CMBR
and the dark matter.  (I shall show below how
the Standard Model vacuum naturally provides a
non-zero cosmological constant; this I will take to be
the dark energy).

\medskip
Let us first consider the time evolution of the Higgs
field by itself, and then consider its interaction with
the $SU(2)_L$ gauge field.  For a general scalar field
Lagrange density of the form $-{1\over2}\partial_\mu
\phi\partial^\mu\phi - V(\phi)$, the equation of
motion for the scalar field will be

$${d\over dt}({1\over2}\dot\phi^2 + V) =
-3H\dot\phi^2$$

\noindent
or

\begin{equation}
\frac{d\rho}{dt} = -3H(\rho + p) 
 \label{eq:scalardensity}
\end{equation} 

\noindent
where

\begin{equation}
\rho = \dot\phi^2 + V(\phi)
 \label{eq:densityphi}
\end{equation} 

\noindent
is the energy density of the scalar field, and 

\begin{equation}
p = \dot\phi^2 - V(\phi)
 \label{eq:pressruephi}
\end{equation}

\noindent
is the pressure of the scalar field.

Turner (1983) has shown that if we regard the scalar field $\phi$
as the sum of a rapidly oscillating part, and a slowly
varying part, then a scalar potential of the form
$V(\phi) = a\phi^2$, which is the approximate form of
the SM Higgs potential in the present epoch, would
give rise to an effective mass density that would drop
of as $R^{-3}$, just as pressureless dust would.  I
conjecture that the combination of the SM Higgs field
coupled to a pure $SU(2)_L$ field would naturally
split into two fields that would appear to evolve
independently, one dropping off as $R^{-3}$, and the
other dropping off as $R^{-4}$.  One would be the CMBR,
and the other would be the dark matter.  Recall that
the Z boson has all the quantum numbers as a photon,
and in fact can be made to form superpositions with
photons.  The interaction strength of the Z with
fermions is stronger than the photon, and the only
reason that the Z boson acts weakly is its large mass.  
Similarly, the candidate I am proposing as the dark
matter will interact only weakly with fermions
because it is basically a Z particle.

\medskip
If this conjecture is correct, then the reason the dark
matter has not been detected is because it must
necessarily always be found in accompanied with
$SU(2)_L$ pseudo-photons, and all the experiments to
detect the dark matter have carefully been designed to
eliminate all photon interactions.

\medskip
Of course, the reason why such a possible dark matter
candidate has heretofore not been considered is that
it has been thought that the rapid oscillations of a SM
Higgs field would quickly decay away (Turner 1983, section IV), into photons.  I would conjecture that this is
indeed what happens; the Higgs field decays into the
$SU(2)_L$ field, which then passes the energy back
into the Higgs field.

\medskip
Let me emphasize (as if it needed emphasizing) that
these are very counter-intuitive conjectures I am
making, and I have given no mathematical evidence
that the combined Higgs coupled to a pure $SU(2)_L$
field could in fact behave this way.  I instead can only
offer an {\it experimental} argument that something
like this scenario must be in operation:  it has been
known for 35 years that ultra high energy cosmic rays
propagate through the CMBR as if the CMBR were not
present, and as I shall demonstrate shortly, this
is possible if --- and if the SM is true, only if ---
the CMBR has the properties of a pure $SU(2)_L$ field. 
And we have no laboratory experimental evidence that
the SM is incorrect.  The SM has passed every test we
have put it through for the past 25 years.

\subsubsection{Why an $SU(2)_L$ CMBR would have no effect on Early Universe Nucleosynthesis} 

\bigskip
The baryons, once created by the mechanism described earlier, would be in a Planck distribution gauge field, with
thermal properties identical to the usual standard
cosmological model.  Recall that the interaction
constants of the charged particles with the radiation
field are relevant only to force the particles to also be
in a thermal distribution like the radiation field. 
Thus, the interaction strength of a pure
$SU(2)_L$ field would at nucleosynthesis temperatures  would have no effect on the distribution of the
particles and thus on nucleosynthesis.  (The same
would be true of the fluctuation spectrum observed in
the acoustic peaks.  As mentioned earlier,
flatness requires a Harrison-Zel'dovich spectrum for
the fluctuations, and the magnitude of the fluctuation
spectrum is fixed by the requirement that the
fluctuations be due entirely from the creation of
baryons.)

\bigskip
\subsubsection{Supressing Early Universe Pair Creation,  Inverse and Doulbe Compton Scattering, and Thermal Bremsstrahlng}

\bigskip

I have argued in previous sections of this paper that in the
beginning, the universe must have contained nothing
but a pure $SU(2)_L$ field.  Even if this were true, one
might think that this pure $SU(2)_L$ field would have thermalized into a
normal EM field long before the de-coupling time around a redshift of
1,000.  I cannot prove that there is {\it no}
mechanism that would have resulted in the
thermalization of the proposed pure $SU(2)_L$ field,
but I can demonstrate that the standard (Lightman 1981, Danese and De Zotti 1982, Burigana {\em et al} 1991, Rephaeli 1995)
three main mechanisms of thermalization in early
universe cosmology, namely pair creation, double
compton scattering, and thermal bremsstrahlung
actually will not thermalize a pure $SU(2)_L$ field.

\medskip
An outline of the proof is simply to write down the
Feynman diagrams for all three processes, (actually
only two; the diagram for pair creation is essentially
the same as the diagram for Bremsstrahlung), and
realize that each ``pseudo-photon'' of the pure
$SU(2)_L$ field can couple {\it only} to left-handed
electrons, and right-handed positrons.  It is quickly
noticed that the diagrams violate conservation of
angular momentum; all of these processes require a
spin flip involving the same particle, and this is
impossible.  The no-go theorem is in all essentials the
same as well-known decay asymmetry of the W boson.  (There is a subtle point that at low energies, chirality and helicity are not the same.  This will be the considered below.)

\subsection {Right-handed electrons Won't Couple to an SU(2)
CBR Component}

\medskip
The main effect of the CBMR being a pure (or as we shall
see, almost pure) $SU(2)_L$ gauge field is that in this
case, {\bf the CMBR will not couple to right-handed
(positive chirality) electrons}, while standard
electomagnetic radiation couples to electrons of both
chiralities with equal strength.  

\medskip
Recall that in the Standard Model, the $U(1)$ gauge
field plays three roles.  First and foremost, it allows
the EM field to couple to right-handed electrons. 
Second, it forces a distinction between the $Z^\mu$
gauge field and the EM field 4-potential $A^\mu$.
Finally, it allows the unit normalizations of the
$U(1)$ and the $SU(2)_L$ fundamental gauge fields
$B^\mu$ and $W_j^\mu$ respectively to be carried over
to the physical gauge fields $Z^\mu$ and $A^\mu$. 
These latter two properties are usually termed the
``orthogonality" and ``normality" properties.  The
orthogonality and normality properties are at risk
when there is no $U(1)$ gauge field at all, so I shall
propose that the actual CMBR contains a small
admixture of $U(1)$ to maintain these key
properties.  I would expect the energy density of the
$U(1)$ component to be of the order of the energy
density of the anisotropic perturbations in the CMBR,
which would be the source of the small $U(1)$
component (recall that in the very early universe, the
radiation field which is the sole matter constituent of
the universe {\it must} be pure $SU(2)_L$).

\medskip
In the Standard Model the gauge fields are related by

\begin{equation}
A^\mu = {{g_2B^\mu + g_1W_3^\mu}\over \sqrt{g_1^2
+ g_2^2}}
\label{eq:Amu}
\end{equation}

\begin{equation}
Z^\mu = {{-g_1B^\mu + g_2W^\mu_3}\over
\sqrt{g_1^2 + g_2^2}}
\label{eq:Zmu}
\end{equation}

\noindent
where $g_1$ and $g_2$ are respectively the $U(1)$
and the $SU(2)_L$ gauge coupling constants.  It is
clear from (\ref{eq:Amu}) and (\ref{eq:Zmu}) that if the fundamental
fields $B^\mu$ are normalized to unity, then so are
$A^\mu$ and $Z^\mu$, and also that the latter two
fields are orthogonal if the former two are orthognal. 
It is also clear that the real reason for the
normalizations is to force the the EM fieldt to couple
with equal strength to both left and right handed
electrons.  But it is this equality that I am proposing
does not exist in the case of the CMBR.

\medskip
The coupling to electrons in the SM Lagrangian is

\begin{equation}
{\bar e_L}\gamma_\mu e_L\left[ {g_1\over2}B^\mu
+ {g_2\over2}W_3^\mu\right] + {\bar
e_R}\gamma_\mu e_Rg_1B^\mu
\label{eq:ecoup}
\end{equation}

Suppose now that we set $B^\mu =0$.  Solving (\ref{eq:Amu}) for $W^\mu_3 = (\sqrt{g_1^2 + g_2^2}/g_1)A^\mu$  --- in which case a normalized $W_3^\mu$ does {\it not}
yield a normalized $A^\mu$ and substituting this into (\ref{eq:ecoup}) gives

\begin{equation}
e_{EM}\left(\alpha_W\over2\alpha\right)A^\mu{\bar
e_L}\gamma_\mu e_l
\label{eq:eEM}
\end{equation}

\noindent
where $\alpha_W = 1/32$ is the $SU(2)_L$ fine
structure constant as before, and $\alpha = 1/137$ is the usual
fine structure constant.  So if we accept the
normalizaton of (\ref{eq:ecoup}), the coupling between
electrons and the pure $SU(2)_L$ field would be
increased relative to the usual $e_{EM}$.  However, I
would argue that either a small admixture of $B^\mu$
would force the usual coupling between the CBMR even
if mainly $SU(2)_L$, or else the appropriate
normalization to use in computing the interaction
between a CMBR which is almost pure $SU(2)_L$ is to
normalize $A^\mu$ even if it is mainly pure
$SU(2)_L$.  But I've gone through this calculation to
point out that there {\it may} be a different coupling
between a CMBR that is almost pure $SU(2)_L$, and the
usual $A^\mu$ field CMBR.  I doubt this possibility,
because a stronger coupling would ruin the early
universe nucleosyntheis results.  The stronger
coupling would also ruin the ultrahigh energy cosmic
ray effect, as I shall discuss in the next section.

	We should also recall that at high energy chirality and helicity are the same, but this is not true at lower energies, for example at the energies of nucleosynthesis and below. The helicity of an electron is a constant of the motion, but the chirality $\gamma^5$ varies with time as

\begin{equation}
\gamma^5(t) = cH^{-1}(\vec{\Sigma}\cdot\vec{p}) + e^{(21Ht/\hbar)}[\gamma^5(0) - cH^{-1}(\vec{\Sigma}\cdot\vec{p})]
\label{eq:chiral}
\end{equation}

Equation (\ref{eq:chiral})  yields the standard relation for the expectation values: 

\begin{equation}
<\gamma^5> = <\Sigma\cdot\hat{p}>\frac{|v|}{c}
\label{eq:expect}
\end{equation}	

\noindent
where $\gamma^5$ is the chirality operator, $\Sigma\cdot\hat{p}$ is the helicity.

\medskip
Although the $SU(2)_L$ field will not couple to right-handed electrons, the $U(1)$ field obviously will, and one might think that once fermions have been created, this interaction would thermalize the $SU(2)_L$ field.  But not at high energy, since the only fermions created will be left-handed.  At lower energy, when equation (\ref{eq:chiral}) shows that the left-handed fermions will oscillate into right-handed ones, I can only suggest that global quantum coherence will not allow the $U(1)$ field to be excited.  An {\it experimental} justification for this suggestion is the existence of ultra high energy cosmic rays, as I shall now demonstrate.

\section{Has an SU(2) CBR Component Already Been Detected?}
 \bigskip
\subsection{ Ultrahigh Energy Cosmic Rays}

\bigskip
\subsubsection{Why the Ultrahigh Energy Cosmic Rays Should Not
Exist But Yet They Do Exist}

\bigskip
In regard to ultra high energy (UHE) cosmic rays ---
particles above $10^{19}$ eV --- Alan Watson of the
University of Leeds recently described the
observations succinctly: ``They $\ldots$ are extremely
hard to understand: put simply --- they should not be
there''  (Watson 2001).   The reason UHE cosmic rays should not be
there is that they are too energetic to be confined to
the galaxy by the galactic magnetic field, yet they
cannot propagate very far in intergalactic space
because their energy would be absorbed by collisions
with CMBR photons (Sciama 1990).

\begin{figure}
\includegraphics[width=5.0in]{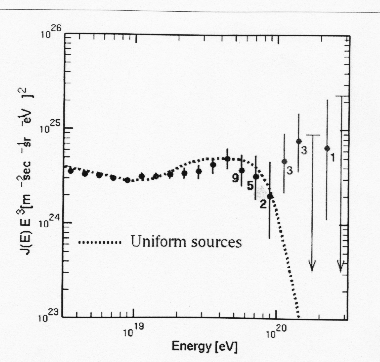}
\caption{\label{fig:UHEspectrum}UHE Energy Spectrum observed
between 1990 and 1997 with the the AGASA detector in
Japan.  The dashed curve is the expected rate with a
uniform cosmological distribution, but with the
expected interaction of protons and the CMBR.  (taken
from (Friedlander 2000, p. 119 and (Watson 2001), p. 819; figure due to M. Takeda {\em et al} (1998) of the Institute for Cosmic Ray Research, University of
Tokyo.)  The upper group of 3 events is 6 $\sigma$ above the theoretical Greisen curve.}
\end{figure}

\medskip
The detailed mechanism preventing the existence of UHE
cosmic rays was discovered by Kenneth Greisen in
1966, shortly after the discovery of the CMBR. 
Greisen pointed out that protons of sufficiently high
energy would interact with the CMBR, producing pions,
resulting in a cut-off to the UHE cosmic ray spectrum. 
Even in his paper of 35 years ago, he pointed out that
``$\cdots$ even the one event recorded at $10^{20}$ eV
appears surprising. $\cdots$ [the CMBR] makes the
observed flattening of the primary spectrum in the range
$10^{18}$ --- $10^{20}$ eV quite remarkable.'' (Greisen 1966, p. 750). 
Since Greisen wrote his paper, the experiments have
become even more inconsistent with the  existence of
the CMBR, as illustrated in Figure \ref{fig:UHEspectrum}.

\medskip
The AGASA array in Japan has detected 461 events
with energies above $10^{19}$ eV, and 7 events above
$10^{20}$ eV.  The Haverah Park array in England has
detected 4 events above $10^{20}$ eV, the Yakutsk
array in Siberia has detected 1 event above
$10^{20}$ eV.  The Volcano Ranch array in New Mexico
has detected 1 event above $10^{20}$ eV (Friedlander 2000, p.118).  So
four different groups over the past 35 years have
repeatedly seen these particles that shouldn't be there.
The consensus of the experimental cosmic ray
physicists is that the Greisen cut-off does not exist
(Watson 2001, p. 818), Cronin 1999).

\begin{figure}
\includegraphics[width=5.0in]{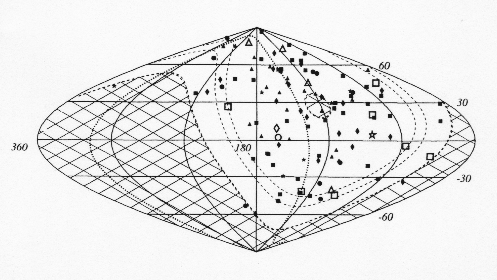}
\caption{\label{fig:UHEdirections}Arrival Directions of 114 events at energies above $4
\times10^{19}$ eV  (Figure taken from Watson (2001), p. 820.)  The arrival directions are essentially isotropic, indicating  a cosmological source.}
\end{figure}

\medskip
At energies above $10^{20}$ eV, there is no
clustering in arrival directions (Friedlander 2000, p.121; Watson 2001, p.
819).  This is illustrated in Figure \ref{fig:UHEdirections}, which gives the
arrival directions of 114 events at energies above $4
\times10^{19}$ eV.   At such energies, the gyromagnetic
radius is comparable to the size of the Galaxy, so UHE
cosmic rays should be extragalactic.  The only obvious
source within 30 Mpc (see below) is the Virgo Cluster,
but there is no clustering in this direction.
(Intergalactic magnetic fields are too low to effect the
arrival direction within 30 Mpc.)
	
\medskip
This blatant inconsistency between the observed
energies of some UHE cosmic rays and the global
existence of the CMBR has lead a number of physicists
to propose modifications in the known physical laws. 
However most physicists, myself included, ``$\cdots$
believe that, within certain well-defined physical
limitations, the laws of physics as we know them can be
trusted,'' to quote the words of Malcolm Longair, the
Astronomer Royal of Scotland (Longair 1994, p. 333).  What I shall
now show is that there is a mechanism, using only the
firmly tested physical laws, whereby UHE protons can
propagate cosmological distances through the CMBR
--- provided the CMBR is not the complete EM field,
but rather only the $SU(2)_L$ part of the EM field.

\bigskip
\subsubsection{How an SU(2) Component to the CBR Would Permit UHE Cosmic Rays to Propagate}

\bigskip
Recall that CMBR blocks the propagation of UHE cosmic
rays via the GZK effect (Sokolsky 1989): protons comprising the UHE
particles collide with a CMBR photon, resulting in pion
production.  The reactions are
	
\begin{equation}
\gamma + p \rightarrow \Delta^+ \rightarrow \pi^0 + p
\label{eq:Delta1}
\end{equation}	

\begin{equation}
\gamma + p \rightarrow \Delta^+ \rightarrow \pi^+ + n
\label{eq:Delta2}
\end{equation}

\begin{equation}
\gamma + p \rightarrow \Delta^{++} + \pi^- \rightarrow \pi^- + \pi^+  + p
\label{eq:Delta3}
\end{equation}

\noindent
where $p$, $n$, and $\pi$ are the proton, neutron, and
pion respectively.  The reaction cross-sections are
dominated by $\Delta$ particle resonances (Nagle {\em et al} 1991, Greisen 1966). 
Of the total cross-section for  (\ref{eq:Delta2}) of 300
microbarns at peak $E_\gamma = 320$ MeV, 270
comes from the $\Delta$ resonance (Nagle {\em et al} 1991, p. 14).  Of the
total cross-section for  (\ref{eq:Delta3}) of 70 microbarns at
peak $E_\gamma = 640$ MeV, virtually all comes from
the $\Delta^{++}$ resonance (Nagle {\em et al} 1991, p. 14).  Of the total
cross-section for (\ref{eq:Delta1}) of 250 microbarns at peak
$E_\gamma = 320$ MeV, 140 comes from the $\Delta$
resonance (Nagle {\em et al} 1991, p. 13).  For (\ref{eq:Delta3}), virtually all the
total cross-section for photon energies less than the
peak also comes from the $\Delta^{++}$ resonance.
For the other reactions, the rest of the total
cross-section arises from a photoelectric term (Nagle {\em et al} 1991, p. 14).					

\medskip
However, if the CMBR consists not of electromagentic
radiation, but is instead a pure $SU(2)_L$ field, then
the $\Delta$ resonance cannot occur.  The reason is
simple: the $\Delta$ particle originates from a proton
by a quark spin flip (Nagle {\em et al} 1991, p. 16), but since a $SU(2)_L$
field couples only to a left-handed quark, it cannot
induce a quark spin flip: a pure $SU(2)_L$ photon would
not couple to a right-handed quark at all, and a
left-handed quark would have the handedness
unchanged by the interaction.  (Notice that at UHE Cosmic Ray energies, there is no difference between the chirality and the helicity, so I shall use these terms interchangeably.)

\medskip
Furthermore, the photoelectric term would be reduced
because only a fraction of the electric charge on the
quarks would interact with a pure $SU(2)_L$ field.  If
for example, a proton were to have only its down
valence quark left handed, then its effective electric
charge would be $-1/3$ rather than $+1$.  Since the
photo cross-sections are proportional to the
$({\rm charge} )^4$ (the square of the classical
electron radius, with the ``electron'' having a charge of
-1/3), the photo cross-section would be reduced by a
factor of $1/81$ from its value for an electromagentic
CMBR.  Even if one of the up quarks were left-handed,
the photo cross-section would be reduced by a factor
of $16/81 \approx 1/5$.

\medskip
The net effect on the total-cross-sections (for the
down valence quark case) is to reduce the
cross-section for pion production from $SU(2)_L$
photons $\sigma_{SU(2)}$ from its value
$\sigma_{EM}$ that we would have if the CMBR were
to be a normal electomagnetic field:

\begin{equation}
\sigma_{SU(2)}^{p\pi^0} =
{1\over150}\sigma_{EM}^{p\pi^0}
\label{eq:sigmaSU2ppi0}
\end{equation}	

\begin{equation}
\sigma_{SU(2)}^{n\pi^+} =
{1\over810}\sigma_{EM}^{n\pi^+}
\label{eq:sigmaSU2npi}
\end{equation}	

\begin{equation}
\sigma_{SU(2)}^{p\pi^+\pi^-} = 0
\label{eq:sigmaSU22pi}
\end{equation}	

The mean free path $L_{MFP}$ for an UHE proton is

$$L_{MFP} = (\sigma^{p\pi^0}N_{photon})^{-1}$$

Using $N_{photon} = 5\times10^8\,{\rm m}^{-3}$ we
would get $L_{MFP} \approx 10^{23}\,{\rm m} \approx
3\, {\rm Mpc}$ (Longair 1994, p. 340) if we used
$\sigma_{EM}^{p\pi^0}$. Using (\ref{eq:sigmaSU2ppi0}), however, we get

\begin{equation}
L_{MFP} = 450\,{\rm Mpc}
\label{eq:MFP}
\end{equation}	

\noindent
which means that UHE protons can propagate through
the intergalactic medium as if the CMBR were not
there.  This becomes even more obvious when we
consider that the fractional energy loss due to pion
creation is $\Delta E/E \approx m_\pi/m_p \approx
{1\over10}$, so the propagation distance would be
more than 4 Gpc, which is truly a cosmological
distance.

\medskip
If pion production is no longer significant, then one
could wonder about the removal of UHE proton via
electron-positron pair production.  As is well-known
(Longair 1994, p. 341), the cross-section for pair production
from the collision of a UHE proton with an EM CMBR
photon is actually greater than the cross-section for
pion production, but the much smaller mass of the pair
means that with EM CMBR photons, the energy loss
per collision is less by a factor of
$(m_\pi/m_e)(\sigma^{p\pi}/\sigma_{pair}) \approx
6$. I have shown in an earlier section of this paper that
pair production is not possible via collision of two
$SU(2)_L$ photons, but it is not necessary to
investigate whether this would be the case for the
collision of such a CMBR photon with a proton.  For the
cross-section for pair production is proportional to
$\alpha_{EM}r^2_e$, and thus the cross-section for
pair production would be also reduced by at least a
factor of $1\over150$ by the effective charge
reduction mechanism that reduced the pion production
cross-section.

\medskip
The energetics of the UHE cosmic rays are completely
consistent with a cosmological source (Longair 1994, pp.
341--343).  The local energy density of cosmic rays
with energies above
$10^{19}\,{\rm eV}$ is $1\,{\rm eV m}^{-3}$. 
Following (Longair 1994, p. 342), let us assume that each source
of UHE cosmic rays generates a total cosmic ray
energy of $E_{CR}$ over the age of the universe, and
that $N$ is the spatial density of these sources.  In
the absence of losses, the energy density of UHE
cosmic rays would be

$$\rho_{CR} = E_{CR}N$$

For strong radio galaxies, $N\approx 10^{-5}\, {\rm
Mpc}^{-3}$, so we would have to have $E_{CR} \approx
5\times10^{53}\,{\rm J}$, or about $3\times
10^7\,{\rm M_\odot}$ of energy in the form of UHE
cosmic rays produced per source over the history of
the universe.  Given that the black hole cores of many
radio galaxies have masses above
$10^{10}\,{\rm M_\odot}$, one would require a
conversion efficiency of mass into UHE cosmic rays of
only 0.6\% (assuming that the source of the UHE
cosmic rays produces these protons with left-handed
and right-handed down valence quarks in equal
numbers), which seems quite reasonable: even ordinary
hydrogen fusion has a 0.7\% efficiency for conversion
of mass into energy, and black hole models can give
mass-energy conversion efficiencies up to 42\%. 

\medskip
The sources for a $3 \times 10^{20} {\rm eV}$
proton which are allowed by the Hillas criterion,
namely that an allowed source must satisfy
$BR \sim 10^{18} {\rm G\, cm}$, where $B$ is the
manetic field and $R$ is the size of the region with
this roughly constant field, are radio galaxy lobes and
AGNs, as is well-known.  However, heretofore, these
sources have been eliminated on the grounds that they
are too far away.  If the CMBR is actually a pure
$SU(2)_L$ field, then such sources are perfectly
acceptable.

\medskip
Cosmic ray physicists have in the past made great
discoveries in fundamental physics: in 1932, the
discovery of the positron (Anderson 1993, reprinted in Hillas 1972); in 1937
the discovery of muons; and in 1947, the discovery of
pions and kaons (Hillas 1972, pp. 50--51).  Positrons were the
first examples of anti-matter, and finding them
deservedly got Carl Anderson the Nobel Prize.  Muons
are elementary particles according to the Standard
Model, as are s-quarks, which made their first
appearance in kaons.  The first s-quark baryons, the
$\Lambda$, the $\Xi^\pm$, and the $\Sigma^+$
particles, were first detected by cosmic ray
physicists, in 1951, in 1952, and in 1953, respectively
(Hillas 1972, pp. 50--51).   But it has been almost half a century
since cosmic ray physicists have made {\it recognized}
fundamental discoveries.  I believe that the discovery
of the UHE cosmic rays are an {\it unrecognized}
discovery of fundamental importance: the observation
of these particles demonstrates that the CMBR is not an
electromagentic field, but rather the pure $SU(2)_L$
component of an electromagnetic field.  

\medskip
The claim of many theorists concerning UHE
cosmic rays, that these particles simply {\it must} be
merely a local phenomena, reminds me of Herzberg's
1950 remark on the observation that CN molecules in
interstellar clouds appeared to be in a heat bath of
around 2.3 K: ``which has of course only a very
restricted meaning.'' (Herzberg 1950, p. 497), by which he meant
that the 2.3 heat bath was merely a phenomena local to
the molecular cloud.
\subsection{Has the $SU(2)_L$ Nature of the CMBF been seen in observations of the Sunyaev Zel'dovich effect?}

The Sunyaev-Zel'dovich effect (SZE) is the slight shift $\Delta T$ in the temperature of the CMBR as the CMBR passes through a cluster of galaxies due to interaction with the free electrons in the cluster.  See Birkinshaw (1999) for a recent review.  If we assume the standard spherical isothermal $\beta$ model in which the free electron density $n_e(r) = n_{e_0}(1+(r^2/r^2_e))^{-2\beta/2}$, the SZE temperature shift will be (Carson {\em et al} 2002)
\begin{equation}
\Delta T= f_{(x,T_e)}T_{CBM}D_A\int d\zeta\, \sigma_Tn_e \frac{k_BT_e}{m_e c^2} =  \Delta T_0\left(1 + \frac{\theta^2}{\theta^2_c}\right)^{(1-3\beta)/2}
\label{eq:SZT},
\end{equation}

The SZE has been seen, and if, as I conjecture, the CMBR is a pure $SU(2)_L$ field, then the temperature shift should be reduced by a factor of 2 since in a galaxy, the CMBR photon will in effect ``see'' only half the electrons that are present.  Only if we can determine the electron density independently of the SZE will it be possible to test my conjecture using the SZE. 

Fortunately, the free electron density can be measured independently of the SZE using X-ray emission.  The X-ray surface brightness of a galaxy is given by

\begin{equation}
    S_X= \frac{1}{4\pi(1+z)^4}D_A\int d\zeta\, n_en_H\Lambda_{eH} = S_{X_0}\left(1 + \frac{\theta^2}{\theta^2_c}\right)^{(1-6\beta)/2}
\label{eq:XraySZ}
\end{equation}

The idea is to solve (\ref{eq:XraySZ}) for $n_e$, and plug this value into (\ref{eq:SZT}) and see if the SZ temperature shift is reduced by the expected factor of 2.  

\par
What I shall do is analyze data in the recent papers on using the SZE to measure Hubble's constant.  It is known that the value of Hubble's constant obtained via the SZE is too low, and I shall suggest (but NOT prove) that this disagreement might be due to the fact that the CMBR is a pure $SU(2)_L$ gauge field.

\par
In a flat universe, the angular diameter distance $D_A$ is related to the Hubble constant $H_0$ and the redshift $z_0$ of a cluster by  (Carroll {\it et al} 1992, p. 511)
\begin{equation}
   H_0D_A = \int_0^{z_0}[(1+z)^2(1+ \Omega_Mz) - z(2+z)\Omega_\Lambda]^{-1/2}dz
\label{eq:HD}
\end{equation}
So if $D_A$ and $z_0$ of the cluster are known, $H_0$ can be determined.  The integral is evaluated by setting $\Omega_\Lambda = 0.73$, and $\Omega_M = \Omega_{DM} + \Omega_b = 0.23 + 0.04 = 0.27$.  Since both equations (\ref{eq:XraySZ}) and (\ref{eq:SZT}) contain $D_A$, we can eliminate $n_e$ in both, and solve for $D_A$, and then use this value in (\ref{eq:HD}) to obtain Hubble's constant.  Since there are in effect two electron densities, one $n_e^X$ generating the X-rays and the other $n_e^{SZ}$ generating the SZE, equations (\ref{eq:XraySZ}) and (\ref{eq:SZT}) are modified accordingly.  I shall assume that these two electron densities are proportional to each other:

\begin{equation}
  n_e^X = An_x^{SZ}
\label{eq:electronD}
\end{equation}

The pure $SU(2)_L$ hypothesis predicts $A = 2$.

The result is

\begin{equation}
D_A = \frac{(\Delta T_0)^2}{S_{X_0}}\left(\frac{m_ec^2}{k_BT_{e_0}}\right)^2 \frac{1}{\theta_c} \frac{\Lambda_{eH_0}(\mu_e/\mu_h)}{[2A^2]\pi^{3/2}f^2_{(x,T_e)}T^2_{CMBR}\sigma_T^2(1+z)^4}
\label{eq:trueDA}
\end{equation}

$$\times\,\,\, \left[\frac{\Gamma(3\beta/2)}{\Gamma(3\beta/2 - 1/2)}\right]^2\frac{\Gamma(3\beta - 1/2)}{\Gamma(3\beta)}$$

Equation (\ref{eq:trueDA}) differs from the  corresponding equation (5) of (Carlson {\em et al} 2002) in that the factor in brackets $[2A^2]$ is replaced by the factor $4$.  In other words, there is a factor of $2$ error in (Carlson {\em et al} 2002) which is not a typo since the error is propagated from a factor of two error in equations (8) and (9) of an earlier paper by the same group (Patel {\em et al} 2000). This means that even if the effect I am expecting were absent, the measurement of $D_A$ and hence Hubble's constant should be off by a factor of 2.  Notice that the error is in {\it exactly} the correct direction that would work to cancel out the effect I am predicting.  This does not show that the effect I predict is present.  There is still a missing factor of $2$ to find if the effect I am predicting is to be seen.  I can only suggest that the SZ data be re-analyzed, this time with the additional factor $A$ as a free variable.  The factor of $2$ error in the analysis does show, however, that the predicted effect {\it may} have been seen in the SZE data.

\section{Simple Experiments to Test the CMBR = $SU(2)_L$ Gauge Field Hypothesis}

\bigskip
\subsection{Detecting an SU(2) Component Using Hans Dehmelt's Penning Trap} 
\bigskip

\begin{figure}
\includegraphics[width=5.0in]{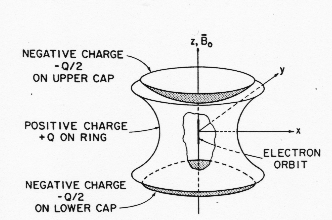}
\caption{\label{fig:Penning1}Basic Structure of the Penning Trap.  Figure is reproduced from Figure 1 of Dehmelt (1990b)}
\end{figure}

Hans Dehmelt's Penning Trap (Dehmelt 1990a,b) is the ideal
instrument to test the idea that the CMBR will not
interact with right-handed electrons.  The basic structure of the Penning Trap is pictured in Figure \ref{fig:Penning1}.

	In a Penning Trap, a single electron (or positron) is
captured by a vertical magnetic field, and an electric
field due to charges on a curved ring and two caps.  In
the Seattle Penning Trap, cap to cap separation is
about $0.8$ cm, the magnetic field $\vec B_0$ was 5
T.  The magnetic field results in a circular cyclotron
motion at $\nu_c = e{\vec B_0}/2\pi m_e = 141\, {\rm
GHz}$, where $e$ and $m_e$ are the electron charge
and electron mass respectively. The charge on the
ring and the caps is adjusted to give a weak
quadrupole field with potential well depth $D= 5\,
{\rm eV}$, yielding an axial oscillation frequency of 64
MHz. (The electron feels a simple harmonic restoring
force with spring constant $k = 2D/Z_0^2$), where
$2Z_0$ is the cap to cap separation.

\medskip
If two turns of nickel wire are wrapped around the ring
electrode, the large applied magnetic field magnetizes
this, and this ``bottle field'' interacts with the
electron's magnetic moment, allowing the spin of the
electron to be continuously measured.  This
``continuous Stern-Gerlach effect'' forces the electron
to be in one of its spin states, and it is possible to
determine which one the electron is in, and to measure
transitions between the two spin states.

\medskip
The energy  of the cyclotron motion of the electron is
quantized, with energy

$$E_n = (n+{1\over2})\,h\nu_c$$

At 4 K, observations give $<n> \approx 0.23$, and for
intervals of about 5 seconds, the electron is observed
in the $m=-1/2$ state or the $m=+1/2$ state (Dehmelt 1990a, p.
543).  With $\nu_z = 64\,{\rm MHz}$, this means that if
the state is chosen to be the $m=-1/2$, the electron
will have positive helicity for one-half the time for 128 million cycles --- while the
electron is moving down, and negative helicity for the other half of the time; that is,
when the electron is moving up. (Note that at the energies of the Penning trap, the distinction between helicity and chirality is important;  I shall deal with this below.)

\medskip
The electron can undergo a spin flip $\Delta n =0, \,m =
+1/2 \rightarrow -1/2$. This is actually the result
of two separate transitions:  the transition
$n= 0 \rightarrow 1$ is induced by the 4 K thermal
radiation, and transition is followed by the transition
$(n=1,\, m=-1/2) \rightarrow (n=0,\, m=+1/2)$ induced
by an applied rf field (Dehmelt 1990a, p. 543).

\medskip
The key point to note that the thermal transition,
were the electron with $m=-1/2$ to be immersed in the
CMBR thermal field and were the CMBR to be a pure
$SU(2)_L$ field, {\bf the ``thermal'' transition rate in a pure
$SU(2)_L$ field would be one-half the transition rate
in a pure electromagnetic heat bath}.  Thus the
Penning Trap can be used to determine whether the
CMBR is indeed pure EM, or instead pure $SU(2)_L$, as
I am claiming.  An experiment to test this would
introduce CMBR radiation via the gap between the ring
electrode and the cap electrodes.

\medskip
In the actual experiment, of course, the radiation
from the cap and ring electrodes would in fact be
thermal EM radiation, and this would induce
transitions at all times in the electron's motion.  If
there were no such radiation, the transition rate
would increase from an average of 5 seconds to 10
seconds, but the actual transition rate would be
proprotional to the ratio of area between the
electrodes to the area of the electrodes that face the
cavity where the electron moves.

\begin{figure}
\includegraphics[width=4.0in]{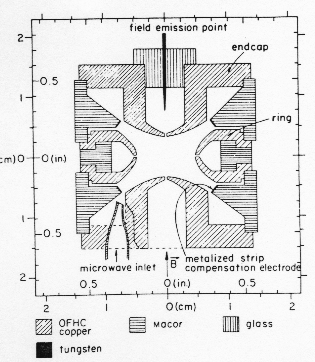}
\caption{\label{fig:Penning2}A More precise drawing of the Penning Trap.  Reproduced from (Brown and Gabrielse 1986, p. 235; Dehmmelt 1988, p. 108). }
\end{figure}

\medskip
From the drawing of the Penning Trap reference above,
one might infer that this ratio would be quite small,
but appearances can be deceiving.  A more precise
drawing of the Penning Trap is reproduced in Figure \ref{fig:Penning2}.

\medskip
In effect the electron is in the center of a spherical
region whose center is the Penning Trap.  Let us approximate the area ratio as
follows.  Take a sphere of radius $a$, and intersect it
with two coaxial cylinders, with the axis of both
passing through the center of the sphere.  Let the radii
of the two cylinders be $r_{in}$ and $r_{out}$, with
$r_{in} < r_{out} < a$.  Then the area of the sphere
between the cylinders is

$$A = 4\pi a^2\left[{ 1 + \sqrt{1-\left(r_{out}\over
a\right)^2} - \sqrt{1-\left(r_{in}\over
a\right)^2}}\right]$$

This area is a good approximation to the gap between
the ring electrode and the cap electrode.  If we feed
the signal from the CMBR through only the gap
between the upper cap and the ring electrode, then the
available surface area would be $1/2$ the above area. 
Making a rough measurement of Figure  \ref{fig:Penning2}), I obtained $A/4\pi a^2 = 0.49$, and if this is
accurate, as much as $1/4$ of the ``thermal'' radiation
inducing a state transition can be a signal from outside
the Penning Trap, assuming only the upper gap is used
(as the terminus of a circular wave guide).

\medskip
In other words, the outside signal will in effect be
transmitted through a coaxial wave guide, for which
the gap between the upper cap and the ring electrode is
the terminus.  Recall that a coaxial wave guide can
transmit TE and TM, as well as TEM waves.  The power
flow through a coaxial wave guide is calculated by
all physics graduate students (Jackson 1975, p. 385) to be

$$P = \left[c\over 4\pi\right]\sqrt{\mu\over\epsilon}
\pi r_{in}^2|H_{in}|^2\ln\left(r_{out}\over r_{in}
\right) = \left[c\over
4\pi\right]\sqrt{\mu\over\epsilon}
\pi r_{out}^2|H_{out}|^2\ln\left(r_{out}\over r_{in}
\right)$$

\noindent
Or, if $|H| = |E|$, as it will be for TEM waves, and we
assume Gaussian units (in which case the factor in
brackets comes into play and $\mu = \epsilon = 1$
for a vacuum or air wave guide), the power passing
through the wave guide will be

$$P = c\rho_{in}\pi r^2_{in}\ln\left(r_{in}\over
r_{out}\right) \approx 2c\rho_{av}A$$

\noindent
where $\rho_{in}$, $\rho_{av}$, and $A$ are the energy
density at the inner radius, the average energy density
in the annulus, and the area of the open annulus
respectively, and I have assumed that $(r_{out} -
r_{in})/r_{in} << 1$.  So the power flow from the
outside is just the flow on would expect through an
opening of the size of the gap; the walls have no
significant effect.

\medskip
Of course, the signal from outside the Penning Trap
will consist of radiation from several sources, only
one of which will be CMBR.  The other radiation
sources will be EM field radiation, and will couple to
right-handed electrons.  The various other sources are
discussed at length in (Partridge 1995, pp. 103--139), and a
shorter introduction to these other sources can be
found in the original papers, e.g. (Boynton {\em et al}, Roll and Wilkinson 1966). 
Remarkably, the main non-CMBR is 300 K radiation
from the ground, and if the detector is shielded from
this radiation --- easy to do with metal reflectors
preventing the ground radiation from reaching the
detector antenna --- then the other radiation sources
all have a radiation temperature of the order of a few
degrees K, and there are methods to measure them
independently, and thus subtract them out.

\medskip
For example, the atmosphere has a zenith temperature
of about 4 K, but as the optical depth depends on the
angle $z$ from the zenith as $\sec z$, the atmosphere
temperature goes as $T_{atm}(z) = T_{atm}(0)\sec z$,
and thus by making a series of measurements of the
total energy received by the antenna at several angles
$z$, the energy of the atmosphere can be subtracted
out (details in Partridge 1995, pp 120--121).

\medskip
Since the transtion probability $(n=0 \rightarrow
n=1)$ depends on the square of the cyclotron
frequency (Brown and Gabrielse 1986), the transition rate due to the CMBR will
be too low unless the frequency looked at is near the
5 T cyclotron frequency  $\nu_c = 141\, {\rm GHz}$. 
This is much higher than the window of 3 to 10 GHz
used in the classical CMBR measurements.  However,
there is an atmospheric window at 0.33 cm, or 91 GHz,
sufficiently near the 5 T cyclotron frequency that the
transition rate would be reduced only by a factor of
$(91/141)^2 = 0.42$, and the required 3.2 T Penning
Trap magnetic field should be easy to achieve.   The
CMBR observation at 91 GHz, however, is best
conducted at high altitudes (the first CMBR
measurement was conducted at the High Altitude
Observatory at Climax Colorado which was at an
altitude of $11,300, {\rm ft}$.   The instrument was
actually tested at Princeton University, where it was
built, but even in the winter, the water vapor at
Princeton made the measurement of the $\sec z$
atmosphere contribution difficult to eliminate (D.T.
Wilkinson, private communication).  But in principle,
the 91 GHz CMBR measurement could be done (though
with difficulty) even at Seattle or Cambridge, MA,
where Penning Traps are in operation.  It would better
done with the operational Penning Trap at Boulder, Colorado,
which is already art a high altitude, and the water
vapor is naturally low.

\medskip
Although I have emphasized that the first effect one
should search for with the Penning Trap is the
reduction in transition rate due to the fact that the
CMBR can interact with the Penning Trap electron only
for 1/2 the time, an even better test would be to
observe that the transition occurs only in that part of
the electron's orbit when the electron is left-handed, that is, when the electron has negative chirality.  I pointed out above that at high energy, chirality and helicity are the same, but this is not true at the low energies of the Penning Trap.  Recall that the helicity of an electron is a constant of the motion, but the chirality $\gamma^5$ varies with time as (\ref{eq:chiral}), yielding an expectation value of (\ref{eq:expect}).  For an electron at 10 eV, $|v|/c = 1/160$.  When a spin down electron is moving up, and when a spin up electron is moving down it will have negative helicity.  So in either of these cases, the electron would be more likely (by a factor of 1/160) to interact with a pure $SU(2)_L$ field than it would with opposite motion or spin.  With positrons, the situation would be reversed, since the
$SU(2)_L$ field can couple only to positive chirality
positrons.  However, such a measurement would be
difficult given the standard voltage between the cap
and ring electrodes, which yield the 64 MHz vertical
motion frequency.  But detecting this effect would be the goal of any Penning Trap CMBR experiment.

	Because it can in principle see the coupling between an $SU(2)_L$ field and negative chirality electrons, the Penning Trap is the ideal
instrument to determine whether or not the CMBR is
indeed a pure $SU(2)_L$ field, or simply an EM field. 
Unfortunately, setting up a Penning Trap to look for
the expected difference is quite an expensive
proposition; Hans Dehmelt indicated to me (private communication) that it would take \$250,000 or more to set up such an experiment, to say nothing of the difficulty of moving the instrument to
the best location, a dry desert high altitude plateau.

\bigskip
\subsection{Detecting an SU(2) Component With the Original
CMBR Detectors with Filters}

\bigskip
It would be nice if a quick and
inexpensive test of the pure $SU(2)_L$ hypothesis
could be found.  In this subsection, I shall outline such
an experiment, but I should caution the reader that the
proposed apparatus depends on estimates on the
behaviour of electrons in conductors and
semi-conductors when an $SU(2)_L$ field interacts
with electrons in such a material, and these estimates
might not be reliable.  So a null result {\it might} not
rule out the pure $SU(2)_L$ hypothesis.  On the plus side, a positive
result {\it would} confirm the $SU(2)_L$ hypothesis,
and the quick and dirty experiment I shall now propose
can be done with a simple modification of the original
apparatus set up by the Princeton group in 1965 to
detect the CMBR.  Even if the original apparatus no
longer exists, it can be re-built for the cost of at
most a few thousand dollars, and a single day's
observation should suffice to see if the effect is
present.

\medskip
The basic physical effect I shall use is as
follows.  Since a pure $SU(2)_L$ CMBR field will not
couple to electrons with positive chirality, a CMBR
wave will penetrate deeper into a conductor than an
ordinary EM wave, since in a conductor at room
temperature the conduction electrons have on the
average zero net chirality: half on the average have
positive chirality and the other half have negative
chirality.  I shall now show how to use this fact to
confirm that the CMBR is a pure $SU(2)_L$ field.

\medskip
The transmission coefficient for EM waves into an
infinite conducting plate with vacuum on either side
has been calcuated by Stratton (1941, pp. 511--516).  I
shall repeat his derivation because it will allow me to
point out some of the problems that may arise using
the filter experiment rather than the Penning Trap to
detect a pure $SU(2)_L$ CMBR.  

\medskip
Let us, following Stratton, imagine that we have three
arbitrary homogeneous media labeled (1), (2), and (3),
with dielectric constants $\epsilon_1$, $\epsilon_2$,
$\epsilon_3$, magnetic permeabilities $\mu_1$,
$\mu_1$, $\mu_1$, and propagation factors $k_1$,
$k_2$, and $k_3$ respectively.  The thickness of the
intermediate medium (2) will be $d$.  In medium (1), we
only have magnitudes of the incident and reflected
waves:

$$E_i = E_0e^{ik_1x - i\omega t}, \,\,\,\,\, H_i =
{k_1\over \omega\mu_1}E_i$$

$$E_r = E_1e^{-ik_1 x - i\omega t}, \,\,\,\,\, H_r =
-{k_1\over \omega\mu_1}E_r$$

The EM field in the middle medium (2) will contain
wave which are moving to the right and waves which
are moving to the left:

$$E_m = (E^+_2e^{ik_2x} + E^-_2e^{-ik_2x})e^{-i\omega
t}$$

$$H_m = {k_2\over \omega \mu_2}(E^+_2e^{ik_2x} -
E^-_2e^{-ik_2x})e^{-i\omega t}$$
	
\noindent
and finally the wave transmitted into medium (3) is 

$$E_t = E_3e^{ik_3x - i\omega t}, \,\,\,\,\, H_i =
{k_3\over \omega\mu_1}E_t$$

From these equations we can see one possible problem in
using a filter rather than a single electron to interact
with an $SU(2)_L$ field: there may be numerous
reflections at the boundaries between the three media,
and these many reflections may cause a pure
$SU(2)_L$ field to be converted into an EM field,
through interactions between left and right handed
electrons in the media themselves.

\medskip
I shall assume that this does not occur.  Stratton
points out that the boundary equations are easier
manipulate in terms of the following quantities:

$$E_j = \pm Z_jH_j , \,\,\,\,\, Z_j \equiv {{\omega
\mu_j}\over k_j} , \,\,\,\,\, Z_{jk} \equiv {Z_j\over
Z_k} = {{\mu_j k_k}\over {\mu_k k_j}}$$

The boundary conditions yield four equations between
five amplitudes:

$$E_0 + E_1 = E^+_2 + E^-_2$$

$$E_0 - E_1 = Z_{12}( E^+_2 - E^-_2)$$

$$E^+_2e^{ik_2d} + E^-_2e^{-ik_2d} = E_3e^{ik_3}d$$

$$E^+_2e^{ik_2d} - E^-_2e^{-ik_2d} =
Z_{23}E_3e^{ik_3}d$$

The transmission coefficent is $ T = |E_3/E_0|^2$, so it
is only necessary to solve for

$${E_3\over E_0} = {{4e^{-ik_3d}}\over
{(1-Z_{12})(1- Z_{23})e^{ik_2d} +
(1+Z_{12})(1+Z_{23})e^{-k_2d}}}$$

I shall simplify by setting $\mu_1 = \mu_2 = \mu_3 =
\mu_0$, where $\mu_0$ is the magnetic permeability
of free space, and assume that $\epsilon_1 =
\epsilon_3 = \epsilon_0$, where $\epsilon_0$ is the
dielectric constant of free space.  We have in this case

$$ k_1 = k_3 = {\omega\over c}$$

\noindent
where $c$ is the speed of light in a vacuum, and

$$k_2 = \alpha + i\beta$$

\noindent
where 

$$\alpha =
{\omega\over c}\left[\left({\mu_2\over\mu_0}\right)
\left({\epsilon_2\over\epsilon_0}\right)\sqrt{1+
\left({{\sigma}\over{\epsilon_2\omega}}\right)^2} +
1\right]^{1/2}$$

\noindent
and

$$\beta =
{\omega\over c}\left[\left({\mu_2\over\mu_0}\right)
\left({\epsilon_2\over\epsilon_0}\right)\sqrt{1+
\left({{\sigma}\over{\epsilon_2\omega}}\right)^2} -
1\right]^{1/2}$$

\noindent
where $\sigma$ is the conductivity of medium (2). 
(The formulae for $\alpha$ and $\beta$ simplify in the
cases of good conductors and bad conductors --- see
(Jackson 1975, pp. 297 --- but with {\it Mathematica}, it's just as easy
to use the general formulae).

\medskip
The electrical conductivity is given by

$$\sigma = {ne^2\tau\over m_e}$$

\noindent
where $n$ is the number of conduction electrons per
unit volume, $e$ is the electron charge, $\tau$ is the
relaxation time, and $m_e$ is the electron mass.  This
formula for the conductivity will be valid unless the
EM frequency is greater than $5\times 10^4$ GHz.  The
key fact I shall use is that as described above,

$$n_{SU(2)_L} = {1\over2}n_{EM}$$

\noindent
since an $SU(2)_L$ field would interact with only half
of the conduction electrons (this will be true even though helicity is not equal to chirality at low energy).

\medskip
For conductors and semi-conductors, almost all of an
incident EM (and $SU(2)_L$) wave will be reflected
unless the thickness $d$ of the filter (medium (2)) is
small relative to the skin depth.

\bigskip
The range of wavelengths for viewing the CMBR at sea
level is 3 cm to 10 cm, or 3 GHz to 10 GHz; the upper
frequency is determined by absorption from water
vapor, and the lower frequency by background sources
in the Galaxy.  The skin depth of Copper is 1.2 microns
at 3 GHz, $\mu_{Cu}/\mu_0 = \epsilon_{Cu}/
\epsilon_0 = 1$, and $\sigma = 5.80\times 10^7\, {\rm
mho/meter}$ (all numbers for $\mu$, $\epsilon$, and $\sigma$ are taken from Corson and Lorrain (1962, Table 10.1, p. 339).   The transmission coefficient for copper at 3 GHz is thus only $T = 0.0013$ even if $d = 10^{-3}\,({\rm skin\,
depth})$, or 12 angstroms --- the size of the copper
atom.  Clearly, no good conductor with a
conductivity comparable in magnitude to copper would
yield a detectable signal.  We want to find a material
with a transmission coefficient greater than a few per
cent for a thickness of at least 10 atoms, so that we
can trust the continuum approximation for medium (2).  I am assured by Tulane University's thin film physicists that 10 atoms is a sufficient thickness for the continuum approximation used here to be valid.

\medskip
The best material for use as a filter is Graphite, with $\mu_{C}/\mu_0 = \epsilon_{C}/
\epsilon_0 = 1$, and $\sigma = 1.0\times 10^5\, {\rm
mho/meter}$.  With a thickness of $d =
10^{-3}\,({\rm skin\, depth})$, or 290 angstroms ---
near the 100 atom thickness --- the transmission
coefficient at 3 Ghz is $T = 0.23$.  The idea would be to allow the CMBR to
pass through the filter, and with this latter thickness,
a graphite filter would transmit 23\% of the CMBR
flux if it were an electromagnetic field.  The test would be to measure the CMBR and a
reference 2.726 K reference EM source with the 
filter, using the above formulae for the relative
conductivities and for $\alpha$ and $\beta$.  More flux
will be detected from the CMBR if the pure $SU(2)_L$
hypothesis is correct.

\medskip
For a thickness of 290 angstroms, we have 

$$T_{EM}^{Graphite}(3\, {\rm GHz}) = 0.228$$

$$T_{SU(2)_L}^{Graphite}(3\, {\rm GHz}) = 0.318$$

which gives 

$${T^{Graphite}_{SU(2)_L}\over T_{EM}^{Graphite}}(3\, {\rm GHz}) = 1.39$$

\noindent
or a 39\% greater flux if the CMBR is a pure $SU(2)_L$
field.  This difference should be easily detectable.

\medskip
The corresponding transmission coefficients at 10 Ghz
is 

$$T_{EM}^{Graphite}(10\, {\rm GHz}) = 0.718$$

$$T_{SU(2)_L}^{Graphite}(10\, {\rm GHz}) = 0.785$$

which gives 

$${T^{Graphite}_{SU(2)_L}\over T^{Graphite}_{EM}}(10\, {\rm GHz}) = 1.09$$

\noindent
or a 9\% greater flux if the CMBR is a pure $SU(2)_L$
field.   This difference should be detectable, and notice that the difference is strongly frequency dependent.  This should make the mechanism clear.

\medskip
Since the filter is an absorber by construction, the filter will also be an emitter of black body radiation, with emissivity that will depend on the particular filter geometry used, and so will have to be determined experimentally, but which should be close to the values of the transmission coefficients which I compute for EM.

\medskip
For a Graphite filter whose thickness is measured in angstroms, the Graphite would have to be fixed on some substrate.  I would suggest using Sapphire.  Sapphire is considered ideal for use as a window through which microwaves must be transmitted (Bates 1992, http://www.tvu.com/LowLossSWinweb.htm).  Sapphire is an insulator (volume resistivity $10^{14}$ ohm-cm and a dielectric constant of 9.4 for electric field perpendicular to c-axis, and 11.5 for electric field paralle to c-axis at 300K).  Sapphire is optically active and birefringent with a large surface reflection unless polished.  Microwave power loss is $\sim \epsilon\tan\delta$.  The MTI Corporation (http://www.mticrystal.com/a2b.html) sells a 4 inch diameter, 0.5 mm thick cylinder, ideal for forming a substrate for a very thin graphite film, polished on both sides, for \$360.  These cylinders have a measured $\tan\delta$ loss at 10 GHz of less than $5\times 10^{-6}$ (see the MTI web site).  I have been advised by the above mentioned thin film physicists that Chemical Vapor Deposition (CVD) (using a hot substrate) is the best method to use to insert the thin Graphite film into the substrate, because this will insure a more uniform thickness.  (This will give a Graphite c-axis perpendicular to the substrate).  Physical Vapor Deposition (PVD) is alternative method.  There are many firms that specialize in CVD and PVD techniques.  A list can be found on the American Vacuum Society web page (www.avs.org) for example.

\medskip
I shall also illustrate the difference in transmission with the material I
think has the second best chance of giving an acceptable
filter, namely the element Germanium, for which
$\sigma = 2.1\, {\rm mho/meter}$ --- this
conductivity is the same at 9.2 GHz as at zero
frequency (Landolt and B\"ornstein (1982), Volume 17a Figure 55 on p. 414) --- and $\mu_{Ge}/\mu_0 = 1$. 
Unfortunately, we have $\epsilon_{Ge}/\epsilon_0 =
16.2$ at 300 K (the ratio is 16.0 at 4.2 K; see Landolt and B\"ornstein (1982), Volume 17a, p. 114)), and this
non-vacuum polarizability needs special
consideration.  According to the simple models of
molecular polarizability given in (Jackson 1975. pp. 152--158),
the small temperature difference in $\epsilon_{GE}$
indicates that most of the molecular polarizability is
due to a distortion in the charge distribution by the
applied electric field generating an induced dipole
moment in each molecule.  Since the electric field will
act on {\it bound} electrons rather than {\it free}
conduction electrons, the distortion on the left handed
electrons will be almost instantaneously transmitted
to the right handed electrons, with the result that the
molecular polarizability would be the same for both an
EM field and for a pure $SU(2)_L$ field, and thus the
dielectric constants would be the same in both cases;
on the conductivities would be different, and the ratio
of the two will be 1/2, as described above.  (Even if
the effective dielectric constant were to be different
for an EM and a pure $SU(2)_L$ field, the discussion on
page 154 of (Jackson 1975) makes it clear that it would vary in a
more complicated way than the conductivities, and
thus the transmission coefficients would be
measurably different for the two types of gauge field. 
For example, equation 4.70 of (Jackson 1975) yields
$N\gamma_{mol} = 1/5$ for Germanium in an EM field;
if $N\gamma_{mol} = 1/10$ in an $SU(2)_L$ field,
equation 4.70 yields $\epsilon/\epsilon_0 = 3.16$, a
substantial change from the value 16.2.)

\medskip
For a thickness of 1.6 millimeters, we have 

$$T_{EM}^{Ge}(3\, {\rm GHz}) = 0.106$$

$$T_{SU(2)_L}^{Ge}(3\, {\rm GHz}) = 0.132$$

which gives 

$${T^{Ge}_{SU(2)_L}\over T_{EM}^{Ge}}(3\, {\rm GHz}) = 1.24$$

\noindent
or a 24\% greater flux if the CMBR is a pure $SU(2)_L$
field.  

\medskip
The corresponding transmission coefficients at 10 Ghz
is 

$$T_{EM}^{Ge}(10\, {\rm GHz}) = 0.176$$

$$T_{SU(2)_L}^{Ge}(10\, {\rm GHz}) = 0.296$$

which gives 

$${T^{Ge}_{SU(2)_L}\over T^{Ge}_{EM}}(10\, {\rm GHz}) = 1.68$$

\noindent
or a 68\% greater flux if the CMBR is a pure $SU(2)_L$
field.  (The fluxes are greater at a higher frequency
than at the lower frequency --- opposite to what we
would expect for a good conductor, and opposite to what we in fact find in Graphite --- because
Germanium is a semi-conductor.)

\medskip
The central sky horn could be covered with the Germanium filter
--- mounting the filter inside the radiometer would
probably work, but there would be the risk of $U(1)$
field generation by second order effects in the wave
guide.  But with a low emissivity wave guide as is standard, this risk is tiny. 
	
\medskip
It is important to note that the above calculations
explain why the effect I am predicting have never
been seen before.  If the parts of the radiometers
designed to absorb CMBR are thicker than the skin
depth for the radiation absorbers --- and indeed all
such absorbers are much thicker --- then all the CBMR
would be absorbed even though the effective conduction
electron density is only 1/2 of the conduction electron
density seen by the usual EM field.  In fact, the CMBR
absorbers are made much thicker than the skin depth
precisely to insure that all incident radiation will be
absorbed, and this thickness hides the effect I am
predicting.  In 1938, the German physicist G. von Droste
bombarded uranium with neutrons, but carefully
covered the uranium with metal foil in order to
eliminate alpha particles which were expected to
occur.  It worked; but the foil also eliminated fission
fragments, which have a shorter range in metal foils
that alpha particles.  In the opinion of historians (e.g. (Graetzer and Anderson 1962, p.
41) the metal foils cost von Droste the Nobel Prize for
the discovery of nuclear fission.  The same
experimental technique also cost (Bernstein 2001, pp. 7--8)  Enrico
Fermi a Nobel for the discovery of nuclear fission. 
Fermi began the bombardment of uranium with neutrons
in 1935, but like Droste he covered his samples with
aluminum foil.  Once again the foil absorbed the fission
fragments that in the absence of the foil, Fermi's
instruments would have clearly seen.  In the end, 
fission was discovered in 1939 by Otto Hahn and Lise
Meitner, who used not standard particle detectors, but
instead the methods of radiochemistry.  All
investigations of the CMBR to date have used too thick
a ``foil'' and thus have missed the important effect I am
predicting.  Actually, as we shall be earlier, there
are measurements of the CMBR that are analogous to
radiochemistry in that the instruments used do not
``cover up'' the effect: these ``instruments'' are
ultrahigh energy cosmic rays and SZE observations, and I have argued that
they have already detected the effect I am predicting.

\medskip
It is possible that some early experiments
detected the expected difference between an EM CMBR
and an $SU(2)_L$ CMBR.  Two groups, one headed by
Gish and the other by Woody and Richards, measured
the CMBR using filters analogous to the filters I have
discussed above, and they detected (Partridge 1995, p. 142) an
excess flux above what was expected for a 2.7 K
blackbody EM field.  The Woody and Richards experiment
also saw a lower flux than a 2.7 at lower frequencies
(below the 2.7 blackbody peak), which is just what we
would expect from an $SU(2)_L$ CMBR field, as I
discussed above.

\bigskip
\subsection{Other Means of Detecting an SU(2) CMBR
Component} 
	
\bigskip
The Penning Trap is not the only way of observing an
interaction between the CMBR and a single electron.  A
Rydberg atom ---which are atoms in states with a
high principal quantum number $n$ --- can undergo
transitions induced by blackbody radiation at liquid
helium temperatures (Gallagher 1994, chapter 5), and hence such
atoms could be used to detect the effect I am
predicting, provided the Rydberg atom can fix its
transition electron in a definite spin state when the
atom's motion is observed.

\medskip
It has been occasionally suggested (e.g., (Ejnisman and Bigelow 1998) that the
devices which allowed the observation of Bose-Einstein
condensation --- the magneto-optical trap (MOT)
--- might be able to observe the CMBR.  A MOT is
designed to excite hyperfine transitions, and the
collective motion of the atoms in the trap is
observable, so in principle the effect I am predicting
{\it might} be observable with a MOT.  The problem
with using a MOT is that the cross-section for the
hyperfine transitions is so low that MOTs usually are
set up at room temperature, and at 300 K, the tail of
the blackbody radiation in the 3 to 10 GHz range is some
two orders of magnitude greater than the 2.7 K
distribution.  The fact that low temperature
experiments can be carried out in MOTs at room
temperature is the reason MOTs are so useful.  But
this very advantage of MOTs for many experiments
makes the MOT useless for observing the CMBR.  The
opinion of Cornell and Wieman (1999, p. 49) is that ``
$\ldots$ it is difficult to imagine that [blackbody
radiation] will ever be important for trapped atoms.''
	
\medskip
If the CMBR temperature were measured more precisely at high redshifts, then an electromagnetic CMBR would be distinguishable from a $SU(2)_L$ CMBR by the atomic transition mechanism described by John Bahcall (2000).  But although the increase in the (pseudo) temperature with redshift has been seen, precision is not yet sufficient to say whether the CMBR is electromagnetic or $SU(2)_L$.  See (Srianand {\em et al} 2000) for a description of the observations of the Bahcall effect.
	
\section{Solution to the Cosmological Constant Problem: The Universe and Life in the Far Future}

I have argued above that the Hawking evaporation
effect plus unitarity prevents the cosmolgical
constant from being exceedingly large, and in fact
requires that the effective cosmological constant, if
ever it becomes small but positive, must eventually
become zero or negative, since otherwise the universe
even if closed would expand forever, resulting in the
evaporation of the black holes which now exist,
violating unitarity.  What I shall now do is describe
the physical mechanism that will eventually
neutralize the observed currently positive effective
cosmological constant.  (See Peebles and Ratta (2003) for a recent review of the observations suggesting that the Dark Energy is in fact an uncanceled cosmological constant.)

\medskip
It is well-known that the mutual consistency of the
particle physics Standard Model and general relativity
requires the existence of a very large positive
cosmological constant.  The reason is simple:  the
non-zero vacuum expectation value for the Higgs 
field yields a vacuum energy density of $\sim - 1.0
\times 10^{26}\, {\rm gm/cm^3}(m_H/246)\, {\rm
GeV}$, where $m_H$ is the Higgs boson mass, which is $114.4\, {\rm GeV}< m_H < 251\, {\rm GeV}$ at the 95\% confidence level (Abazov {\em et al} 2004).  Since
this is a negative vacuum energy, it is accompanied by
a positive pressure of equal magnitude, and both the
pressure and energy yield a negative cosmological
constant.  Since the closure density is $1.88\times
10^{-29}\,\Omega_{tot}h^2{\rm gm/cm^3}$, and
observations (Spergel {\em et al} 2003) indicate that  $\Omega_{tot} = 1.02 \pm0.02$ and $h=
0.71 \pm0.04$, there must be a fundamental {\it positive}
cosmological constant to cancel out the negative
cosmological constant coming from the Higgs field. 
What we observe accelerating the universe today is the
sum of the fundamental positive cosmological
constant, and the negative Higgs field cosmological
constant;  this sum is the ``effective'' cosmological
constant.

\medskip
What we would expect is that these two cosmological
constant would exactly cancel, and what must be
explained is why they do not: the vacuum energy
coming from the Higgs field --- more generally, the
 sum of the vacuum energies of all the physical fields
--- is observed to be slightly less in magnitude than
the magnitude of the fundamental positive cosmological
constant.  What must be explained therefore, is why
the vacuum energy sum is slighly less than expected.

\medskip
I have argued above that the instanton tunnelling that to generate a net baryon
number also results in the universe being in a false
vacuum slightly above the true vacuum, where, as
expected, the fundamental cosmological constant and
the vacuum energy densities of all the physical fields
do indeed cancel.

\medskip
Recall that the instanton tunnelling works by
non-perturbatively changing the global winding number
of the $SU(2)_L$ field; the winding number is equal to
the number of fermions in the universe.  There is also
a winding number associated with the $SU(3)$ color
force, and the color vacuum --- the
$\theta$-vacuum --- is a weighed sum over all the
winding numbers:  $|\theta> = \sum_n
e^{-in\theta}|n>$.  The fact that $\theta$ is
observed to be essentially  zero is of course the
``strong CP problem'' which I resolved above.

\medskip
There is no necessary connection between the winding
numbers of $SU(2)_L$ and color $SU(3)_L$, but in
fact $\pi_3(G) = Z$ for any compact connected Lie
group $G$, where $\pi_3(G)$ is the third homotopy
group of $G$, expressing that there are non-trivial
mapping of the three-sphere into $G$.  There are thus
{\it three} 3-spheres in cosmology and the Standard
Model:  (1) electroweak $SU(2)_L$ itself, (3) subgroups
of color $SU(3)$ and (3) the spatial 3-sphere.  I
propose that the non-zero winding number due to
mapping of $SU(2)_L$ into itself gives rise to a false
vacuum in one or all of these three, and that the true
vacuum corresponds to a winding number of zero.

\medskip
This means that as long as the number of fermions
minus anti-fermions remains constant on the spatial
3-sphere, the physical fields will remain in the false
vacuum, the effective cosmological constant will
remain positive, and the universe will continue to
accelerate.  Conversely, if instanton tunnelling occurs
in reverse, so that the fermion number of the universe
decreases, then the false vacuum will decrease to the
true vacuum, a state which I have assumed has an
energy density which cancels the positive fundamental
cosmological constant. In the present epoch of
universal history, the winding number remains
constant --- the tunneling probability is very small in
the present epoch --- and thus the sum of the false
vacuum energy density and the fundamental
cosmological constant, this sum being the {\it dark
energy} since it is seen only gravitationally, is
constant.

\medskip
But in the long run, it cannot remain constant, since an
unchanging positive dark energy would cause the
universe to accelerate forever, violating unitarity
when the observed black holes evaporate if we accept Hawking's argument.  In addition, the mechanism I described above for stabilizing QFT and forcing quantum gravity to be finite, would not work if the universe expanded forever.  The stabilization mechanism requires a final singularity.  Since the proton lifetime due to electroweak instanton
tunnelling is greater than the solar mass black hole
lifetime, something must act in the future to speed up
the tunnelling probability.

\medskip
I propose that life itself acts to annihilate protons
and other fermions via induced instanton tunnelling. 
Barrow and I have established that the main source of
energy for information processing in the expanding phase of the far future
will be the conversion of the mass of fermions into
energy.  Baryon number conservation prevents this
process from being 100\% efficient, but since the
Standard Model allows baryon non-conservation via
instanton tunnelling, I assume that some means can
and will be found in the far future to allow life to
speed up this process.  So once again the existence of
intelligent life in the far future is required for the
consistency of the laws of physics, since in the
absence of life acting to speed up fermion
annihilation, the universe would accelerate forever,
violating unitarity, forcing quantum field theory to diverge, and incidentally, extinguishing life.

But if life goes on forever, with the computer capacity increasing without limit, then life in the far future could emulate the present state of the universe down to the quantum state, and indeed in such an emulation, the structure of the world would be a mere manifestation of pure number.

\section{Conclusion}

	The attempt to infer the structure of the world from pure number has led us to consider the converse, namely perhaps pure number is a consequence of the structure of the world.  I have argued that the problems of the Standard Model disappear if the unique quantum gravity theory which results from applying QFT to general relativity's Hilbert action is used, together with the cosmological boundary conditions that make these two theories consistent.  I have outlined how these two theories naturally answer most of the objections that have been raised against these two theories, for instance, why do the constants of the SM have the values they are observed to have:  all possible values of the constants are obtained in some universe of the multiverse at some time.  I have also shown that the flatness, isotropy, and horizon problems are naturally resolved, and that there is a natural mechanism that gives a scale free perturbation spectrum on super-horizon scales.  I have pointed out that some of the consequences of the theory may have already been observed in the SZE, and the existence of UHE cosmic rays may be yet another manifestation of the consistency of the SM with quantum gravity.  I have described how a simple modification of the original devices used to detect the CMBR could be used to confirm the theory, and how one could use a Penning trap to test the theory.  There is some life remaining in the ``old'' SM and quantum gravity based on the metric of spacetime (and perhaps some life even in the far future of the universe).  In fact, these theories may be able to explain ALL observations, even the dark matter and dark energy, and why there is more matter than antimatter.  We may have already discovered the long sought Theory of Everything!

\section{Acknowledgements}

I am very grateful for helpful conversations with John Barrow, James Bryan, Hans Dehmelt, Ulrike Diebold, Keith Devlin, Paul Frampton, Morris Kalka, Anthony Lasenby, George Rosensteel,  David Raub, Martin Rees and the late David Wilkinson.

\end{document}